# Defeasible Inclusions in Low-Complexity DLs

**Piero A. Bonatti**                                                    BONATTI@NA.INFN.IT
**Marco Faella**                                                        MFAELLA@NA.INFN.IT
**Luigi Sauro**                                                         SAURO@NA.INFN.IT
*Dipartimento di Scienze Fisiche,*
*Università di Napoli "Federico II"*

## Abstract

Some of the applications of OWL and RDF (e.g. biomedical knowledge representation and semantic policy formulation) call for extensions of these languages with nonmonotonic constructs such as inheritance with overriding. Nonmonotonic description logics have been studied for many years, however no practical such knowledge representation languages exist, due to a combination of semantic difficulties and high computational complexity. Independently, low-complexity description logics such as DL-lite and $\mathcal{EL}$ have been introduced and incorporated in the OWL standard. Therefore, it is interesting to see whether the syntactic restrictions characterizing DL-lite and $\mathcal{EL}$ bring computational benefits to their nonmonotonic versions, too. In this paper we extensively investigate the computational complexity of Circumscription when knowledge bases are formulated in DL-lite$_R$, $\mathcal{EL}$, and fragments thereof. We identify fragments whose complexity ranges from P to the second level of the polynomial hierarchy, as well as fragments whose complexity raises to PSPACE and beyond.

## 1. Introduction

The ontologies at the core of the semantic web — as well as ontology languages such as RDF, OWL, and related Description Logics (DLs) — are founded on fragments of first-order logic and inherit strengths and weaknesses of this well-established formalism. Limitations include monotonicity, and the consequent inability to design knowledge bases (KBs) by describing prototypes whose general properties can be later refined with suitable exceptions. This natural, iterative approach is commonly used by biologists and calls for an extension of DLs with defeasible inheritance with overriding (a mechanism normally supported by object-oriented languages). Some workarounds have been devised for particular cases; however, no general solutions are currently available (Rector, 2004; Stevens, Aranguren, Wolstencroft, Sattler, Drummond, Horridge, & Rector, 2007). Another motivation for nonmonotonic DLs stems from the recent development of policy languages based on DLs (Uszok, Bradshaw, Jeffers, Suri, Hayes, Breedy, Bunch, Johnson, Kulkarni, & Lott, 2003; Finin, Joshi, Kagal, Niu, Sandhu, Winsborough, & Thuraisingham, 2008; Zhang, Artale, Giunchiglia, & Crispo, 2009; Kolovski, Hendler, & Parsia, 2007). DLs nicely capture role-based policies and facilitate the integration of semantic web policy enforcement with reasoning about semantic metadata (which is typically necessary in order to check policy conditions). However, in order to formulate standard default policies such as *open* and *closed* policies,[1] and support common policy language features such as authorization inheritance with exceptions (which is meant to facilitate incremental policy formulation), it is necessary to adopt a nonmonotonic semantics; Bonatti and Samarati (2003) provide further details on the matter.

---

1. If no explicit authorization has been specified for a given access request, then an open policy permits the access while a closed policy denies it.

Given the increasing size of semantic web ontologies and RDF bases, the complexity of reasoning is an influential factor that may either foster or prevent the adoption of a knowledge representation language. Accordingly, OWL2 introduces *profiles* that adopt syntactic restrictions (compatible with application requirements) in order to make reasoning tractable. Two of such profiles are based on the following families of DLs: *DL-lite* (Calvanese, De Giacomo, Lembo, Lenzerini, & Rosati, 2005), that formalizes RDFS, and $\mathcal{EL}$ (Baader, 2003; Baader, Brandt, & Lutz, 2005), that extensively covers important biomedical ontologies such as GALEN and SNOMED. Unfortunately, in general, nonmonotonic DL reasoning can be highly complex and reach NExpTime$^{\text{NP}}$ and even 3-ExpTime (Donini, Nardi, & Rosati, 1997, 2002; Bonatti, Lutz, & Wolter, 2009). A natural question, in this context, is whether restrictions such as those adopted by DL-lite and $\mathcal{EL}$ help in reducing the complexity of nonmonotonic DL reasoning, too.

Answering this question is the main goal of this paper. We extensively investigate the complexity of reasoning in DL-lite and $\mathcal{EL}$. The nonmonotonic semantics adopted is Circumscription (McCarthy, 1980), whose main appealing properties (discriminating Circumscription from other nonmonotonic DL semantics proposed in the literature) are summarized below:

1. Circumscription is compatible with all the interpretation domains supported by classical DLs; there is no need for adopting a fixed domain of standard names;

2. In circumscribed DLs, nonmonotonic inferences apply to all individuals, including those that are not denoted by any constants and are implicitly asserted by existential quantifiers;

3. Circumscription naturally supports priorities among conflicting nonmonotonic axioms and can easily simulate specificity-based overriding.

As an attempt to simplify the usage of circumscribed DLs and simultaneously remove potential sources of computational complexity, we do not support the usage of *abnormality predicates* (McCarthy, 1986) in their full generality; we rather hide them within *defeasible inclusions* (Bonatti, Faella, & Sauro, 2009). Defeasible inclusions are expressions $C \sqsubseteq_n D$ whose intuitive meaning is: *an instance of $C$ is normally an instance of $D$*. Such inclusions can be prioritized to resolve conflicts. Priorities can be either explicit or automatically determined by the inclusion's *specificity*, i.e. a defeasible inclusion $C_1 \sqsubseteq_n D_1$ may override $C_2 \sqsubseteq_n D_2$ if $C_1$ is classically subsumed by $C_2$. In this framework, we prove that restricting the syntax to DL-lite inclusions suffices—in almost all cases—to reduce complexity to the second level of the polynomial hierarchy. On the contrary, circumscribed $\mathcal{EL}$ is still ExpTime-hard and further restrictions are needed to confine complexity within the second level of the polynomial hierarchy. Syntactic restrictions will be analyzed in conjunction with other semantic parameters, such as the kind of priorities adopted (explicit or specificity-based), and which predicates may or may not be affected by Circumscription (i.e., fixed and variable predicates, in Circumscription's jargon).

The paper is organized as follows: First, the basics of low-complexity description logics and their extension based on circumscription are recalled in Section 2 and Section 3, respectively. Then, some reductions that can be used to eliminate language features and work on simpler frameworks are illustrated in Section 4. After an undecidability result caused by fixed roles (Section 5), the paper focuses on variable roles: The complexity of circumscribed DL-lite$_R$ and $\mathcal{EL}/\mathcal{EL}^{\perp}$ are investigated in Section 6 and Section 7, respectively. A section on related work and a final discussion conclude the paper.

| Name | Syntax | Semantics |
|------|--------|-----------|
| inverse role | $R^-$ | $(R^-)^{\mathcal{I}} = \{(d,e) \mid (e,d) \in R^{\mathcal{I}}\}$ |
| nominal | $\{a\}$ | $\{a^{\mathcal{I}}\}$ |
| negation | $\neg C$ | $\Delta^{\mathcal{I}} \setminus C^{\mathcal{I}}$ |
| conjunction | $C \sqcap D$ | $C^{\mathcal{I}} \cap D^{\mathcal{I}}$ |
| existential restriction | $\exists R.C$ | $\{d \in \Delta^{\mathcal{I}} \mid \exists (d,e) \in R^{\mathcal{I}} : e \in C^{\mathcal{I}}\}$ |
| top | $\top$ | $\top^{\mathcal{I}} = \Delta^{\mathcal{I}}$ |
| bottom | $\bot$ | $\bot^{\mathcal{I}} = \emptyset$ |

Figure 1: Syntax and semantics of some DL constructs.

## 2. Preliminaries

In DLs, *concepts* are inductively defined with a set of *constructors*, starting with a set $\mathsf{N_C}$ of *concept names*, a set $\mathsf{N_R}$ of *role names*, and (possibly) a set $\mathsf{N_I}$ of *individual names* (all countably infinite). We use the term *predicates* to refer to elements of $\mathsf{N_C} \cup \mathsf{N_R}$. Hereafter, letters $A$ and $B$ will range over $\mathsf{N_C}$, $P$ will range over $\mathsf{N_R}$, and $a, b, c$ will range over $\mathsf{N_I}$. The concepts of the DLs dealt with in this paper are formed using the constructors shown in Figure 1. There, the inverse role constructor is the only role constructor, whereas the remaining constructors are concept constructors. Letters $C, D$ will range over concepts and letters $R, S$ over (possibly inverse) roles.

The semantics of the above concepts is defined in terms of *interpretations* $\mathcal{I} = (\Delta^{\mathcal{I}}, \cdot^{\mathcal{I}})$. The *domain* $\Delta^{\mathcal{I}}$ is a non-empty set of individuals and the *interpretation function* $\cdot^{\mathcal{I}}$ maps each concept name $A \in \mathsf{N_C}$ to a set $A^{\mathcal{I}} \subseteq \Delta^{\mathcal{I}}$, each role name $P \in \mathsf{N_R}$ to a binary relation $P^{\mathcal{I}}$ on $\Delta^{\mathcal{I}}$, and each individual name $a \in \mathsf{N_I}$ to an individual $a^{\mathcal{I}} \in \Delta^{\mathcal{I}}$. The extension of $\cdot^{\mathcal{I}}$ to inverse roles and arbitrary concepts is inductively defined as shown in the third column of Figure 1. An interpretation $\mathcal{I}$ is called a *model* of a concept $C$ if $C^{\mathcal{I}} \neq \emptyset$. If $\mathcal{I}$ is a model of $C$, we also say that $C$ is *satisfied* by $\mathcal{I}$.

A *(strong) knowledge base* is a finite set of (i) *concept inclusions (CIs)* $C \sqsubseteq D$ where $C$ and $D$ are concepts, (ii) *concept assertions* $A(a)$ and *role assertions* $P(a, b)$, where $a, b$ are individual names, $P \in \mathsf{N_R}$, and $A \in \mathsf{N_C}$, (iii) *role inclusions (RIs)* $R \sqsubseteq R'$. An interpretation $\mathcal{I}$ *satisfies* (i) a CI $C \sqsubseteq D$ if $C^{\mathcal{I}} \subseteq D^{\mathcal{I}}$, (ii) an assertion $C(a)$ if $a^{\mathcal{I}} \in C^{\mathcal{I}}$, (iii) an assertion $P(a, b)$ if $(a^{\mathcal{I}}, b^{\mathcal{I}}) \in P^{\mathcal{I}}$, and (iv) a RI $R \sqsubseteq R'$ iff $R^{\mathcal{I}} \subseteq R'^{\mathcal{I}}$. Then, $\mathcal{I}$ is a *model* of a strong knowledge base $\mathcal{S}$ iff $\mathcal{I}$ satisfies all the elements of $\mathcal{S}$. We write $C \sqsubseteq_{\mathcal{S}} D$ iff for all models $\mathcal{I}$ of $\mathcal{S}$, $\mathcal{I}$ satisfies $C \sqsubseteq D$.

*Terminologies* are particular strong knowledge bases consisting of *definitions*, i.e. axioms such as $A \equiv C$, that abbreviate the inclusions $A \sqsubseteq C$ and $C \sqsubseteq A$. If a terminology $\mathcal{T}$ contains the above definition, then we say that $A$ is *defined* in $\mathcal{T}$ and that $C$ is the *definition of A*. Each $A$ defined in $\mathcal{T}$ must have a unique definition. A concept name $A$ *directly depends* on $B$ (in $\mathcal{T}$) if $B$ occurs in $A$'s definition; moreover, $A$ *depends* on $B$ (in $\mathcal{T}$) if there is a chain of such direct dependencies leading from $A$ to $B$. A terminology $\mathcal{T}$ is *acyclic* if no $A$ depends on itself in $\mathcal{T}$. Terminologies are conservative extensions, and the concept names defined in an acyclic terminology $\mathcal{T}$ can be eliminated by *unfolding* them w.r.t. $\mathcal{T}$, that is, by exhaustively replacing the concepts defined in $\mathcal{T}$ with their definition. For all expressions (i.e., concepts or inclusions) $E$, we denote with $\mathsf{unf}(E, \mathcal{T})$ the unfolding of $E$ w.r.t. $\mathcal{T}$.

The logic DL-lite$_R$ (Calvanese et al., 2005) restricts concept inclusions to expressions $C_L \sqsubseteq C_R$, where

$$C_L \quad ::= \quad A \mid \exists R \qquad R \quad ::= \quad P \mid P^- \qquad C_R \quad ::= \quad C_L \mid \neg C_L$$

(as usual, $\exists R$ abbreviates $\exists R.\top$).

The logic $\mathcal{EL}$ (Baader, 2003; Baader et al., 2005) restricts knowledge bases to assertions and concept inclusions built from the following constructs:

$$C \quad ::= \quad A \mid \top \mid C_1 \sqcap C_2 \mid \exists P.C$$

(note that inverse roles are not supported). The extension of $\mathcal{EL}$ with $\bot$, role hierarchies, and nominals (respectively) are denoted by $\mathcal{EL}^\bot$, $\mathcal{ELH}$, and $\mathcal{ELO}$. Combinations are allowed: for example $\mathcal{ELHO}$ denotes the extension of $\mathcal{EL}$ supporting role hierarchies and nominals. Finally, $\mathcal{EL}^{\neg A}$ denotes the extension where negation can be applied to concept names.

## 3. Defeasible Knowledge

A *general defeasible inclusion* (GDI) is an expression $C \sqsubseteq_n D$ whose intended meaning is: *C's elements are normally in D.*

**Example 3.1** (Bonatti et al., 2009) The sentences: *"in humans, the heart is usually located on the left-hand side of the body; in humans with situs inversus, the heart is located on the right-hand side of the body"* (Rector, 2004; Stevens et al., 2007) can be formalized with the $\mathcal{EL}^\bot$ axioms and GDIs:

$$\texttt{Human} \sqsubseteq_n \exists\texttt{has\_heart}.\exists\texttt{has\_position}.\texttt{Left};$$
$$\texttt{Situs\_Inversus} \sqsubseteq \exists\texttt{has\_heart}.\exists\texttt{has\_position}.\texttt{Right};$$
$$\exists\texttt{has\_heart}.\exists\texttt{has\_position}.\texttt{Left} \sqcap$$
$$\exists\texttt{has\_heart}.\exists\texttt{has\_position}.\texttt{Right} \sqsubseteq \bot.$$

$\square$

A *defeasible knowledge base* (DKB) in a logic $\mathcal{DL}$ is a pair $\langle \mathcal{K}, \prec \rangle$, where $\mathcal{K} = \mathcal{K}_S \cup \mathcal{K}_D$, $\mathcal{K}_S$ is a strong $\mathcal{DL}$ KB, $\mathcal{K}_D$ is a set of GDIs $C \sqsubseteq_n D$ such that $C \sqsubseteq D$ is a $\mathcal{DL}$ inclusion, and $\prec$ is a strict partial order (a priority relation) over $\mathcal{K}_D$. In the following, by $C \sqsubseteq_{[n]} D$ we denote an inclusion that is either classical or defeasible. Moreover, for a DKB $\mathcal{KB} = \langle \mathcal{K} \cup \mathcal{T}, \prec \rangle$, where $\mathcal{T}$ is a (classical) acyclic terminology, we denote by $\mathsf{unf}(\mathcal{KB}) = \langle \mathcal{K}', \prec' \rangle$ the DKB where $\mathcal{K}'$ is the unfolding of all inclusions in $\mathcal{K}$ w.r.t. $\mathcal{T}$, and, for all DIs $\delta, \delta'$ in $\mathcal{K}$, the relation $\mathsf{unf}(\delta, \mathcal{T}) \prec' \mathsf{unf}(\delta', \mathcal{T})$ holds if and only if $\delta \prec \delta'$.

As priority relation we shall often adopt the *specificity* relation $\prec_{\mathcal{K}}$ which is determined by classically valid inclusions. Formally, for all GDIs $\delta_1 = (C_1 \sqsubseteq_n D_1)$ and $\delta_2 = (C_2 \sqsubseteq_n D_2)$, let

$$\delta_1 \prec_{\mathcal{K}} \delta_2 \text{ iff } C_1 \sqsubseteq_{\mathcal{K}_S} C_2 \text{ and } C_2 \not\sqsubseteq_{\mathcal{K}_S} C_1.$$

**Example 3.2** The access control policy: "*Normally users cannot read project files; staff can read project files; blacklisted staff is not granted any access*" can be encoded with:

$$\texttt{Staff} \sqsubseteq \texttt{User}$$
$$\texttt{Blacklisted} \sqsubseteq \texttt{Staff}$$
$$\texttt{UserRequest} \sqsubseteq \exists\texttt{subj.User} \sqcap \exists\texttt{target.Proj} \sqcap \exists\texttt{op.Read}$$
$$\texttt{StaffRequest} \sqsubseteq \exists\texttt{subj.Staff} \sqcap \exists\texttt{target.Proj} \sqcap \exists\texttt{op.Read}$$
$$\texttt{UserRequest} \sqsubseteq_n \exists\texttt{decision.\{Deny\}}$$
$$\texttt{StaffRequest} \sqsubseteq_n \exists\texttt{decision.\{Grant\}}$$
$$\exists\texttt{subj.Blacklisted} \sqsubseteq \exists\texttt{decision.\{Deny\}}$$
$$\exists\texttt{decision.\{Grant\}} \sqcap \exists\texttt{decision.\{Deny\}} \sqsubseteq \bot \,.$$

Staff members cannot simultaneously satisfy the two defeasible inclusions (due to the last inclusion above). With specificity, the second defeasible inclusion *overrides* the first one and yields the intuitive inference that non-blacklisted staff members are indeed allowed to access project files. More formally, the subsumption

$$\exists\texttt{subj.(Staff} \sqcap \neg\texttt{Blacklisted)} \sqcap \exists\texttt{target.Proj} \sqcap \exists\texttt{op.Read} \sqsubseteq \exists\texttt{decision.\{Grant\}}$$

holds in all the models of the above knowledge base (as defined below). □

Intuitively, a model of $\langle \mathcal{K}, \prec \rangle$ is a model of $\mathcal{K}_S$ that maximizes the set of individuals satisfying the defeasible inclusions in $\mathcal{K}_D$, resolving conflicts by means of the priority relation $\prec$ whenever possible. In formalizing the notion of model, one should specify how to deal with the predicates occurring in the knowledge base: is their extension allowed to vary in order to satisfy defeasible inclusions? A discussion of the effects of letting predicates vary vs. fixing their extension can be found in the work of Bonatti, Lutz and Wolter (2006); they conclude that the appropriate choice is application dependent. So, in general, the set of predicates $\mathsf{N_C} \cup \mathsf{N_R}$ can be arbitrarily partitioned into two sets $F$ and $V$ containing fixed and varying predicates, respectively; we denote this semantics by $\mathsf{Circ}_F^*$.

However, in Section 5 it is shown that fixed roles cause undecidability issues, so most of our results concern a specialized framework in which all role names are varying predicates, that is, $F \subseteq \mathsf{N_C}$. We use the notation $\mathsf{Circ}_F$ (rather than $\mathsf{Circ}_F^*$) to indicate that $F \subseteq \mathsf{N_C}$.

The set $F$, the GDIs $\mathcal{K}_D$, and the priority relation $\prec$ induce a strict partial order over interpretations. As we move down the ordering we find interpretations that are more and more normal w.r.t. $\mathcal{K}_D$. For all $\delta = (C \sqsubseteq_n D)$ and all interpretations $\mathcal{I}$ let the set of individuals *satisfying* $\delta$ be:

$$\mathsf{sat}_{\mathcal{I}}(\delta) = \{x \in \Delta^{\mathcal{I}} \mid x \notin C^{\mathcal{I}} \text{ or } x \in D^{\mathcal{I}}\} \,.$$

**Definition 3.3** Let $\mathcal{KB} = \langle \mathcal{K}, \prec \rangle$ be a DKB. For all interpretations $\mathcal{I}$ and $\mathcal{J}$, and all $F \subseteq \mathsf{N_C} \cup \mathsf{N_R}$, let $\mathcal{I} <_{\mathcal{KB},F} \mathcal{J}$ iff:

1. $\Delta^{\mathcal{I}} = \Delta^{\mathcal{J}}$;

2. $a^{\mathcal{I}} = a^{\mathcal{J}}$, for all $a \in \mathsf{N_I}$;

3. $A^{\mathcal{I}} = A^{\mathcal{J}}$, for all $A \in F \cap \mathsf{N_C}$, and $P^{\mathcal{I}} = P^{\mathcal{J}}$, for all $P \in F \cap \mathsf{N_R}$;

4. for all $\delta \in \mathcal{K}_{\mathrm{D}}$, if $\mathsf{sat}_{\mathcal{I}}(\delta) \not\supseteq \mathsf{sat}_{\mathcal{J}}(\delta)$ then there exists $\delta' \in \mathcal{K}_{\mathrm{D}}$ such that $\delta' \prec \delta$ and $\mathsf{sat}_{\mathcal{I}}(\delta') \supset \mathsf{sat}_{\mathcal{J}}(\delta')$;

5. there exists a $\delta \in \mathcal{K}_{\mathrm{D}}$ such that $\mathsf{sat}_{\mathcal{I}}(\delta) \supset \mathsf{sat}_{\mathcal{J}}(\delta)$.

The subscript $\mathcal{KB}$ will be omitted when clear from context. Now a *model* of a DKB can be defined as a maximally preferred model of its strong (i.e. classical) part.

**Definition 3.4** Let $\mathcal{KB} = \langle \mathcal{K}, \prec \rangle$ and $F \subseteq \mathsf{N}_{\mathsf{C}} \cup \mathsf{N}_{\mathsf{R}}$. An interpretation $\mathcal{I}$ is a *model* of $\mathsf{Circ}^*_F(\mathcal{KB})$ iff $\mathcal{I}$ is a (classical) model of $\mathcal{K}_{\mathrm{S}}$ and for all models $\mathcal{J}$ of $\mathcal{K}_{\mathrm{S}}$, $\mathcal{J} \not\prec_F \mathcal{I}$.

**Remark 3.5** This semantics is a special case of the circumscribed DLs introduced by Bonatti et al. (2006). The correspondence can be seen by (i) introducing for each GDI $C \sqsubseteq_n D$ a fresh atomic concept $Ab$, playing the role of an abnormality predicate; (ii) replacing $C \sqsubseteq_n D$ with $C \sqcap \neg Ab \sqsubseteq D$; (iii) minimizing all the predicates $Ab$ introduced above.

In order to enhance readability, we will use the following notation for the special cases in which all concept names are varying and the case in which they are all fixed: $<_{\mathsf{var}}$ and $\mathsf{Circ}_{\mathsf{var}}$ stand for $<_{\emptyset}$ and $\mathsf{Circ}_{\emptyset}$, respectively; $<_{\mathsf{fix}}$ and $\mathsf{Circ}_{\mathsf{fix}}$ stand respectively for $<_{\mathsf{N}_{\mathsf{C}}}$ and $\mathsf{Circ}_{\mathsf{N}_{\mathsf{C}}}$. For a DKB $\mathcal{KB} = \langle \mathcal{K}_{\mathrm{S}} \cup \mathcal{K}_{\mathrm{D}}, \prec \rangle$, we say that an interpretation $\mathcal{I}$ is a *classical model* of $\mathcal{KB}$ in case $\mathcal{I}$ is a model of $\mathcal{K}_{\mathrm{S}}$.

In this paper, we consider the following standard reasoning tasks over defeasible DLs:

**Knowledge base consistency**  Given a DKB $\mathcal{KB}$, decide whether $\mathsf{Circ}^*_F(\mathcal{KB})$ has a model.

**Concept consistency**  Given a concept $C$ and a DKB $\mathcal{KB}$, check whether $C$ is *satisfiable w.r.t. $\mathcal{KB}$*, that is, whether a model $\mathcal{I}$ of $\mathsf{Circ}^*_F(\mathcal{KB})$ exists such that $C^{\mathcal{I}} \neq \emptyset$.

**Subsumption**  Given two concepts $C$, $D$ and a DKB $\mathcal{KB}$, check whether $\mathsf{Circ}^*_F(\mathcal{KB}) \models C \sqsubseteq D$, that is, whether for all models $\mathcal{I}$ of $\mathsf{Circ}^*_F(\mathcal{KB})$, $C^{\mathcal{I}} \subseteq D^{\mathcal{I}}$.

**Instance checking**  Given $a \in \mathsf{N}_{\mathsf{I}}$, a concept $C$, and a DKB $\mathcal{KB}$, check whether $\mathsf{Circ}^*_F(\mathcal{KB}) \models C(a)$, that is, whether for all models $\mathcal{I}$ of $\mathsf{Circ}^*_F(\mathcal{KB})$, $a^{\mathcal{I}} \in C^{\mathcal{I}}$.

The following example illustrates most of the above tasks as well as the main difference between $\mathsf{Circ}_{\mathsf{var}}$ and $\mathsf{Circ}_{\mathsf{fix}}$.

**Example 3.6**  Consider the following simplification of Example 3.2:

> ```
> User ⊑ₙ ¬∃decision.{Grant}
> Staff ⊑ User
> Staff ⊑ₙ ∃decision.{Grant}
> BlacklistedStaff ⊑ Staff ⊓ ¬∃decision.{Grant} .
> ```

Extend the knowledge base with the assertion `Staff(John)`, and let the priority relation be $\prec_{\mathcal{K}}$ (i.e., priorities are determined by specificity). Denote the resulting knowledge base with $\mathcal{KB}$. Due to the inclusion `Staff ⊑ User`, the GDI for `Staff` (third line) has higher priority than the GDI for `User` (first line). Therefore, in all models $\mathcal{I}$ of $\mathsf{Circ}_{\mathsf{var}}(\mathcal{KB})$, John belongs to $\exists\mathtt{decision}.\{\mathtt{Grant}\}$ and hence the following entailments hold:

$$\mathsf{Circ}_{\mathsf{var}}(\mathcal{KB}) \quad\models\quad \{\mathtt{John}\} \sqsubseteq \exists\mathtt{decision}.\{\mathtt{Grant}\} \quad\text{(subsumption)} \tag{1}$$

$$\mathsf{Circ}_{\mathsf{var}}(\mathcal{KB}) \quad\models\quad \exists\mathtt{decision}.\{\mathtt{Grant}\}(\mathtt{John}) \quad\text{(instance checking)} \tag{2}$$

Interestingly, John does not belong to $\mathtt{BlacklistedStaff}$, because this is the only way of satisfying the top-priority GDI for $\mathtt{Staff}$. Analogously, in all models $\mathcal{I}$ of $\mathsf{Circ}_{\mathsf{var}}(\mathcal{KB})$, John is the only member of $\mathtt{Staff}$ because this setting maximizes the number of individuals satisfying both GDIs (as all the individuals in $\neg\mathtt{Staff}$ vacuously satisfy the GDI for $\mathtt{Staff}$ for all values of $\mathtt{decision}$). More generally, as a side effect of the maximization of all $\mathsf{sat}_{\mathcal{I}}(\delta)$, $\mathsf{Circ}_{\mathsf{var}}$ induces a sort of closed-world assumption over all concepts with exceptional properties (w.r.t. some larger concept). Consequently, $\mathtt{BlacklistedStaff}$ is *not* satisfiable w.r.t. $\mathcal{KB}$, and the following subsumption holds:

$$\mathsf{Circ}_{\mathsf{var}}(\mathcal{KB}) \models \mathtt{Staff} \sqsubseteq \{\mathtt{John}\}\,. \tag{3}$$

On the contrary, under $\mathsf{Circ}_{\mathsf{fix}}$, $\mathtt{User}$ and $\mathtt{Staff}$ may contain any number of individuals (other than zero) because $\mathsf{Circ}_{\mathsf{fix}}$ is not allowed to change the extension of any atomic concept, even if this would satisfy more GDIs. Similarly, there exist models of $\mathsf{Circ}_{\mathsf{fix}}(\mathcal{KB})$ where John does not belong to $\exists\mathtt{decision}.\{\mathtt{Grant}\}$ because John belongs to $\mathtt{BlacklistedStaff}$ and $\mathsf{Circ}_{\mathsf{fix}}$ does not allow to change its extension to satisfy more GDIs. As a consequence, it can be easily verified that $\mathtt{BlacklistedStaff}$ is satisfiable w.r.t. $\mathsf{Circ}_{\mathsf{fix}}(\mathcal{KB})$, and that (1), (2), and (3) do not hold if $\mathsf{Circ}_{\mathsf{var}}$ is replaced by $\mathsf{Circ}_{\mathsf{fix}}$. We only have inferences such as:

$$\mathsf{Circ}_{\mathsf{fix}}(\mathcal{KB}) \quad\models\quad \mathtt{User} \sqcap \neg\mathtt{Staff} \sqsubseteq \neg\exists\mathtt{decision}.\{\mathtt{Grant}\}\,, \tag{4}$$

$$\mathsf{Circ}_{\mathsf{fix}}(\mathcal{KB}) \quad\models\quad \mathtt{Staff} \sqcap \neg\mathtt{BlacklistedStaff} \sqsubseteq \exists\mathtt{decision}.\{\mathtt{Grant}\}\,. \tag{5}$$

$\square$

Note that in $\mathsf{Circ}_{\mathsf{var}}$ one can obtain nominals (cf. $\mathtt{Staff}$ in (3)) without using nominals explicitly in the knowledge base. If other axioms do not interfere, then an assertion $A(a)$ and a GDI $A \sqsubseteq_n \bot$ suffice to make $A$ a singleton. This idea will be used in some reductions later on.

The next example deals with multiple inheritance, and in particular with parent concepts with conflicting properties.

**Example 3.7** Let $\mathcal{KB}$ consist of the axioms:

$$
\begin{aligned}
\mathtt{Whale} &\sqsubseteq \mathtt{Mammal} \sqcap \mathtt{SeaAnimal} \\
\mathtt{Mammal} &\sqsubseteq_n \exists\mathtt{has\_organ}.\mathtt{Lungs} \\
\mathtt{SeaAnimal} &\sqsubseteq_n \exists\mathtt{has\_organ}.\mathtt{Gills} \\
\exists\mathtt{has\_organ}.\mathtt{Lungs} \sqcap &\ \exists\mathtt{has\_organ}.\mathtt{Gills} \sqsubseteq \bot\,,
\end{aligned}
$$

where the priority relation is specificity. Note that mammals and sea animals have conflicting default properties. In all models of $\mathsf{Circ}_{\mathsf{var}}(\mathcal{KB})$ $\mathtt{Whale}$ is empty, because this is the only way of having both GDIs satisfied by *all* individuals. Now let $\mathcal{KB}' = \mathcal{KB} \cup \{\mathtt{Whale(Moby)}\}$. In each model of $\mathsf{Circ}_{\mathsf{var}}(\mathcal{KB}')$, Moby satisfies as many GDIs as possible, that is, exactly one of the two GDIs of $\mathcal{KB}'$. As a consequence, we have the reasonable inference:[2]

$$\mathsf{Circ}_{\mathsf{var}}(\mathcal{KB}') \models \{\mathtt{Moby}\} \sqsubseteq \exists\mathtt{has\_organ.Lungs} \sqcup \exists\mathtt{has\_organ.Gills}\,.$$

The conflict between the two default properties inherited by Moby can be settled by adding a simple axiom like Whale $\sqsubseteq \exists\mathtt{has\_organ.Lungs}$, that overrides the property of sea animals. In this specific example a strong axiom is appropriate; note, however, that the corresponding GDI would have the same effect under $\prec_K$; for instance, we have:

$$\mathsf{Circ}_{\mathsf{var}}(\mathcal{KB}' \cup \{\mathtt{Whale} \sqsubseteq_n \exists\mathtt{has\_organ.Lungs}\}) \models \{\mathtt{Moby}\} \sqsubseteq \exists\mathtt{has\_organ.Lungs}\,.$$

Consider $\mathsf{Circ}_{\mathsf{fix}}$ now. The expected subsumptions Mammal $\sqsubseteq \exists\mathtt{has\_organ.Lungs}$ and SeaAnimal $\sqsubseteq \exists\mathtt{has\_organ.Gills}$ are *not* entailed by $\mathsf{Circ}_{\mathsf{fix}}(\mathcal{KB})$, because Lungs and Gills are empty in some models (as $\mathsf{Circ}_{\mathsf{fix}}$ cannot change their extension to satisfy more GDIs). The two GDIs could be enabled by forcing Lungs and Gills to be nonempty. This can be done in several ways, e.g. via assertions such as Lungs(L) or inclusions such as $\top \sqsubseteq \exists\mathtt{aux.Lungs}$ (where aux is a new auxiliary role). Let $\mathcal{KB}''$ denote such an extension. Then

$$\mathsf{Circ}_{\mathsf{fix}}(\mathcal{KB}'') \models \{\mathtt{Moby}\} \sqsubseteq \exists\mathtt{has\_organ.Lungs} \sqcup \exists\mathtt{has\_organ.Gills}\,,$$

(similarly, the aforementioned expected consequences are entailed by $\mathsf{Circ}_{\mathsf{fix}}(\mathcal{KB}'')$). The conflict between the properties inherited from Mammal and SeaAnimal can be settled as discussed above. □

The impossibility of forcing existentials with GDIs in $\mathsf{Circ}_{\mathsf{fix}}$, illustrated by the above example, can be exploited to check whether a concept is always nonempty. It suffices to introduce a fresh role $aux$ (in order to prevent interference with the other axioms of the knowledge base) and a GDI $\top \sqsubseteq_n \exists aux.C$. Clearly, the subsumption $\top \sqsubseteq \exists aux.C$ is entailed iff $C$ is nonempty in all models of $\mathsf{Circ}_{\mathsf{fix}}(\mathcal{KB})$. Similar ideas will be used in the rest of the paper.

The next example is artificial. It is a convenient way of illustrating the interplay of specificity and multiple inheritance.

**Example 3.8** Let $\mathcal{KB}$ be the following set of axioms:

$$
\begin{array}{llr@{\qquad}llr}
A_1 & \sqsubseteq & A_1' & (6) & A_1 \sqsubseteq_n \exists R_1 & (12) \\
A_2 & \sqsubseteq & A_2' & (7) & A_2 \sqsubseteq_n \exists R_2 & (13) \\
B & \sqsubseteq & A_1 \sqcap A_2 & (8) & A_1' \sqsubseteq_n \exists R_1' & (14) \\
\exists R_1 & \sqcap & \exists R_1' \sqsubseteq \bot & (9) & A_2' \sqsubseteq_n \exists R_2' & (15) \\
\exists R_2 & \sqcap & \exists R_2' \sqsubseteq \bot & (10) & & \\
\exists R_1 & \sqcap & \exists R_2 \sqsubseteq \bot & (11) & &
\end{array}
$$

For all sets of concept names $F$, and for all models $\mathcal{I}$ of $\mathsf{Circ}_F(\mathcal{KB})$, each member $x$ of $B^{\mathcal{I}}$ (if any) satisfies exactly one of the top priority GDIs (12) and (13), that are conflicting due to (11). If $x$ does not satisfy (12) then it can satisfy the conflicting GDI (14); symmetrically, if $x$ does not satisfy (13) then $x$ can satisfy (15). Consequently, we have:

$$\mathsf{Circ}_{\mathsf{fix}}(\mathcal{KB}) \models B \sqsubseteq (\exists R_1 \sqcap \exists R_2') \sqcup (\exists R_2 \sqcap \exists R_1').$$

---

2. The symbol $\sqcup$ is description logics' equivalent of disjunction. Formally, $(C \sqcup D)^{\mathcal{I}} = C^{\mathcal{I}} \cup D^{\mathcal{I}}$.

□

The last two examples show that GDIs and disjointness constraints together can express disjoint unions. Similar techniques will be used later on to simulate the law of excluded middle, negation, and 3-valued logic.

Subsumption, instance checking, and the complement of concept satisfiability can be reduced to each other, as in the classical setting:

**Theorem 3.9** *Let $\mathcal{DL}$ range over DL-lite$_R$ and $\mathcal{EL}^\perp$; let $X = \mathsf{var}, \mathsf{fix}, F$. In $\mathsf{Circ}_X(\mathcal{DL})$, subsumption, instance checking, and concept unsatisfiability can be reduced to each other in constant time.*

The proof is not completely standard, due to the limited expressiveness of DL-lite$_R$ and $\mathcal{EL}^\perp$, as well as the peculiarities of nonmonotonic reasoning.[3]

**Proof.** First we focus on $\mathsf{Circ}_{\mathsf{var}}$ and $\mathsf{Circ}_F$, where $F$ consists of concept names occurring in a given $\mathcal{KB}$.

*From unsatisfiability to subsumption.* Checking unsatisfiability of a concept $C$ can be reduced to checking the subsumption $C \sqsubseteq \perp$. DL-lite$_R$ does not support $\perp$ explicitly, however an equivalent concept $A_\perp$ can be easily defined with the inclusion $A_\perp \sqsubseteq \neg A_\perp$.

*From subsumption to unsatisfiability.* Conversely, a subsumption $C \sqsubseteq D$ can be reduced to the unsatisfiability of $C \sqcap \neg D$. In DL-lite$_R$ conjunction is not supported, so the subsumption must be reduced to the unsatisfiability of a fresh variable concept $A$ axiomatized by $A \sqsubseteq C$ and $A \sqsubseteq \neg D$. In $\mathcal{EL}^\perp$ conjunction is supported while negation is not; therefore the given subsumption can be reduced to the unsatisfiability of $C \sqcap \bar{D}$ where $\bar{D}$ is a fresh variable concept axiomatized with $\bar{D} \sqcap D \sqsubseteq \perp$.

*From instance checking to subsumption.* An instance checking query $C(a)$ can be reduced to subsumption as follows: Introduce a fresh variable concept $A$ and assert $A(a)$; then minimize $A$ with $A \sqsubseteq_n \perp$; now in all models $\mathcal{I}$ of $\mathsf{Circ}_{\mathsf{var}}$, $A^\mathcal{I} = \{a^\mathcal{I}\}$. It follows that $C(a)$ holds iff the subsumption $A \sqsubseteq C$ holds.

*From unsatisfiability to instance checking.* Finally, the unsatisfiability of a concept $C$ is equivalent to the validity of the instance checking problem $\neg C(a)$, where $a$ is a fresh individual constant. In $\mathcal{EL}^\perp$, $\neg C$ must be suitably axiomatized with a fresh concept name $\bar{C}$ and the inclusions $\bar{C} \sqcap C \sqsubseteq \perp$, $\top \sqsubseteq_n C$, and $\top \sqsubseteq_n \bar{C}$ (these three axioms entail the subsumption $\top \sqsubseteq C \sqcup \bar{C}$, thereby enforcing the law of the excluded middle). In order to preserve the semantics of the knowledge base, $\top \sqsubseteq_n C$ and $\top \sqsubseteq_n \bar{C}$ must be given a priority strictly smaller than the priority of all the other defeasible inclusions in the KB. This ensures that the new GDIs cannot block the application of any of the original GDIs. Clearly, the two new GDIs must have the same priority.

---

3. For example, in classical logic a subsumption $C \sqsubseteq D$ is a logical consequence of $\mathcal{KB}$ iff for any fresh individual $a$, $D(a)$ is a logical consequence of $\mathcal{KB} \cup \{C(a)\}$. This approach is not correct for Circumscription. The models of $\mathsf{Circ}_F(\mathcal{KB})$ can be quite different from the models of $\mathsf{Circ}_F(\mathcal{KB} \cup \{C(a)\})$; for instance, consider the example in which nonmonotonic reasoning makes `Whale` empty and the assertion `Whale(Moby)` overrides this inference.

This completes the proof for $\mathsf{Circ}_{\mathsf{var}}$ and $\mathsf{Circ}_F$. The proof for $\mathsf{Circ}_{\mathsf{fix}}$ can be obtained by replacing the fresh variable concept names $A$, $\bar{C}$, and $\bar{D}$ with a corresponding (variable) concept $\exists R$, where $R$ is a fresh role. □

Note that the above reductions still apply if priorities are specificity-based ($\prec_{\mathcal{K}}$), with the exception of the reduction of concept unsatisfiability to instance checking in $\mathcal{EL}^{\perp}$. For this case, one can use Theorem 4.6 below to eliminate general priorities, and get a reduction for $\mathsf{Circ}_{\mathsf{fix}}$.

# 4. Complexity Preserving Features

In some cases, nonmonotonic inferences and language features—e.g. variable predicates and explicit priorities—do not affect complexity. In this section several such results (and related lemmata) are collected; the reader is warned that, in general, they may not apply to all reasoning tasks and all language fragments. We start by observing that the logics we deal with enjoy the finite model property.

**Lemma 4.1** *Let $\mathcal{KB} = \langle \mathcal{K}, \prec \rangle$ be a DKB in DL-lite$_R$ or $\mathcal{ELHO}^{\perp,\neg}$. For all $F \subseteq \mathsf{N_C}$, $\mathsf{Circ}_F(\mathcal{KB})$ has a model only if $\mathsf{Circ}_F(\mathcal{KB})$ has a finite model whose size is exponential in the size of $\mathcal{KB}$.*

**Proof.** A simple adaptation of a result for $\mathcal{ALCIO}$ (Bonatti et al., 2006), taking role hierarchies into account. □

As a consequence, these logics preserve classical consistency (because all descending chains of models originating from a finite model must be finite):

**Theorem 4.2** *Let $\mathcal{KB} = \langle \mathcal{K}_{\mathrm{D}} \cup \mathcal{K}_{\mathrm{S}}, \prec \rangle$ be a DKB in DL-lite$_R$ or $\mathcal{ELHO}^{\perp,\neg}$. For all $F \subseteq \mathsf{N_C}$, $\mathcal{K}_{\mathrm{S}}$ is (classically) consistent iff $\mathsf{Circ}_F(\mathcal{KB})$ has a model.*

**Remark 4.3** Obviously, a similar property holds for all circumscribed DLs with the finite model property, including $\mathcal{ALCIO}$ and $\mathcal{ALCQO}$.

Since knowledge base consistency is equivalent to its classical version, it will not be discussed in this paper any further.

Next, we prove that under mild assumptions, $\mathsf{Circ}_F$ is not more expressive than $\mathsf{Circ}_{\mathsf{fix}}$ (which is a special case of the former), that is, variable concept names do not increase the expressiveness of the logic and can be eliminated.[4]

**Theorem 4.4** *If $\mathcal{DL}$ is a description logic fully supporting unqualified existential restrictions ($\exists R$),[5] then, for all $F \subseteq \mathsf{N_C}$, concept consistency, subsumption, and instance checking in $\mathsf{Circ}_F(\mathcal{DL})$ can be reduced in linear time to concept consistency, subsumption, and instance checking (respectively) in $\mathsf{Circ}_{\mathsf{fix}}(\mathcal{DL})$.*

---

4. The standard techniques for eliminating variable predicates (Cadoli, Eiter, & Gottlob, 1992) use connectives that are not fully supported in DL-lite$_R$ and $\mathcal{EL}$, therefore an ad-hoc proof is needed.

5. We say that $\mathcal{DL}$ fully supports unqualified restrictions if they can occur wherever a concept name could.

**Proof.** Let $\mathcal{KB}$ be any given DKB in the language $\mathcal{DL}$. Introduce a new role name $R_A$ for each (variable) concept name $A \notin F$. Then, replace each occurrence of $A$ in $\mathcal{KB}$ with $\exists R_A$ and call $\mathcal{KB}'$ the resulting KB. Recall that in $\mathsf{Circ}_{\mathsf{fix}}(\mathcal{DL})$ all concept names are fixed and all roles are variable. Hence, the newly added roles $R_A$ behave in $\mathsf{Circ}_{\mathsf{fix}}(\mathcal{KB}')$ exactly in the same way as concepts $A \notin F$ do in $\mathsf{Circ}_F(\mathcal{KB})$. Formally, there is a bijection between the models of $\mathsf{Circ}_F(\mathcal{KB})$ and the models of $\mathsf{Circ}_{\mathsf{fix}}(\mathcal{KB}')$, which preserves the interpretation of all role and concept names, except that the extension of concept names $A \notin F$ in a model of $\mathsf{Circ}_F(\mathcal{KB})$ coincides with the domain of the corresponding role $R_A$ in the corresponding model of $\mathsf{Circ}_{\mathsf{fix}}(\mathcal{KB}')$. As a consequence, the consistency of a concept $C$ w.r.t. $\mathsf{Circ}_F(\mathcal{KB})$ is equivalent to the consistency of $C'$ w.r.t. $\mathsf{Circ}_{\mathsf{fix}}(\mathcal{KB}')$, where $C'$ is obtained from $C$ by replacing each occurrence of $A \notin F$ with the corresponding $\exists R_A$. Similarly for subsumption and instance checking. $\qquad\square$

Symmetrically, the next theorem proves that in $\mathcal{EL}^{\perp}$ fixed predicates can be eliminated using general priorities. The reduction adapts the classical encoding of fixed predicates to the limited expressiveness of $\mathcal{EL}^{\perp}$.

**Theorem 4.5** *For all* $F \subseteq \mathsf{N_C}$, *concept consistency, subsumption, and instance checking in* $\mathsf{Circ}_F(\mathcal{EL}^{\perp})$ *can be reduced in linear time to concept consistency, subsumption, and instance checking (respectively) in* $\mathsf{Circ}_{\mathsf{var}}(\mathcal{EL}^{\perp})$ *with general priorities.*

**Proof.** Let $\mathcal{K} = \langle \mathcal{K}_S \cup \mathcal{K}_D, \prec \rangle$ be a given $\mathcal{EL}^{\perp}$ DKB. Fixed predicates are removed through the following transformation. For each concept name $A \in F$ introduce a new concept name $\bar{A}$ (representing $\neg A$). Let $\mathcal{K}_S^*$ be the set of all disjointness axioms $A \sqcap \bar{A} \sqsubseteq \perp$, for each $A \in F$. Let $\mathcal{K}_D^*$ be the set of all defeasible inclusions $\top \sqsubseteq_n A$ and $\top \sqsubseteq_n \bar{A}$, for each $A \in F$. Finally, let $\prec'$ be the minimal extension of $\prec$ such that for all $\delta^* \in \mathcal{K}_D^*$ and all $\delta \in \mathcal{K}_D$, $\delta^* \prec' \delta$. Define

$$\mathcal{K}' = \langle \mathcal{K}_S \cup \mathcal{K}_S^* \cup \mathcal{K}_D \cup \mathcal{K}_D^*, \prec' \rangle .$$

*Claim 1. Let $\mathcal{J}$ and $\mathcal{J}'$ be two classical models of $\mathcal{K}_S \cup \mathcal{K}_S^*$ such that $\mathcal{J}' <_{\mathcal{K}',\mathsf{var}} \mathcal{J}$ and for all $A \in F$, $\bar{A}^{\mathcal{J}} = \neg A^{\mathcal{J}}$. Then (i) for all $\delta \in \mathcal{K}_D^*$, $\mathsf{sat}^{\mathcal{J}'}(\delta) = \mathsf{sat}^{\mathcal{J}}(\delta)$ and (ii) $\mathcal{J}$ and $\mathcal{J}'$ agree on all $A \in F$.*

Proof of Claim 1: By definition of $\prec'$, the members of $\mathcal{K}_D^*$ have maximal priority, therefore, for all $\delta \in \mathcal{K}_D^*$, $\mathsf{sat}^{\mathcal{J}'}(\delta) \supseteq \mathsf{sat}^{\mathcal{J}}(\delta)$. If there exists $A \in F$ such that $\mathsf{sat}^{\mathcal{J}'}(\top \sqsubseteq_n A) \supset \mathsf{sat}^{\mathcal{J}}(\top \sqsubseteq_n A)$, then $A^{\mathcal{J}'} \supset A^{\mathcal{J}}$, and hence $\bar{A}^{\mathcal{J}'} \subset \bar{A}^{\mathcal{J}}$; consequently $\mathsf{sat}^{\mathcal{J}'}(\top \sqsubseteq_n \bar{A}) \subset \mathsf{sat}^{\mathcal{J}}(\top \sqsubseteq_n \bar{A})$ (a contradiction). Symmetrically, the assumption that $\mathsf{sat}^{\mathcal{J}'}(\top \sqsubseteq_n \bar{A}) \supset \mathsf{sat}^{\mathcal{J}}(\top \sqsubseteq_n \bar{A})$ leads to a contradiction. This proves (i); (ii) is a straightforward consequence of (i).

*Claim 2. Every model $\mathcal{I}$ of $\mathsf{Circ}_F(\mathcal{K})$ can be extended to a model $\mathcal{J}$ of $\mathsf{Circ}_{\mathsf{var}}(\mathcal{K}')$.*

To prove this claim, extend $\mathcal{I}$ to $\mathcal{J}$ by setting $\bar{A}^{\mathcal{J}} = \neg A^{\mathcal{I}}$, for all concept names $A \in F$. Suppose that $\mathcal{J}$ is not a model of $\mathsf{Circ}_{\mathsf{var}}(\mathcal{K}')$. Since $\mathcal{J}$ satisfies $\mathcal{K}_S \cup \mathcal{K}_S^*$ by construction, there must be a $\mathcal{J}'$ that satisfies $\mathcal{K}_S \cup \mathcal{K}_S^*$ and such that $\mathcal{J}' <_{\mathcal{K}',\mathsf{var}} \mathcal{J}$. By Claim 1.(ii), $\mathcal{J}$ and $\mathcal{J}'$ agree on all $A \in F$; by Claim 1.(i), the improvement of $\mathcal{J}'$ over $\mathcal{J}$ concerns the GDIs in $\mathcal{K}_D$. It follows that $\mathcal{I}' <_{\mathcal{K},F} \mathcal{I}$, where $\mathcal{I}'$ is the restriction of $\mathcal{J}'$ to the language of $\mathcal{K}$. This contradicts the assumption that $\mathcal{I}$ is a model of $\mathsf{Circ}_F(\mathcal{K})$.

*Claim 3. For all models $\mathcal{J}$ of $\mathsf{Circ}_{\mathsf{var}}(\mathcal{K}')$, the restriction of $\mathcal{J}$ to the language of $\mathcal{K}$ is a model of $\mathsf{Circ}_F(\mathcal{K})$.*

Let $\mathcal{I}$ be the restriction of $\mathcal{J}$ to the language of $\mathcal{K}$. Clearly $\mathcal{I}$ satisfies $\mathcal{K}_S$. Now suppose that $\mathcal{I}$ is not a model of $\mathsf{Circ}_F(\mathcal{K})$, which means that there exists $\mathcal{I}'$ satisfying $\mathcal{K}_S$ and such that $\mathcal{I}' <_{\mathcal{K},F} \mathcal{I}$.

Extend $\mathcal{I}'$ to $\mathcal{J}'$ by setting $\bar{A}^{\mathcal{J}'} = \neg A^{\mathcal{I}'}$, for all concept names $A \in F$. Note that $\mathcal{I}$ and $\mathcal{I}'$ must agree on all $A \in F$, therefore $\mathcal{J}$ and $\mathcal{J}'$ agree on them, too. Consequently, for all $\delta \in \mathcal{K}_{\mathrm{D}}^*$, $\mathsf{sat}^{\mathcal{J}'}(\delta) = \mathsf{sat}^{\mathcal{J}}(\delta)$. Moreover, $\mathcal{I}'$ improves $\mathcal{I}$ over the GDIs of $\mathcal{K}_{\mathrm{D}}$, therefore $\mathcal{J}'$ improves $\mathcal{J}$ over $\mathcal{K}_{\mathrm{D}}$, too. It follows that $\mathcal{J}' <_{\mathcal{K}',\mathsf{var}} \mathcal{J}$ (a contradiction).

The Theorem is now a straightforward consequence of Claims 2 and 3. $\qquad\square$

Now consider priority relations and GDIs. We are going to prove that if the language fragment is sufficiently rich, then they can be simulated with the specificity-based relation $\prec_{\mathcal{K}}$ and normalized defeasible inclusions $A \sqsubseteq_n C$ (whose left-hand side is a concept name), respectively.

Let $\mathcal{KB} = \langle \mathcal{K}, \prec \rangle$ be any given DKB in $\mathcal{EL}^{\perp}$. First we need to define a new fixed concept $Dom$ that encodes the domain without being equivalent to $\top$. This requires the following transformation:

$$A^* = Dom \sqcap A \qquad\qquad (\exists R.C)^* = Dom \sqcap \exists R.(C^*)$$
$$\top^* = Dom \qquad\qquad (C \sqcap D)^* = C^* \sqcap D^*$$
$$\perp^* = \perp \qquad\qquad (C \sqsubseteq_{[n]} D)^* = C^* \sqsubseteq_{[n]} D^*$$

Obtain $\mathcal{K}^*$ from $\mathcal{K}$ by transforming all inclusions in $\mathcal{K}$ and by adding a nonemptiness axiom $\top \sqsubseteq \exists aux.Dom$ ($aux$ a fresh role) plus an assertion $Dom(a)$ for each $a \in \mathsf{N}_{\mathsf{I}}$ occurring in $\mathcal{K}$. It is not hard to see that the restrictions to $Dom$ of the classical models of $\mathcal{K}^*$ correspond to the classical models of $\mathcal{K}$. More precisely, let $\mathcal{I}^*$ be a classical model of $\mathcal{K}^*$, we obtain a classical model $\mathcal{I}$ of $\mathcal{K}$ by setting $A^{\mathcal{I}} = A^{\mathcal{I}^*} \cap Dom^{\mathcal{I}^*}$ and $R^{\mathcal{I}} = R^{\mathcal{I}^*} \cap (Dom^{\mathcal{I}^*} \times Dom^{\mathcal{I}^*})$, for each concept name $A$ and role name $R$ occurring in $\mathcal{K}$. Notice that it is necessary for $Dom^{\mathcal{I}^*}$ to be non-empty for this to work. Conversely, given a classical model $\mathcal{I}$ of $\mathcal{K}$, it is sufficient to set $Dom^{\mathcal{I}} = \Delta^{\mathcal{I}}$ and $aux^{\mathcal{I}} = \Delta^{\mathcal{I}} \times \Delta^{\mathcal{I}}$ to make $\mathcal{I}$ a classical model of $\mathcal{K}^*$.

Now we have to remove general priorities and GDIs. For all GDIs $\delta = (C \sqsubseteq_n D) \in \mathcal{K}^*$, add two fresh predicates $A_\delta, P_\delta$ and replace $\delta$ with the following axiom schemata:

$$Dom \sqsubseteq A_\delta \qquad\qquad A_{\delta'} \sqsubseteq A_\delta \quad \text{for all } \delta' \prec \delta \qquad (16)$$
$$A_\delta \sqsubseteq_n \exists P_\delta \qquad\qquad \exists P_\delta \sqcap C \sqsubseteq D \qquad\qquad\qquad (17)$$

Call the new DKB $\mathcal{KB}' = \langle \mathcal{K}', \prec_{\mathcal{K}'} \rangle$. By (16), the specificity-based relation $\prec_{\mathcal{K}'}$ prioritizes the new GDIs according to the original priorities. Moreover, by (17), a defeasible inclusion $A_\delta \sqsubseteq_n \exists P_\delta$ is satisfied by an individual if and only if the same individual satisfies the corresponding GDI $\delta$. Then it is not difficult to verify that all reasoning tasks such that none of the new predicates $A_\delta$ and $P_\delta$ occur in the query yield the same answer in $\langle \mathcal{K}^*, \prec^* \rangle$ and $\mathcal{KB}'$. As a consequence of the above discussion, by combining the transformation $\cdot^*$ with (16) and (17), and by observing that the reduction makes use of $\mathcal{EL}^{\perp}$ constructs only, we have:

**Theorem 4.6** *Reasoning in* $\mathsf{Circ}_{\mathsf{fix}}(\mathcal{EL}^{\perp})$ *with explicit priorities and GDIs can be reduced in polynomial time to reasoning in* $\mathsf{Circ}_{\mathsf{fix}}(\mathcal{EL}^{\perp})$ *with only specificity-based priority and defeasible inclusions of the form* $A \sqsubseteq_n \exists P$.

Finally, by Theorem 4.4, the above result can be extended to all of $\mathsf{Circ}_F$:

**Corollary 4.7** *Reasoning in* $\mathsf{Circ}_F(\mathcal{EL}^{\perp})$ *with explicit priorities and GDIs can be reduced in polynomial time to reasoning in* $\mathsf{Circ}_{\mathsf{fix}}(\mathcal{EL}^{\perp})$ *with only specificity-based priority and defeasible inclusions of the form* $A \sqsubseteq_n \exists P$.

## 5. Undecidability of $\mathcal{EL}$ with Fixed Roles

In Circ* both concept names and roles can be fixed; however, as we show in this section, fixed roles in general make reasoning tasks undecidable. To this end, the *model conservative extension problem* defined by Lutz and Wolter (2010) is reduced to the subsumption problem.[6]

Some preliminary definitions are needed; given a signature $\Sigma \subseteq N_C \cup N_R$ and two interpretations $\mathcal{I}$ and $\mathcal{J}$, we say that $\mathcal{I}$ and $\mathcal{J}$ *coincide on* $\Sigma$ if and only if $\Delta^{\mathcal{I}} = \Delta^{\mathcal{J}}$ and for all predicates $X \in \Sigma$, $X^{\mathcal{I}} = X^{\mathcal{J}}$. Then, let $\mathcal{T}_1 \subseteq \mathcal{T}_2$ be classical $\mathcal{EL}$ TBoxes, $\mathcal{T}_2$ is a *model conservative extension* of $\mathcal{T}_1$ if and only if for every model $\mathcal{I}$ of $\mathcal{T}_1$, there exists a model $\mathcal{J}$ of $\mathcal{T}_2$ such that $\mathcal{I}$ and $\mathcal{J}$ coincide on the signature of $\mathcal{T}_1$.

Lutz and Wolter (2010) prove (see Lemma 40) that there exists a class $\mathcal{C}$ of $\mathcal{EL}$ TBoxes such that the problem of checking whether a TBox in $\mathcal{C}$ is a model conservative extension of another TBox in $\mathcal{C}$ is undecidable. Moreover, the following property holds.

**Lemma 5.1** *Given two TBoxes $\mathcal{T}_1 \subseteq \mathcal{T}_2$ in $\mathcal{C}$, if $\mathcal{T}_2$ is not a model conservative extension of $\mathcal{T}_1$ then there exists a model $\mathcal{I}$ of $\mathcal{T}_1$ and an interpretation $\mathcal{J}$ of $\mathcal{T}_2$ such that $\mathcal{I}$ and $\mathcal{J}$ coincide on the signature of $\mathcal{T}_1$ and the set of individuals in $\mathcal{J}$ that violate at least one concept inclusion from $\mathcal{T}_2$ is finite.*

For a DKB $\mathcal{KB} = \langle \mathcal{K}_S \cup \mathcal{K}_D, \prec \rangle$ and an interpretation $\mathcal{I}$ of $\mathcal{KB}$, we denote by $ab_{\mathcal{KB}}(\mathcal{I})$ (for *abnormal*) the total number of individuals $x \in \Delta^{\mathcal{I}}$ such that $x \notin \mathsf{sat}_{\mathcal{I}}(\delta)$ for some $\delta \in \mathcal{K}_D$.

**Lemma 5.2** *Let $\mathcal{KB} = \langle \mathcal{K}_S \cup \mathcal{K}_D, \emptyset \rangle$ be a DKB and $\mathcal{I}$ be a classical model of $\mathcal{KB}$ s.t. $ab_{\mathcal{KB}}(\mathcal{I})$ is finite. For all $F \subseteq \mathsf{N}_C \cup \mathsf{N}_R$, either $\mathcal{I}$ is a model of $\mathsf{Circ}_F(\mathcal{KB})$ or there exists a model $\mathcal{J}$ of $\mathsf{Circ}_F(\mathcal{KB})$ such that $\mathcal{J} <_{\mathcal{KB},F} \mathcal{I}$.*

**Proof.** It suffices to show that each $<_{\mathcal{KB},F}$-chain descending from $\mathcal{I}$ is finite. Since DIs are incomparable with each other, each step in the $<_{\mathcal{KB},F}$-chain must improve at least one DI and leave all other DIs unchanged. Formally, if $\mathcal{I}' <_{\mathcal{KB},F} \mathcal{I}$ then there exists at least a DI $\delta \in \mathcal{K}_D$ such that $\mathsf{sat}_{\mathcal{I}}(\delta) \subset \mathsf{sat}_{\mathcal{I}'}(\delta)$ and for all $\delta' \in \mathcal{K}_D$ it holds $\mathsf{sat}_{\mathcal{I}}(\delta') \subseteq \mathsf{sat}_{\mathcal{I}'}(\delta')$. Hence, $ab_{\mathcal{KB}}(\mathcal{I}') < ab_{\mathcal{KB}}(\mathcal{I})$ and the thesis follows. □

Assume that $\mathcal{T}_1, \mathcal{T}_2 \in \mathcal{C}$ are given, where $\mathcal{T}_1 \subseteq \mathcal{T}_2$, and let $\Sigma$ be the signature of $\mathcal{T}_1$. Let $F = \Sigma$ and $\mathcal{KB} = \langle \mathcal{K}, \emptyset \rangle$ where $\mathcal{K} = \mathcal{T}_1 \cup \{C \sqsubseteq_n D \mid C \sqsubseteq D \in \mathcal{T}_2 \setminus \mathcal{T}_1\}$.

**Lemma 5.3** *For all $\mathcal{T}_1, \mathcal{T}_2 \in \mathcal{C}$, $\mathcal{T}_2$ is a model conservative extension of $\mathcal{T}_1$ iff $\mathsf{Circ}_F^*(\mathcal{KB}) \models C \sqsubseteq D$, for all $C \sqsubseteq D \in \mathcal{T}_2$.*

**Proof.** [*if*] Suppose by contradiction that $\mathcal{T}_2$ is not a model conservative extension of $\mathcal{T}_1$ and $\mathsf{Circ}_F^*(\mathcal{KB}) \models C \sqsubseteq D$, for all $C \sqsubseteq D \in \mathcal{T}_2$. By Lemma 5.1, we can consider a model $\mathcal{I}$ of $\mathcal{T}_1$ and an extension $\mathcal{J}$ of $\mathcal{I}$ on the signature of $\mathcal{T}_2$, such that the set of individuals that violate in $\mathcal{J}$ at least one inclusion of $\mathcal{T}_2$ is finite. Since $\mathcal{J}$ is a classical model of $\mathcal{KB}$ and $ab_{\mathcal{KB}}(\mathcal{J})$ is finite, by Lemma 5.2, there exists a model $\mathcal{J}'$ of $\mathsf{Circ}_F^*(\mathcal{KB})$ such that either $\mathcal{J}' = \mathcal{J}$ or $\mathcal{J}' <_{\mathcal{KB},F} \mathcal{J}$. Since $\mathsf{Circ}_F^*(\mathcal{KB})$ entails $\mathcal{T}_2$ and $F = \Sigma$, $\mathcal{J}'$ is a (classical) model of $\mathcal{T}_2$ that coincides with $\mathcal{J}$ and $\mathcal{I}$ on $\Sigma$ (*absurdum*).

---

6. The sketch of this proof has been kindly provided by Frank Wolter in a personal communication. Any imprecision in the proof is due to the authors.

[*only if*] Suppose by contradiction that $\mathcal{T}_2$ is a model conservative extension of $\mathcal{T}_1$, and for some $\hat{C} \sqsubseteq \hat{D} \in \mathcal{T}_2$, it holds $\mathsf{Circ}_F^*(\mathcal{KB}) \not\models \hat{C} \sqsubseteq \hat{D}$. Since $\mathsf{Circ}_F^*(\mathcal{KB}) \not\models \hat{C} \sqsubseteq \hat{D}$, there exists a model $\mathcal{I}$ of $\mathsf{Circ}_F^*(\mathcal{KB})$ such that $\mathsf{sat}_{\mathcal{I}}(\hat{C} \sqsubseteq_n \hat{D}) \subset \Delta^{\mathcal{I}}$. Since $\mathcal{I}$ is a model of $\mathcal{T}_1$ and $\mathcal{T}_2$ is a model conservative extension of $\mathcal{T}_1$, there exists an interpretation $\mathcal{J}$ that *(i)* coincides with $\mathcal{I}$ on $\Sigma$ and *(ii)* is a model of $\mathcal{T}_2$, i.e., for all defeasible inclusions $C \sqsubseteq_n D$ in $\mathcal{KB}$, $\mathsf{sat}_{\mathcal{J}}(C \sqsubseteq_n D) = \Delta^{\mathcal{J}}$. Therefore, $\mathcal{J} <_{\mathcal{KB},F} \mathcal{I}$ (*absurdum*). □

Since checking whether a TBox is a model conservative extension of another one has been proved to be undecidable for $\mathcal{C} \subseteq \mathcal{EL}$ (Lutz & Wolter, 2010), it immediately follows that subsumption in $\mathsf{Circ}_F^*(\mathcal{EL})$ is undecidable. Moreover, since subsumption can be reduced to concept unsatisfiability or instance checking (Theorem 3.9), the latter reasoning tasks are undecidable as well.

**Theorem 5.4** *In $\mathsf{Circ}_F^*(\mathcal{EL})$, subsumption, concept consistency and instance checking are undecidable.*

# 6. Complexity of Circumscribed DL-lite$_R$

In this section we focus on DL-lite$_R$ DKBs. We first prove that in $\mathsf{Circ}_{\mathsf{var}}(\text{DL-lite}_R)$ the reasoning tasks are complete for the second level of the polynomial hierarchy. From this, according to Theorem 4.4, we immediately obtain an hardness result for $\mathsf{Circ}_{\mathsf{fix}}(\text{DL-lite}_R)$ too. Then, the membership for $\mathsf{Circ}_{\mathsf{fix}}(\text{DL-lite}_R)$ to second level of the polynomial hierarchy is shown for the fragment of DKBs with *left-fixed* defeasible inclusions, i.e. defeasible inclusions of type $A \sqsubseteq_n C$.

## 6.1 Complexity of $\mathsf{Circ}_{\mathsf{var}}(\textbf{DL-lite}_R)$

In this section we prove that $\mathsf{Circ}_{\mathsf{var}}(\text{DL-lite}_R)$ subsumption, concept unsatisfiability (co-sat) and instance checking are complete for $\Pi_2^p$.

Our membership results rely on the possibility of extracting a small (polynomial) model from any model of a circumscribed DKB.

**Lemma 6.1** *Let $\mathcal{KB}$ be a DL-lite$_R$ DKB. For all models $\mathcal{I}$ of $\mathsf{Circ}_{\mathsf{var}}(\mathcal{KB})$ and all $x \in \Delta^{\mathcal{I}}$ there exists a model $\mathcal{J}$ of $\mathsf{Circ}_{\mathsf{var}}(\mathcal{KB})$ such that (i) $\Delta^{\mathcal{J}} \subseteq \Delta^{\mathcal{I}}$, (ii) $x \in \Delta^{\mathcal{J}}$, (iii) for all DL-lite$_R$ concepts $C$, $x \in C^{\mathcal{I}}$ iff $x \in C^{\mathcal{J}}$ (iv) $|\Delta^{\mathcal{J}}|$ is polynomial in the size of $\mathcal{KB}$.*

**Proof.** Assume that $\mathcal{KB} = \langle \mathcal{K}_{\mathsf{S}} \cup \mathcal{K}_{\mathsf{D}}, \prec \rangle$, $\mathcal{I}$ is a model of $\mathsf{Circ}_{\mathsf{var}}(\mathcal{KB})$, and $x \in \Delta^{\mathcal{I}}$. Let $\mathsf{cl}(\mathcal{KB})$ be the set of all concepts occurring in $\mathcal{KB}$. Choose a minimal set $\Delta \subseteq \Delta^{\mathcal{I}}$ containing: (i) $x$, (ii) all $a^{\mathcal{I}}$ such that $a \in \mathsf{N}_{\mathsf{I}} \cap \mathsf{cl}(\mathcal{KB})$, (iii) for each concept $\exists R$ in $\mathsf{cl}(\mathcal{KB})$ satisfied in $\mathcal{I}$, a node $y_R$ such that for some $z \in \exists R^{\mathcal{I}}$, $(z, y_R) \in R^{\mathcal{I}}$.

Now define $\mathcal{J}$ as follows: (i) $\Delta^{\mathcal{J}} = \Delta$, (ii) $a^{\mathcal{J}} = a^{\mathcal{I}}$ (for $a \in \mathsf{N}_{\mathsf{I}} \cap \mathsf{cl}(\mathcal{KB})$), (iii) $A^{\mathcal{J}} = A^{\mathcal{I}} \cap \Delta$ ($A \in \mathsf{N}_{\mathsf{C}} \cap \mathsf{cl}(\mathcal{KB})$), and (iv) $P^{\mathcal{J}} = \{(z, y_P) \mid z \in \Delta \text{ and } z \in \exists P^{\mathcal{I}}\} \cup \{(y_{P^-}, z) \mid z \in \Delta \text{ and } z \in \exists P^{-\mathcal{I}}\}$ ($P \in \mathsf{N}_{\mathsf{R}}$).

Note that by construction, for all $z \in \Delta^{\mathcal{J}}$ and for all $C \in \mathsf{cl}(\mathcal{KB})$, $z \in C^{\mathcal{J}}$ iff $z \in C^{\mathcal{I}}$; consequently, $\mathcal{J}$ is a classical model of $\mathcal{S}$. Moreover, the cardinality of $\Delta^{\mathcal{J}}$ is linear in the size of $\mathcal{KB}$ (by construction). So we are only left to show that $\mathcal{J}$ is a $<_{\mathcal{K}_{\mathsf{D}},\mathsf{var}}$-minimal model of $\mathcal{KB}$.

Suppose not, and consider any $\mathcal{J}' <_{\mathcal{K}_{\mathsf{D}},\mathsf{var}} \mathcal{J}$. Define $\mathcal{I}'$ as follows: (i) $\Delta^{\mathcal{I}'} = \Delta^{\mathcal{I}}$, (ii) $a^{\mathcal{I}'} = a^{\mathcal{I}}$, (iii) $A^{\mathcal{I}'} = A^{\mathcal{J}'}$, (iv) $P^{\mathcal{I}'} = P^{\mathcal{J}'}$. Note that the elements in $\Delta^{\mathcal{I}} \setminus \Delta^{\mathcal{J}'}$ satisfy no left-hand side of any DL-lite$_R$ inclusion (be it classical or defeasible), therefore all inclusions are vacuously satisfied.

Moreover, the restriction of $\mathcal{I}'$ to $\Delta^{\mathcal{J}'}$ is $<_{\mathcal{K}_{\mathrm{D}},\mathsf{var}}$-smaller than the corresponding restriction of $\mathcal{I}$ in the interpretation ordering. It follows that $\mathcal{I}' <_{\mathcal{K}_{\mathrm{D}},\mathsf{var}} \mathcal{I}$, and hence $\mathcal{I}$ cannot be a model of $\mathsf{Circ}_{\mathsf{var}}(\mathcal{KB})$ (a contradiction). □

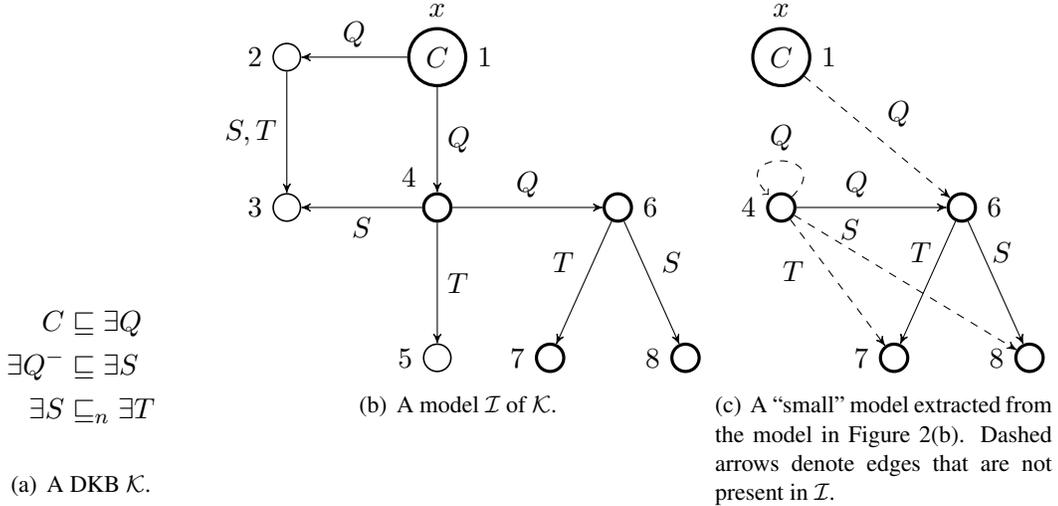

$C \sqsubseteq \exists Q$

$\exists Q^- \sqsubseteq \exists S$

$\exists S \sqsubseteq_n \exists T$

(a) A DKB $\mathcal{K}$.

(b) A model $\mathcal{I}$ of $\mathcal{K}$.

(c) A "small" model extracted from the model in Figure 2(b). Dashed arrows denote edges that are not present in $\mathcal{I}$.

Figure 2: Illustrating Lemma 6.1.

To illustrate Lemma 6.1, consider the DKB $\mathcal{KB}$ in Figure 2(a) and the model $\mathcal{I}$ in Figure 2(b). Note that all individuals in $\mathcal{I}$ satisfy the defeasible inclusion in $\mathcal{KB}$. The "small" model $\mathcal{J}$, depicted in Figure 2(c), is obtained as follows. First, it contains the designated individual $x$; then, for each concept $\exists R$ that occurs in $\mathcal{KB}$ and is satisfied in $\mathcal{I}$ (where $R$ is a possibly inverse role), it contains a representative $y_R$ that receives role $R$ in $\mathcal{I}$. In our case, assume that the chosen representatives are: $y_Q = 6$, $y_{Q^-} = 4$, $y_S = 8$, and $y_T = 7$. Hence, $\Delta^{\mathcal{J}} = \{x, 6, 4, 8, 7\}$. The roles in $\mathcal{J}$ are obtained by connecting each individual $z$ that satisfies a concept $\exists P$ in $\mathcal{I}$ to the chosen representative $y_P$. For instance, since 4 satisfies $\exists S$ in $\mathcal{I}$, we have the edge $(4, 8) \in S^{\mathcal{J}}$. Moreover, the representative for an inverse role $P^-$ is connected to all nodes that satisfy the concept $\exists P^-$ in $\mathcal{I}$. In our case, since 4 is the representative for $Q^-$ and 6 satisfies $\exists Q^-$, we have the edge $(6, 4) \in Q^{\mathcal{J}}$. Besides, since 4 itself satisfies $\exists Q^-$, we also have $(4, 4) \in Q^{\mathcal{J}}$. It can be verified by inspection that $\mathcal{J}$ is a model of $\mathsf{Circ}_{\mathsf{var}}(\mathcal{KB})$, as its individuals satisfy all classical and defeasible inclusions in $\mathcal{KB}$.

**Theorem 6.2** *Concept consistency over $\mathsf{Circ}_{\mathsf{var}}(DL\text{-}lite_R)$ DKBs is in $\Sigma_2^p$. Subsumption and instance checking are in $\Pi_2^p$.*

**Proof.** By Lemma 6.1, it suffices to guess a polynomial size model that provides an answer to the given reasoning problem. Then, with an NP oracle, it is possible to check that the model is minimal w.r.t. $<_{\mathsf{var}}$. □

The complexity upper bounds proved by Theorem 6.2 are in fact tight, as stated by Theorem 6.6. The proof of hardness is based on the reduction of the minimal-entailment problem of *positive disjunctive logic programs* — which has been proved to be $\Pi_2^p$-hard (Eiter & Gottlob, 1995).

A *clause* is a formula $l_1 \vee \cdots \vee l_h$, where the $l_i$ are literals over a set of propositional variables $PV = \{p_1, \ldots, p_n\}$. A *positive disjunctive logic program* (PDLP for short) is a set of clauses $S = \{c_1, \ldots, c_m\}$ where each $c_j$ contains at least one positive literal. A *truth valuation* for $S$ is a set $I \subseteq PV$, containing the propositional variables which are true. A truth valuation is a *model* of $S$ if it satisfies all clauses in $S$. For a literal $l$, we write $S \models_{\min} l$ if and only if every minimal[7] model of $S$ satisfies $l$. The *minimal-entailment problem* can be then defined as follows: given a PDLP $S$ and a literal $l$, determine whether $S \models_{\min} l$.

For each propositional variable $p_i$, $1 \le i \le n$, we introduce two concept names $P_i$ and $\bar{P}_i$, where the latter encodes $\neg p_i$. We denote by $L_j$, $1 \le j \le 2n$, a generic $P_i$ or $\bar{P}_i$. For each clause $c_j \in S$ we introduce the concept name $C_j$. Then, two other concept names *True* and *False* represent the set of true and false literals, respectively. We employ roles $RL_i$, $RTrueC_j$, $RFalse\bar{P}_i$, $RTrue\bar{P}_i$, and $TL_i$.

In the following, defeasible inclusions $\delta$ are assigned a numerical priority $h(\delta)$, with the intended meaning that $\delta_1 \prec \delta_2$ iff $h(\delta_1) < h(\delta_2)$.

The first step consists in reifying all the propositional literals, i.e., we want each of them to correspond to an individual. Therefore, we introduce the axioms:

$$NonEmpty(a) \tag{18}$$

$$NonEmpty \sqsubseteq \neg L_i \qquad (1 \le i \le 2n) \tag{19}$$

$$NonEmpty \sqsubseteq \exists RL_i \qquad (1 \le i \le 2n) \tag{20}$$

$$\exists RL_i^- \sqsubseteq L_i \qquad (1 \le i \le 2n) \tag{21}$$

$$L_i \sqsubseteq \neg L_j \qquad (1 \le i < j \le 2n) \tag{22}$$

$$L_i \sqsubseteq_n \neg L_i \qquad (1 \le i \le 2n) \qquad [\text{priority: } 0] \tag{23}$$

Axioms (18-21) force the literal encodings $L_i$ to be non empty. Axioms (22) make literal encodings pairwise disjoint. Finally, defeasible inclusions (23) are used to *reduce* the $L_i$ to singletons.
Then, we represent the set of clauses $S$ by adding for each clause $c_j = l_{j1} \vee \cdots \vee l_{jk}$, $1 \le j \le m$, the following axioms.

$$L_{ji} \sqsubseteq C_j \qquad (1 \le i \le k) \tag{24}$$

$$C_j \sqsubseteq_n \neg C_j \qquad [\text{priority: } 0] \tag{25}$$

$$NonEmpty \sqsubseteq \exists RTrueC_j \tag{26}$$

$$\exists RTrueC_j^- \sqsubseteq TrueC_j \tag{27}$$

$$TrueC_j \sqsubseteq True \tag{28}$$

$$TrueC_j \sqsubseteq C_j \tag{29}$$

Axioms (24–25) ensure that each (encoding of a) clause $C_j$ is the union of its literals $L_{ji}$. Axioms (26–29) assure that each clause contains at least one true literal. In order to model the concepts

---

7. With respect to set inclusion.

*True* and *False* and the correct meaning of complementary literals we add the following axioms.

$$True \sqsubseteq \neg False \tag{30}$$

$$TrueP_i \sqsubseteq \exists RFalse\bar{P}_i \quad (1 \leq i \leq n) \tag{31}$$

$$\exists RFalse\bar{P}_i^- \sqsubseteq False\bar{P}_i \quad (1 \leq i \leq n) \tag{32}$$

$$TrueP_i \sqsubseteq P_i \quad (1 \leq i \leq n) \tag{33}$$

$$TrueP_i \sqsubseteq True \quad (1 \leq i \leq n) \tag{34}$$

$$False\bar{P}_i \sqsubseteq False \quad (1 \leq i \leq n) \tag{35}$$

$$False\bar{P}_i \sqsubseteq \bar{P}_i \quad (1 \leq i \leq n) \tag{36}$$

The previous schemata regard $TrueP_i$ only; analogous schemata are defined for $FalseP_i$. The following inclusions ensure that the truth of a given literal is locally visible in the individual $a$, through the auxiliary roles $TL_i$.

$$TrueL_i \sqsubseteq \exists TL_i^- \quad (1 \leq i \leq 2n) \tag{37}$$

$$TrueL_i \sqsupseteq \exists TL_i^- \quad (1 \leq i \leq 2n) \tag{38}$$

$$\exists TL_i \sqsubseteq NonEmpty \quad (1 \leq i \leq 2n) \tag{39}$$

The axioms defined so far encode the classical semantics of $S$. To represent only minimal models we add the following axioms.

$$P_i \sqsubseteq_n FalseP_i \quad (1 \leq i \leq n) \quad [\text{priority: } 1] \tag{40}$$

$$P_i \sqsubseteq_n TrueP_i \quad (1 \leq i \leq n) \quad [\text{priority: } 2] \tag{41}$$

Given a PDLP $S$, we call the KB defined above $\mathcal{KB}_S$.

Given a truth assignment $I \subseteq PV$ and a domain $\Delta = \{a, d_1, \ldots, d_{2n}\}$, we define a corresponding interpretation, denoted by $model(S, I, \Delta)$, whose structure mirrors $I$. Formally, $model(S, I, \Delta)$ is the interpretation $\mathcal{I} = (\Delta, \cdot^{\mathcal{I}})$ such that:

I. $a^{\mathcal{I}} = a$;

II. $NonEmpty^{\mathcal{I}} = \{a\}$;

III. $RL_i^{\mathcal{I}} = \{(a, d_i)\}$ and $TL_i^{\mathcal{I}} = \{(a, d_i) \mid I \models l_i\}$;

IV. for each $1 \leq i \leq 2n$, $L_i^{\mathcal{I}} = \{d_i\}$;

V. for each $1 \leq j \leq m$, $C_j^{\mathcal{I}} = \{L_{j1}^{\mathcal{I}}, \ldots, L_{jh}^{\mathcal{I}}\}$ where $c_j = l_{j1} \vee \cdots \vee l_{jh}$;

VI. for each $1 \leq i \leq 2n$, $d_i \in TrueL_i^{\mathcal{I}}$ (resp. $d_i \in FalseL_i^{\mathcal{I}}$) iff $I \models l_i$ (resp. $I \not\models l_i$);

VII. $(XY)^{\mathcal{I}} = X^{\mathcal{I}} \cap Y^{\mathcal{I}}$, where $XY$ is a concept name obtained by concatenating two other concept names $X$ and $Y$ (for instance, concept name $TrueC_j$ is obtained by concatenating concept names $True$ and $C_j$); in other words, juxtaposition represents conjunction;

VIII. $(RXY)^{\mathcal{I}} = \Delta^{\mathcal{I}} \times (XY)^{\mathcal{I}}$.

The following lemma, proved in the Appendix, states the relationship between $I$ and $model(S, I, \Delta)$.

**Lemma 6.3** *Given a PDLP $S$ over $PV = \{p_1, \ldots, p_n\}$ and a truth assignment $I \subseteq PV$, $I$ is a minimal model of $S$ iff the interpretation $model(S, I, \Delta)$ is a model of $\mathsf{Circ}_{\mathsf{var}}(\mathcal{KB}_S)$, for all domains $\Delta$ with $|\Delta| = 2n + 1$.*

The following result, also proved in the Appendix, shows that any model of $\mathsf{Circ}_{\mathsf{var}}(\mathcal{KB}_S)$ in fact corresponds to a minimal model of $S$.

**Lemma 6.4** *If $\mathcal{I}$ is a model of $\mathsf{Circ}_{\mathsf{var}}(\mathcal{KB}_S)$, then there exist a minimal model $I$ of $S$ such that $p_i \in I$ iff $P_i^{\mathcal{I}} \subseteq True^{\mathcal{I}}$ iff $\bar{P}_i^{\mathcal{I}} \subseteq False^{\mathcal{I}}$, for all $i = 1, \ldots, n$.*

**Lemma 6.5** *Given a PDLP $S$ and a literal $l$ represented by concept name $L$, the following are equivalent:*

**(minimal entailment)** $S \models_{min} l$;

**(subsumption)** $\mathsf{Circ}_{\mathsf{var}}(\mathcal{KB}_S) \models L \sqsubseteq True$;

**(co-sat)** $False L$ *is not satisfiable w.r.t* $\mathsf{Circ}_{\mathsf{var}}(\mathcal{KB}_S)$;

**(instance checking)** $\mathsf{Circ}_{\mathsf{var}}(\mathcal{KB}_S) \models (\exists T L)(a)$.

**Proof.** The three inference problems on $\mathcal{KB}_S$ represent the fact that for all models $\mathcal{I}$ of $\mathsf{Circ}_{\mathsf{var}}(\mathcal{KB}_S)$ we have that $L^{\mathcal{I}} \cap True^{\mathcal{I}}$ is not empty — co-sat in particular relies on the fact that in all $\mathsf{Circ}_{\mathsf{var}}(\mathcal{KB}_S)$ models $True$ and $False$ are a partition of the individuals belonging to the literal concepts. Therefore, it suffices to prove that $l$ is true in all minimal models of $S$ **iff** $L^{\mathcal{I}} \cap True^{\mathcal{I}} \neq \emptyset$ in all models of $\mathsf{Circ}_{\mathsf{var}}(\mathcal{KB}_S)$.

Lemma 6.3 establishes a bijection between minimal models $I$ of $S$ and certain models $\mathcal{I} = model(S, I, \Delta)$ of $\mathsf{Circ}_{\mathsf{var}}(\mathcal{KB}_S)$, such that the truth of a literal $l$ in $I$ corresponds to the inclusion of $L^{\mathcal{I}}$ into $True^{\mathcal{I}}$ in $\mathcal{I}$ (see rule VI in the definition of $model$). Therefore, the right-to-left direction is immediately satisfied. For the left-to-right direction, assume that $l$ is true in all minimal models of $S$ and let $\mathcal{I}$ be a model of $\mathsf{Circ}_{\mathsf{var}}(\mathcal{KB}_S)$. By Lemma 6.4, there is a minimal model $I$ of $S$ such that $p_i \in I$ iff $P_i^{\mathcal{I}} \subseteq True^{\mathcal{I}}$. If $l = p_i$, we conclude $L^{\mathcal{I}} \subseteq True^{\mathcal{I}}$ and the thesis. Similarly, for $l = \bar{p}_i$. $\square$

The following theorem provides complexity lower-bounds for the main decision problems of both $\mathsf{Circ}_{\mathsf{var}}(\text{DL-lite}_R)$ and $\mathsf{Circ}_{\mathsf{fix}}(\text{DL-lite}_R)$. The result for $\mathsf{Circ}_{\mathsf{var}}(\text{DL-lite}_R)$ follows immediately from Lemma 6.5, and it extends to $\mathsf{Circ}_{\mathsf{fix}}(\text{DL-lite}_R)$ due to Theorem 4.4.

**Theorem 6.6** *Subsumption, co-sat and instance checking over circumscribed DL-lite$_R$ DKBs with general priorities are $\Pi_2^p$-hard.*

## 6.2 Upper Bound of $\mathsf{Circ}_{\mathsf{fix}}(\text{DL-lite}_R)$ with Restrictions

We develop the same argument used for $\mathsf{Circ}_{\mathsf{var}}(\text{DL-lite}_R)$ to prove similar upper bounds for $\mathsf{Circ}_{\mathsf{fix}}(\text{DL-lite}_R)$ DKBs with left-fixed DIs (i.e., their left-hand side is fixed or—equivalently—a concept name) or empty priority relations. Whether the same upper bounds apply to $\mathsf{Circ}_{\mathsf{fix}}(\text{DL-lite}_R)$ without any restriction is left as an open question.

**Lemma 6.7** *Let $\mathcal{KB}$ be a DL-lite$_R$ knowledge base whose DIs are left-fixed. For all models $\mathcal{I}$ of $\mathsf{Circ}_{\mathsf{fix}}(\mathcal{KB})$ and all $x \in \Delta^{\mathcal{I}}$ there exists a model $\mathcal{J}$ of $\mathsf{Circ}_{\mathsf{fix}}(\mathcal{KB})$ such that (i) $\Delta^{\mathcal{J}} \subseteq \Delta^{\mathcal{I}}$, (ii) $x \in \Delta^{\mathcal{J}}$, (iii) for all DL-lite$_R$ concepts $C$, $x \in C^{\mathcal{I}}$ iff $x \in C^{\mathcal{J}}$, and (iv) $|\Delta^{\mathcal{J}}|$ is polynomial in the size of $\mathcal{KB}$.*

**Proof.** Assume that $\mathcal{I}$ is a model of $\mathsf{Circ}_{\mathsf{fix}}(\mathcal{KB})$, with $\mathcal{KB} = \langle \mathcal{K}_{\mathsf{S}} \cup \mathcal{K}_{\mathsf{D}}, \prec \rangle$, and $x \in \Delta^{\mathcal{I}}$. Let $\mathsf{cl}(\mathcal{KB})$ be the set of all concepts occurring in $\mathcal{KB}$. Choose a minimal set $\Delta \subseteq \Delta^{\mathcal{I}}$ containing: (i) $x$, (ii) all $a^{\mathcal{I}}$ such that $a \in \mathsf{N}_{\mathsf{I}} \cap \mathsf{cl}(\mathcal{KB})$, (iii) for each concept $\exists R$ in $\mathsf{cl}(\mathcal{KB})$ satisfied in $\mathcal{I}$, a node $y_R$ such that $y_R \in (\exists R^-)^{\mathcal{I}}$ (where $\exists P^{--}$ is considered equivalent to $\exists P$), and finally (iv) for all inclusions $C \sqsubseteq_{[n]} \exists R$ in $\mathcal{KB}$ such that $(C \sqcap \exists R)^{\mathcal{I}} \neq \emptyset$, a node $z \in (C \sqcap \exists R)^{\mathcal{I}}$.

Now define $\mathcal{J}$ as follows: (i) $\Delta^{\mathcal{J}} = \Delta$, (ii) $a^{\mathcal{J}} = a^{\mathcal{I}}$ (for $a \in \mathsf{N}_{\mathsf{I}} \cap \mathsf{cl}(\mathcal{KB})$), (iii) $A^{\mathcal{J}} = A^{\mathcal{I}} \cap \Delta$ ($A \in \mathsf{N}_{\mathsf{C}} \cap \mathsf{cl}(\mathcal{KB})$), and (iv) $P^{\mathcal{J}} = \{(z, y_P) \mid z \in \Delta$ and $z \in \exists P^{\mathcal{I}}\} \cup \{(y_P, z) \mid z \in \Delta$ and $z \in \exists P^{-\mathcal{I}}\}$ ($P \in \mathsf{N}_{\mathsf{R}}$).

Note that by construction, for all $z \in \Delta^{\mathcal{J}}$ and for all $C \in \mathsf{cl}(\mathcal{KB})$, $z \in C^{\mathcal{J}}$ iff $z \in C^{\mathcal{I}}$; consequently, $\mathcal{J}$ is a classical model of $\mathcal{KB}$. Moreover, the cardinality of $\Delta^{\mathcal{J}}$ is linear in the size of $\mathcal{KB}$ (by construction). So we are only left to show that $\mathcal{J}$ is a $<_{\mathsf{fix}}$-minimal model of $\mathcal{KB}$.

Suppose not, and consider any $\mathcal{J}' <_{\mathsf{fix}} \mathcal{J}$. Define $\mathcal{I}'$ as follows: (a) $\Delta^{\mathcal{I}'} = \Delta^{\mathcal{I}}$, (b) $a^{\mathcal{I}'} = a^{\mathcal{I}}$, (c) $A^{\mathcal{I}'} = A^{\mathcal{I}}$, (d) each $R^{\mathcal{I}'}$ is a minimal set such that (d1) $R^{\mathcal{I}'} \supseteq R^{\mathcal{J}'}$, (d2) for all $z \in \Delta^{\mathcal{I}} \setminus \Delta^{\mathcal{J}}$, and for all inclusions $C \sqsubseteq \exists R$ or $C \sqsubseteq_n \exists R$ in $\mathcal{KB}$ such that $z \in (C \sqcap \exists R)^{\mathcal{I}}$, if $R^{\mathcal{J}'}$ contains a pair $(v, w)$, then $(z, w) \in R^{\mathcal{I}'}$; finally, (d3) each $P^{\mathcal{I}'}$ is closed under the role inclusion axioms of $\mathcal{KB}$. Note that, by construction,

(*) for all $z \in \Delta^{\mathcal{I}} \setminus \Delta^{\mathcal{J}}$, $z \in \exists R^{\mathcal{I}'}$ only if $z \in \exists R^{\mathcal{I}}$;

(**) for all $z \in \Delta^{\mathcal{I}} \setminus \Delta^{\mathcal{J}}$, $z \in \exists R^{\mathcal{I}'}$ only if there exists $v \in \Delta^{\mathcal{J}'}$ such that $v \in \exists R^{\mathcal{J}'}$.

Now we prove that $\mathcal{I}'$ is a classical model of $\mathcal{KB}$. By construction, the edges $(z, w)$ introduced in (d2) do not change the set of existential restrictions satisfied by the members of $\Delta^{\mathcal{J}}$; as a consequence — and since $\mathcal{J}'$ is a model of $\mathcal{KB}$— the members of $\Delta^{\mathcal{J}}$ satisfy all the classical CIs of $\mathcal{KB}$.

Now consider an arbitrary element $z \in \Delta^{\mathcal{I}} \setminus \Delta^{\mathcal{J}}$ and any CI $\gamma$ of $\mathcal{K}_{\mathsf{S}}$. If $\gamma$ is an inclusion without existential quantifiers, then $\mathcal{I}$ and $\mathcal{I}'$ give the same interpretation to $\gamma$ by definition, therefore $z$ satisfies $\gamma$. If $\gamma$ is $\exists R \sqsubseteq A$, $\exists R \sqsubseteq \neg A$, $\exists R \sqsubseteq \neg \exists S$, or $A \sqsubseteq \neg \exists R$ (and considering that $\mathcal{I}$ satisfies $\gamma$) $z$ fails to satisfy $\gamma$ only if for some $R' \in \{R, S\}$, $z \notin (\exists R')^{\mathcal{I}}$ and $z \in (\exists R')^{\mathcal{I}'}$; this is impossible by (*). Next, suppose $\gamma$ is $\exists R \sqsubseteq \exists S$. If $z \in (\exists R)^{\mathcal{I}'}$, then by (**) there exists a $v \in \Delta^{\mathcal{J}'}$ satisfying $(\exists R)^{\mathcal{J}'}$ and hence $(\exists S)^{\mathcal{J}'}$ (as $\mathcal{J}'$ is a model of $\mathcal{K}_{\mathsf{S}}$), therefore $z \in (\exists S)^{\mathcal{I}'}$ (by d2). We are only left to consider $\gamma = A \sqsubseteq \exists R$: If $z \in A^{\mathcal{I}'} = A^{\mathcal{I}}$, then there exists $w_A \in A^{\mathcal{J}'}$ (by construction of $\Delta$) and $w_A \in (\exists R)^{\mathcal{J}'}$ because $\mathcal{J}'$ is a model of $\mathcal{KB}$. Then $z \in (\exists R)^{\mathcal{I}'}$ by (d2). Therefore, in all possible cases, $z$ satisfies $\gamma$.

This proves that $\mathcal{I}'$ satisfies all the strong CIs of $\mathcal{KB}$. It is not hard to verify that $\mathcal{I}'$ satisfies also all role inclusions of $\mathcal{KB}$. Therefore, in order to derive a contradiction, we are left to show that $\mathcal{I}' <_{\mathsf{fix}} \mathcal{I}$ (which implies that $\mathcal{I}$ is not a model of $\mathsf{Circ}_{\mathsf{fix}}(\mathcal{KB})$). Since by assumption $\mathcal{J}' <_{\mathsf{fix}} \mathcal{J}$, it suffices to prove the following claim: if $\mathsf{sat}_{\mathcal{J}}(\delta) \subseteq \mathsf{sat}_{\mathcal{J}'}(\delta)$ (resp. $\mathsf{sat}_{\mathcal{J}}(\delta) \subset \mathsf{sat}_{\mathcal{J}'}(\delta)$), then $\mathsf{sat}_{\mathcal{I}}(\delta) \subseteq \mathsf{sat}_{\mathcal{I}'}(\delta)$ (resp. $\mathsf{sat}_{\mathcal{I}}(\delta) \subset \mathsf{sat}_{\mathcal{I}'}(\delta)$).

In $\Delta^{\mathcal{J}}$, $\mathcal{I}$ and $\mathcal{J}$ (resp. $\mathcal{I}'$ and $\mathcal{J}'$) satisfy the same concepts, therefore we only need to show that for all $z \in \Delta^{\mathcal{I}} \setminus \Delta^{\mathcal{J}}$, if $z \in \mathsf{sat}_{\mathcal{I}}(\delta)$ then $z \in \mathsf{sat}_{\mathcal{I}'}(\delta)$. In all cases but those in which the right-hand side of $\delta$ is $\exists R$, the proof is similar to the proof for strong CIs (it exploits (*) and the fact that concept names are fixed).

Let $\delta$ be $A \sqsubseteq_n \exists R$ and consider an arbitrary $z \in \Delta^{\mathcal{I}} \setminus \Delta^{\mathcal{J}}$ such that $z \in \mathsf{sat}_{\mathcal{I}}(\delta)$. Since concept names are fixed, the only interesting case is that $z$ actively satisfies $\delta$, i.e. $z \in (A \sqcap \exists R)^{\mathcal{I}}$. By construction, $\Delta$ contains an individual $v \in (A \sqcap \exists R)^{\mathcal{J}}$. Since by hypothesis $\mathsf{sat}_{\mathcal{J}}(\delta) \subseteq \mathsf{sat}_{\mathcal{J}'}(\delta)$, $v \in (A \sqcap \exists R)^{\mathcal{J}'}$ and hence, by (d2), $z \in (\exists R)^{\mathcal{I}'}$, that is $z \in \mathsf{sat}_{\mathcal{I}'}(\delta)$. $\qquad\square$

To prove the same lemma under the assumption that the priority relation is empty we need some preliminary notions. Given a KB $\mathcal{KB} = \langle \mathcal{K}_{\mathrm{S}} \cup \mathcal{K}_{\mathrm{D}}, \prec \rangle$, an interpretation $\mathcal{I}$ and an individual $z \in \Delta^{\mathcal{I}}$, we denote with $\mathcal{KB}^{[z]}$ the classical knowledge base:

$$\mathcal{KB}^{[z]} = \mathcal{K}_{\mathrm{S}} \cup \big\{ C \sqsubseteq D \mid (C \sqsubseteq_n D) \in \mathcal{K}_{\mathrm{D}} \text{ and } z \in \mathsf{sat}_{\mathcal{I}}(C \sqsubseteq_n D) \big\}$$

Then, the *support* of a concept $C$ in $\mathcal{I}$, $\mathsf{supp}_{\mathcal{I}}(C)$, is the set of individuals $z \in \Delta^{\mathcal{I}}$ such that, for some $A$, $z \in A^{\mathcal{I}}$ and $A \sqsubseteq_{\mathcal{KB}^{[z]}} C$. If $z \in \mathsf{supp}_{\mathcal{I}}(C)$ we say that $z$ *supports* $C$ in $\mathcal{I}$.

**Lemma 6.8** *Let $\mathcal{KB} = \langle \mathcal{K}, \emptyset \rangle$ be a DL-lite$_R$ knowledge base. For all models $\mathcal{I}$ of $\mathsf{Circ}_{\mathsf{fix}}(\mathcal{KB})$ and all $x \in \Delta^{\mathcal{I}}$ there exists a model $\mathcal{J}$ of $\mathsf{Circ}_{\mathsf{fix}}(\mathcal{KB})$ such that (i) $\Delta^{\mathcal{J}} \subseteq \Delta^{\mathcal{I}}$, (ii) $x \in \Delta^{\mathcal{J}}$, (iii) for all DL-lite$_R$ concepts $C$, $x \in C^{\mathcal{I}}$ iff $x \in C^{\mathcal{J}}$, and (iv) $|\Delta^{\mathcal{J}}|$ is polynomial in the size of $\mathcal{KB}$.*

**Proof.** Assume that $\mathcal{I}$ is a model of $\mathsf{Circ}_{\mathsf{fix}}(\mathcal{KB})$ and let $\Delta \subseteq \Delta^{\mathcal{I}}$ be defined as in the above proof of Lemma 6.7, except for case (iv), which is replaced by: (iv') for all inclusions $C \sqsubseteq_{[n]} \exists R$ in $\mathcal{KB}$ such that $\mathsf{supp}_{\mathcal{I}}(\exists R) \neq \emptyset$, a node $w_{\exists R} \in \mathsf{supp}_{\mathcal{I}}(\exists R)$. That is, for each inclusion whose RHS is variable, we pick a witness that is in the support of the RHS, if such a witness exists.

Next, define $\mathcal{J}$ as in the proof of Lemma 6.7. As before, $\mathcal{J}$ is a classical model of $\mathcal{KB}$ and the cardinality of $\Delta^{\mathcal{J}}$ is linear in the size of $\mathcal{KB}$. We are left to show that $\mathcal{J}$ is $<_{\mathsf{fix}}$-minimal. Suppose not, and consider any $\mathcal{J}' <_{\mathsf{fix}} \mathcal{J}$. Since the priority relation is empty, for all DIs $\delta$ in $\mathcal{KB}$, $\mathsf{sat}_{\mathcal{J}}(\delta) \subseteq \mathsf{sat}_{\mathcal{J}'}(\delta)$. Hence, for all concepts $C$ it holds $\mathsf{supp}_{\mathcal{J}}(C) \subseteq \mathsf{supp}_{\mathcal{J}'}(C)$. Define $\mathcal{I}'$ as in the proof of Lemma 6.7, except for case (d2), which is replaced by: (d2') for all $z \in \Delta^{\mathcal{I}} \setminus \Delta^{\mathcal{J}}$, and for all inclusions $C \sqsubseteq_{[n]} \exists R$ in $\mathcal{KB}$ such that $z \in \mathsf{supp}_{\mathcal{I}}(\exists R)$, if $R^{\mathcal{J}'}$ contains a pair $(v, w)$, then $(z, w) \in R^{\mathcal{I}'}$. We prove that $\mathcal{I}' <_{\mathsf{fix}} \mathcal{I}$, contradicting the hypothesis that $\mathcal{I}$ is a model of $\mathsf{Circ}_{\mathsf{fix}}(\mathcal{KB})$. The only non-trivial case consists in proving that the individuals in $\Delta^{\mathcal{I}'} \setminus \Delta^{\mathcal{J}'}$ satisfy in $\mathcal{I}'$ the same inclusions of type $C \sqsubseteq_{[n]} \exists S$ that they satisfy in $\mathcal{I}$.

Assume that $z \in \Delta^{\mathcal{I}'} \setminus \Delta^{\mathcal{J}'}$ satisfies $C \sqsubseteq_{[n]} \exists S$ in $\mathcal{I}$; we distinguish two cases. First, if $z \in \mathsf{supp}_{\mathcal{I}}(\exists S)$, we have that $\Delta^{\mathcal{J}'}$ contains a witness $w_{\exists S}$ s.t. $w_{\exists S} \in \mathsf{supp}_{\mathcal{J}}(\exists S) \subseteq \mathsf{supp}_{\mathcal{J}'}(\exists S)$. Therefore, there exists a pair $(w_{\exists S}, y) \in S^{\mathcal{J}'}$ and, by (d2'), $(z, y) \in S^{\mathcal{I}'}$. Second, assume that $z \notin \mathsf{supp}_{\mathcal{I}}(\exists S)$. Since $z \in sat_{\mathcal{I}}(C \sqsubseteq_{[n]} D)$, we have that $z \notin \mathsf{supp}_{\mathcal{I}}(C)$. Therefore, if $C = A$, then $z \notin A^{\mathcal{I}} = A^{\mathcal{I}'}$, whereas if $C = \exists R$, then by (d2') $z \notin (\exists R)^{\mathcal{I}'}$. In both cases, $z$ vacuously satisfies $\delta$ in $\mathcal{I}'$. $\qquad\square$

**Theorem 6.9** *Let $\mathcal{KB}$ be a DKB with left-fixed DIs or an empty priority relation. Concept consistency in $\mathsf{Circ}_{\mathsf{fix}}(\mathcal{KB})$ is in $\Sigma_2^p$. Subsumption and instance checking are in $\Pi_2^p$.*

**Proof.** Similar to the proof of Theorem 6.2. $\qquad\square$

## 7. Complexity of Circumscribed $\mathcal{EL}$ and $\mathcal{EL}^\perp$

Recall that reasoning in circumscribed $\mathcal{EL}$ is undecidable when roles can be fixed. Here we analyze the other cases, were $F \subseteq \mathsf{N_C}$.

In $\mathcal{EL}$, that cannot express any contradictions, defeasible inclusions cannot be possibly blocked under $\mathsf{Circ_{var}}$, and circumscription collapses to classical reasoning:

**Theorem 7.1** *Let $\mathcal{KB} = \langle \mathcal{K}_\mathsf{S} \cup \mathcal{K}_\mathsf{D}, \prec \rangle$ be an $\mathcal{EL}$ DKB. Then $\mathcal{I}$ is a model of $\mathsf{Circ_{var}}(\mathcal{KB})$ iff $\mathcal{I}$ is a model of $\mathcal{K}_\mathsf{S} \cup \hat{\mathcal{K}}_\mathsf{D}$, where $\hat{\mathcal{K}}_\mathsf{D} = \{A \sqsubseteq C \mid (A \sqsubseteq_n C) \in \mathcal{K}_\mathsf{D}\}$.*

**Proof.** Let $\mathcal{I}$ be any model of $\mathsf{Circ_{var}}(\mathcal{KB})$, and let $\mathcal{J}$ be the interpretation such that (i) $\Delta^\mathcal{J} = \Delta^\mathcal{I}$, (ii) for all $a \in \mathsf{N_I}$, $a^\mathcal{J} = a^\mathcal{I}$, (iii) for all $A \in \mathsf{N_C}$, $A^\mathcal{J} = \Delta^\mathcal{J}$, and (iv) for all $P \in \mathsf{N_R}$, $P^\mathcal{J} = \Delta^\mathcal{J} \times \Delta^\mathcal{J}$. It can be easily verified by structural induction that for all $\mathcal{EL}$ concepts $C$, $C^\mathcal{J} = \Delta^\mathcal{J}$ and hence each domain element of $\mathcal{J}$ satisfies all $\mathcal{EL}$ inclusions (strong and defeasible). Then, clearly, $\mathcal{J}$ is a model of $\mathsf{Circ_{var}}(\mathcal{KB})$. Consequently, for all $\delta \in \mathcal{K}_\mathsf{D}$, $\mathsf{sat}_\mathcal{I}(\delta) = \Delta^\mathcal{I}$, otherwise $\mathcal{J} <_\mathsf{var} \mathcal{I}$ (a contradiction). It follows that $\mathcal{I}$ is a classical model of $\mathcal{K}_\mathsf{S} \cup \hat{\mathcal{K}}_\mathsf{D}$. $\qquad\square$

By the results of the work of Baader et al. (2005), it follows that in $\mathsf{Circ_{var}}(\mathcal{EL})$, concept satisfiability is trivial, subsumption and instance checking are in P.

**Remark 7.2** Clearly, the same argument and the same result apply to $\mathsf{Circ_{var}}(\mathcal{ELHO})$.

If we make $\mathcal{EL}$ more interesting by adding $\perp$ as a source of inconsistency, then complexity increases significantly.

**Theorem 7.3** *In $\mathsf{Circ_{var}}(\mathcal{EL}^\perp)$, concept satisfiability, instance checking, and subsumption are ExpTime-hard. These results still hold if knowledge bases contain no assertion.*[8]

**Proof.** Let $\mathcal{EL}^{\neg A}$ be the extension of $\mathcal{EL}$ where atomic concepts can be negated. We first reduce TBox satisfiability in $\mathcal{EL}^{\neg A}$ (which is known to be ExpTime-hard, see Baader et al., 2005) to the complement of subsumption in $\mathsf{Circ_{var}}(\mathcal{EL}^\perp)$. Let $\mathcal{T}$ be a TBox (i.e., a set of CIs) in $\mathcal{EL}^{\neg A}$. First introduce for each concept name $A$ occurring in $\mathcal{T}$ a fresh concept name $\bar{A}$ whose intended meaning is $\neg A$. Obtain $\mathcal{T}'$ from $\mathcal{T}$ by replacing each literal $\neg A$ with $\bar{A}$. Let $\mathcal{KB} = \langle \mathcal{K}, \prec_\mathcal{K} \rangle$ be the DKB obtained by extending $\mathcal{T}'$ with the following inclusions, where $U$ and $U_A$ — for all $A$ occurring in $\mathcal{T}$ — are fresh concept names (representing undefined truth values), and $R$ is a fresh role name:

$$
\begin{array}{llcl@{\qquad\qquad}llcl}
A \sqcap \bar{A} & \sqsubseteq & \perp & (42) & \top & \sqsubseteq_n & A & (46) \\
A \sqcap U_A & \sqsubseteq & \perp & (43) & \top & \sqsubseteq_n & \bar{A} & (47) \\
\bar{A} \sqcap U_A & \sqsubseteq & \perp & (44) & \top & \sqsubseteq_n & U_A & (48) \\
U_A & \sqsubseteq & U & (45) & \top & \sqsubseteq_n & \exists R.U_A & (49)
\end{array}
$$

We prove that $\mathcal{T}$ is satisfiable iff in some model of $\mathsf{Circ_{var}}(\mathcal{KB})$ all $U_A$'s are empty, which holds iff $\mathsf{Circ_{var}}(\mathcal{KB}) \not\models \top \sqsubseteq \exists R.U$. Consequently, subsumption in $\mathsf{Circ_{var}}(\mathcal{EL}^\perp)$ is ExpTime-hard. Assume that $\mathcal{T}$ is satisfiable and $\mathcal{I}$ is a model of $\mathcal{T}$ with domain $\Delta^\mathcal{I}$. From $\mathcal{I}$ we define an interpretation $\mathcal{J}$ that is a model of $\mathsf{Circ}_{\mathsf{var}}(\mathcal{KB})$ such that $U^{\mathcal{J}} = \emptyset$, thus proving that $\mathsf{Circ}_{\mathsf{var}}(\mathcal{KB}) \not\models \top \sqsubseteq \exists R.U$. $\mathcal{J}$ has the same domain as $\mathcal{I}$, and all concepts and roles occurring in $\mathcal{T}$ have the same interpretation as in $\mathcal{I}$; we only need to define the interpretation of the newly introduced concepts $\bar{A}$, $U_A$, and $U$, and of the role $R$. We set $\bar{A}^{\mathcal{J}} = \Delta^{\mathcal{I}} \setminus A^{\mathcal{I}}$ and $U_A^{\mathcal{J}} = U^{\mathcal{J}} = R^{\mathcal{J}} = \emptyset$.

---

8. Equivalently, in DL's terminology: *ABoxes are empty.*

By construction $\mathcal{J}$ is a model of the classical inclusions in $\mathcal{KB}$, in particular CIs (42)–(45). It remains to prove that $\mathcal{J}$ is minimal w.r.t. $<_{\mathsf{var}}$, i.e., it is not possible to improve any DI $\delta$ without violating another DI that is either incomparable with $\delta$, or has a higher priority than $\delta$. Notice that defeasible inclusions (46) (resp., (47)) are violated by all individuals not in $A^{\mathcal{J}}$ (resp., all individuals in $A^{\mathcal{J}}$). DIs (48) and (49) are violated by all individuals. Moreover, notice that DIs (46)–(49) are mutually incomparable according to specificity.

Each DI of type (46) or (47) can only be improved at the expenses of the corresponding DI of the other type. Moreover, improving DIs (48) or (49) requires setting $U_A^{\mathcal{J}} \neq \emptyset$, which, due to rules (43) and (44) would damage the incomparable DIs (46) and/or (47). This proves that $\mathcal{J}$ is a model of $\mathsf{Circ}_{\mathsf{var}}(\mathcal{KB})$, and hence $\mathsf{Circ}_{\mathsf{var}}(\mathcal{KB}) \not\models \top \sqsubseteq \exists R.U$.

Conversely, assume that $\mathsf{Circ}_{\mathsf{var}}(\mathcal{KB}) \not\models \top \sqsubseteq \exists R.U$, and let $\mathcal{I}$ be a model of $\mathsf{Circ}_{\mathsf{var}}(\mathcal{KB})$ with an individual $x \in \Delta^{\mathcal{I}}$ such that $x \notin (\exists R.U)^{\mathcal{I}}$. Due to rule (45), $x \notin (\exists R.U_A)^{\mathcal{I}}$ for all atomic concepts $A$. Hence, $x$ violates all DIs of type (49). If there exists a concept $U_A$ such that $(U_A)^{\mathcal{I}}$ is not empty, then the model obtained from $\mathcal{I}$ by adding an $R$-edge from $x$ to an individual in $(U_A)^{\mathcal{I}}$ is smaller than $\mathcal{I}$ according to $<_{\mathsf{var}}$, which is a contradiction. Therefore, all concepts $U_A$ are empty in $\mathcal{I}$.

Next, we show that for all atomic concepts $A$ and all individuals $y \in \Delta^{\mathcal{I}}$, either $y \in A^{\mathcal{I}}$ or $y \in \bar{A}^{\mathcal{I}}$. Assume the contrary, i.e., there exists an individual $y$ which belongs neither to $A^{\mathcal{I}}$ nor to $\bar{A}^{\mathcal{I}}$. Then, $y$ violates all DIs (46)–(49). Consider the interpretation $\mathcal{I}'$, obtained from $\mathcal{I}$ by setting $(U_A)^{\mathcal{I}'} = U^{\mathcal{I}'} = \{y\}$. By construction $\mathcal{I}'$ satisfies all CIs in $\mathcal{KB}$. Compared to $\mathcal{I}$, the status of the DIs is the same, except that in $\mathcal{I}'$ the individual $y$ satisfies (48). Hence, $\mathcal{I}' <_{\mathsf{var}} \mathcal{I}$, which is a contradiction. Since each individual belongs to either $A^{\mathcal{I}}$ or $\bar{A}^{\mathcal{I}}$, we can convert $\mathcal{I}$ into a classical model of $\mathcal{T}$, thus showing that $\mathcal{T}$ is satisfiable.

Similarly, for any given $a \in \mathsf{N_I}$, $\mathcal{T}$ is satisfiable iff there exists a model $\mathcal{I}$ of $\mathsf{Circ}_{\mathsf{var}}(\mathcal{KB})$ such that $a^{\mathcal{I}} \notin (\exists R.U)^{\mathcal{I}}$. Therefore, instance checking in $\mathsf{Circ}_{\mathsf{var}}(\mathcal{EL})$ is ExpTime-hard as well.

Finally, add a fresh concept name $B$ and all the inclusions $B \sqcap \exists R.U_A \sqsubseteq \bot$; call the new DKB $\mathcal{KB}'$. Note that $\mathcal{T}$ is satisfiable iff in some model of $\mathsf{Circ}_{\mathsf{var}}(\mathcal{KB})$ all $U_A$'s are empty, which holds iff $B$ is satisfiable w.r.t. $\mathsf{Circ}_{\mathsf{var}}(\mathcal{KB}')$. Consequently, concept satisfiability in $\mathsf{Circ}_{\mathsf{var}}(\mathcal{EL}^{\bot})$ is ExpTime-hard. □

Since $\mathsf{Circ}_{\mathsf{var}}$ is a special case of $\mathsf{Circ}_F$, and by Theorem 4.4, the above theorem applies to $\mathsf{Circ}_F$ and $\mathsf{Circ}_{\mathsf{fix}}$, too:

**Corollary 7.4** *For $X = F$,* fix, *concept satisfiability checking, instance checking, and subsumption in $\mathsf{Circ}_X(\mathcal{EL}^{\bot})$ are ExpTime-hard. These results still hold if ABoxes are empty (i.e. assertions are not allowed).*

Fixed concept names can play a role similar to $\bot$, so that the above proof can be adapted to $\mathsf{Circ}_F(\mathcal{EL})$.

**Theorem 7.5** *Instance checking and subsumption are ExpTime-hard both in $\mathsf{Circ}_F(\mathcal{EL})$ and in $\mathsf{Circ}_{\mathsf{fix}}(\mathcal{EL})$. The same holds in the restriction of $\mathcal{EL}$ not supporting $\top$.*

**Proof.** We reduce satisfiability of an $\mathcal{EL}^{\neg A}$ TBox $\mathcal{T}$ to the complement of subsumption in $\mathsf{Circ}_F(\mathcal{EL})$. First we have to introduce a new concept name $D$ representing $\top$ and translate each concept $C$ in $\mathcal{EL}^{\neg A}$ into a corresponding $C^*$ in $\mathcal{EL}$, as follows:

- $C^* = C$ if $C$ is a concept name;

- $C^* = \bar{A}$ if $C$ is $\neg A$ (for all $A$, $\bar{A}$ is a new concept name);

- $C^* = D \sqcap \exists R.(C_1^* \sqcap D)$ if $C$ is $\exists R.C_1$;

- $C^* = C_1^* \sqcap C_2^*$ if $C$ is $C_1 \sqcap C_2$.

Each $C_1 \sqsubseteq C_2$ in $\mathcal{T}$ is translated into $C_1^* \sqsubseteq C_2^*$. Then we extend the translated TBox with the following inclusions, where $Bot$ (representing $\bot$), all $U_A$'s, and $Bad$ are new concept names and $R$ is a new role name:

$$
\begin{array}{rcll \qquad rcll}
A & \sqsubseteq & D & (50) & D & \sqsubseteq_n & U_A & (58) \\
\bar{A} & \sqsubseteq & D & (51) & D & \sqsubseteq & D' & (59) \\
U_A & \sqsubseteq & D & (52) & D' & \sqsubseteq_n & \exists R.U_A & (60) \\
A \sqcap \bar{A} & \sqsubseteq & Bot & (53) & D' & \sqsubseteq_n & \exists R.Bot & (61) \\
A \sqcap U_A & \sqsubseteq & Bot & (54) & \exists R.U_A & \sqsubseteq & Bad & (62) \\
\bar{A} \sqcap U_A & \sqsubseteq & Bot & (55) & \exists R.Bot & \sqsubseteq & Bad & (63) \\
D & \sqsubseteq_n & A & (56) & & & & \\
D & \sqsubseteq_n & \bar{A} & (57) & & D(a) & \text{(ABox assertion)} & (64)
\end{array}
$$

Let $\mathcal{KB} = \langle \mathcal{K}, \prec_{\mathcal{K}} \rangle$ be the resulting DKB, and set $F = \{D, Bot\}$. We prove that the following three properties are equivalent: *(i)* $\mathcal{T}$ is satisfiable, *(ii)* $\mathsf{Circ}_F(\mathcal{KB}) \not\models D \sqsubseteq Bad$, and *(iii)* $\mathsf{Circ}_F(\mathcal{KB}) \not\models Bad(a)$.

Let us prove that *(ii)* implies *(i)*. Let $\mathcal{I}$ be a model of $\mathsf{Circ}_F(\mathcal{KB})$ with an individual $x$ s.t. $x \in D^{\mathcal{I}}$ and $x \notin Bad^{\mathcal{I}}$. By (62)–(63), $x \notin (\exists R.U_A)^{\mathcal{I}}$ for all $A$, and $x \notin (\exists R.Bot)^{\mathcal{I}}$. By (59), $x \in (D')^{\mathcal{I}}$. Hence, $x$ violates DIs (60) and (61). Assume that at least one concept $U_A$ is not empty in $\mathcal{I}$ (resp., $Bot$ is not empty in $\mathcal{I}$). Then, $\mathcal{I}$ can be improved (according to $<_F$) by connecting with an $R$-edge the individual $x$ with the non-empty concept $U_A$ (resp., $Bot$), and then adding $x$ to $Bad$ (notice that $Bad$ is a variable concept). This being a contradiction, we conclude that $Bot$ and all $U_A$'s are empty in $\mathcal{I}$. Then, we prove that the restriction of $\mathcal{I}$ to the domain $D^{\mathcal{I}}$ is a model of $\mathcal{T}$.

Inclusions (50)–(51) ensure that all individuals satisfying either $A$ or $\bar{A}$ are in $D^{\mathcal{I}}$. DIs (56)–(57), together with the fact that all $U_A$'s are empty, guarantee that each individual in $D^{\mathcal{I}}$ satisfies either $A$ or $\bar{A}$, for all concept names $A$. Rules (53), together with the fact that $Bot$ is empty, guarantee that no individual satisfies both $A$ and $\bar{A}$. The translation from $C$ to $C^*$ completes our claim.

Next, we show that *(i)* implies *(ii)*. Let $\mathcal{I}$ be a model of $\mathcal{T}$. We extend $\mathcal{I}$ to become a model $\mathcal{J}$ of $\mathsf{Circ}_F(\mathcal{KB})$ such that $D^{\mathcal{J}} \not\subseteq Bad^{\mathcal{J}}$, because $D^{\mathcal{J}} \neq \emptyset$ and $Bad^{\mathcal{J}} = \emptyset$. For each $A$, set $\bar{A}^{\mathcal{J}} = \Delta^{\mathcal{I}} \setminus A^{\mathcal{I}}$ and $U_A^{\mathcal{J}} = \emptyset$. Then, set $D^{\mathcal{J}} = (D')^{\mathcal{J}} = \Delta^{\mathcal{J}} = \Delta^{\mathcal{I}}$ and $Bot^{\mathcal{J}} = Bad^{\mathcal{J}} = \emptyset$. It is easy to verify that $\mathcal{J}$ satisfies all CIs of $\mathcal{KB}$. It remains to prove that $\mathcal{J}$ is minimal w.r.t. $<_F$.

| Name | Restrictions |
|------|-------------|
| full left local ($LL_f$) | no qualified existentials on the LHS of inclusions |
| almost left local ($aLL$) | union of an $LL_f$ KB and a classical acyclic terminology, s.t. unfolding the former w.r.t. the latter produces a $LL_f$ KB |
| left local ($LL$) | only the following schemata: $\quad A \sqsubseteq_{[n]} \exists P.B \quad A_1 \sqcap A_2 \sqsubseteq B$ $\qquad\qquad\qquad\qquad\qquad\quad \exists P \sqsubseteq B \qquad \exists P_1 \sqsubseteq \exists P_2.B$ (no nesting; no conflicts between DIs in $\mathsf{Circ}_{\mathsf{fix}}$) |
| $LL_2$ | only the following schemata: $\quad A \sqsubseteq_{[n]} \exists P.B \quad \exists P_1 \sqcap \exists P_2 \sqsubseteq \exists P_3.B$ $\qquad\qquad\qquad\qquad\qquad\quad \exists P \sqsubseteq B$ (no nesting; potential conflicts between DIs even in $\mathsf{Circ}_{\mathsf{fix}}$) |

Figure 3: Fragments of $\mathcal{EL}^{\perp}$ considered in Section 7.

Since $Bot$ is a fixed concept, inclusions (53)–(55) ensure that $A$, $\bar{A}$ and $U_A$ are mutually exclusive, for all concepts $A$. Hence, each DI of type (56)–(58) can only be improved at the expenses of another DI of incomparable priority, which does not count as an improvement according to $<_F$. DI (61) cannot be improved because $Bot$ is empty and fixed. Finally, suppose one tries to improve one of the DIs of type (60). To do so, at least one individual $x$ must enter the concept $U_A$. Due to the mutual exclusion property described earlier, $x$ needs to exit from $A$ (resp. $\bar{A}$), thus violating DI (56) (resp., (57)), which has a higher priority than (60) due to (59).

The equivalence between *(i)* and *(iii)* can be proved along similar lines. Just notice that the fact (64) makes $Bad(a)$ equivalent to the inclusion $D \sqsubseteq Bad$. The thesis for $\mathsf{Circ}_{\mathsf{fix}}$ is obtained as a consequence of Theorem 4.4. □

Concept consistency is simpler, instead. Call an interpretation $\mathcal{I}$ *complete* iff for all $A \in \mathsf{N_C}$, $A^{\mathcal{I}} = \Delta^{\mathcal{I}}$, and for all $P \in \mathsf{N_R}$, $P^{\mathcal{I}} = \Delta^{\mathcal{I}} \times \Delta^{\mathcal{I}}$. It is not hard to verify that all $\mathcal{EL}$ concepts and all $\mathcal{EL}$ inclusions (both classical and defeasible) are satisfied by all $x \in \Delta^{\mathcal{I}}$, therefore complete models are always models of $\mathsf{Circ}_F(\mathcal{KB})$, for all DKBs $\mathcal{KB}$ and all $F \subseteq \mathsf{N_C}$. As a consequence we have that concept consistency is trivial:

**Theorem 7.6** *For all $\mathcal{EL}$ concepts $C$, DKBs $\mathcal{KB}$, and $F \subseteq \mathsf{N_C}$, $C$ is satisfied by some model of $\mathsf{Circ}_F(\mathcal{KB})$.*

### 7.1 Left Local $\mathcal{EL}^{\perp}$ and $\mathsf{Circ}_{\mathsf{var}}$

In this subsection and in the next one, we prove that qualified existentials in the left-hand side of inclusions are responsible for the higher complexity of $\mathcal{EL}^{\perp}$ w.r.t. DL-lite$_R$. In particular, qualified existentials in the left-hand side make the proof strategy of Lemma 6.7 fail: when the target of an edge which starts in $x$ is redirected, the individual $x$ may satisfy a qualified existential restriction that it did not satisfy before. If so, the truth value of inclusions may be affected. By limiting the occurrences of qualified existential restrictions in the left-hand side of inclusions, it is possible to reduce significantly the complexity of instance checking and subsumption in circumscribed $\mathcal{EL}^{\perp}$. Figure 3 summarizes the syntactic fragments of $\mathcal{EL}^{\perp}$ that we consider. We start with the following class of knowledge bases:

**Definition 7.7** A defeasible knowledge base $\langle \mathcal{K}, \prec \rangle$ is in the *full left local* ($LL_f$) fragment of $\mathcal{EL}^{\perp}$ iff the left-hand sides of the inclusions of $\mathcal{K}$ contain no qualified existential restrictions.

Note that this restriction rules out all the acyclic terminologies containing a qualified existential restriction, and hence most of the existing ontologies of practical interest. Therefore, we introduce the following relaxation of $LL_f \mathcal{EL}^\perp$:

**Definition 7.8** An $\mathcal{EL}^\perp$ knowledge base $\mathcal{KB} = \langle \mathcal{K}, \prec \rangle$ is *almost LL (aLL* for short) iff (i) $\mathcal{K} = \mathcal{K}_{LL} \cup \mathcal{K}_a$, (ii) $\mathcal{K}_{LL}$ is in $LL_f$, (iii) $\mathcal{K}_a$ is a classical acyclic terminology, and (iv) if a concept name $A$ defined in $\mathcal{K}_a$ occurs in the left-hand side of an inclusion in $\mathcal{K}_{LL}$, then $A$ does not depend (in $\mathcal{K}_a$) on any qualified existential restriction.

In other words, by unfolding $\mathcal{K}_{LL}$ with respect to $\mathcal{K}_a$, we obtain a $LL_f$ knowledge base.

**Example 7.9** Example 3.1 can be reformulated in aLL $\mathcal{EL}^\perp$:

> Human $\sqsubseteq_n$ ∃has_lhs_heart ;
> ∃has_lhs_heart $\sqcap$ ∃has_rhs_heart $\sqsubseteq$ $\perp$ ;
> Situs_Inversus $\equiv$ Human $\sqcap$ ∃has_rhs_heart .

Here $\mathcal{K}_{LL}$ consists of the first two axioms and $\mathcal{K}_a$ consists of the third axiom. Note that, in general, a concept name $A$ occurring in a terminology $\mathcal{T}$ can be extended with default properties by means of an inclusion $A \sqsubseteq_n C$ in the following cases: $A$ can be a primitive concept (with no definition in $\mathcal{T}$), or a concept partially defined by a one-way inclusion (e.g. Human $\sqsubseteq$ Mammal), or even a concept with a complete definition $A \equiv D$ in $\mathcal{T}$, provided that $A$ does not depend on any qualified existentials. Accordingly, in this example, we could add a defeasible inclusion like Situs_Inversus $\sqsubseteq_n$ $C$, that would not be permitted if the definition of situs inversus depended on qualified existential restrictions as in

> Situs_Inversus $\equiv$ Human $\sqcap$ ∃has_heart.∃has_position.Right .     □

A small model property similar to Lemma 6.7 can be proved for $\mathsf{Circ}_{\mathsf{var}}(aLL\,\mathcal{EL}^\perp)$ provided that the right-hand side of subsumption queries has bounded quantifier depth. It is convenient to split the proof into a proof for $LL_f \mathcal{EL}^\perp$ and later extend it to *aLL* $\mathcal{EL}^\perp$ .

Since in $LL_f \mathcal{EL}^\perp$ the RHS of an inclusion may have nested qualified existential restrictions, it is difficult to prove the small model property when considering the entire language. For this reason, we prove it *indirectly*: first we show how to transform a knowledge base $\mathcal{KB}$ into another $\mathcal{KB}^*$ that yields the following properties: *(i)* no nested formulas occur, *(ii)* defeasible inclusions are only of type $A \sqsubseteq_n B$, *(iii)* every model $\mathcal{I} \in \mathsf{Circ}_{\mathsf{var}}(\mathcal{KB})$ can be extended to a model of $\mathsf{Circ}_{\mathsf{var}}(\mathcal{KB}^*)$ on the same domain and *(iv)* every model of $\mathsf{Circ}_{\mathsf{var}}(\mathcal{KB}^*)$ is a model of $\mathsf{Circ}_{\mathsf{var}}(\mathcal{KB})$. Then, we prove a small model property for the fragment with no nesting and, thanks to properties *(iii)* and *(iv)*, we recover the small model property for the entire language.

Each $LL_f \mathcal{EL}^\perp$ inclusion $C \sqsubseteq_{[n]} D$ is transformed in three steps. Note that $C$'s shape is:

$$A_1 \sqcap \ldots \sqcap A_n \sqcap \exists R_1 \sqcap \ldots \sqcap \exists R_m .$$

In the **first step**, $C$ is replaced by a fresh concept name $F_0$ (for convenience, we later refer to $F_0$ also as $F_C$) and the following axioms are added:

$$F_i \sqsubseteq A_{i+1} \sqcap \ldots \sqcap A_n \sqcap \exists R_1 \sqcap \ldots \sqcap \exists R_m \quad (0 \le i \le n-1) \tag{65}$$

$$A_{i+1} \sqcap F_{i+1} \sqsubseteq F_i \quad (0 \le i \le n-2) \tag{66}$$

if $m = 0$, i.e., there are no existentials in $C$, add the inclusion:

$$A_n \sqsubseteq F_{n-1} \tag{67}$$

Otherwise, add the following inclusions:

$$A_n \sqcap G_0 \sqsubseteq F_{n-1} \tag{68}$$

$$G_j \sqsubseteq \exists R_{j+1} \sqcap \ldots \sqcap \exists R_m \quad (0 \le j \le m-1) \tag{69}$$

$$B_{j+1} \sqcap G_{j+1} \sqsubseteq G_j \quad (0 \le j \le m-2) \tag{70}$$

$$B_m \sqsubseteq G_{m-1} \tag{71}$$

$$B_j \sqsubseteq \exists R_j \quad (1 \le j \le m) \tag{72}$$

$$\exists R_j \sqsubseteq B_j \quad (1 \le j \le m) \tag{73}$$

where the $F_i$, $G_j$ and $B_j$ are fresh concept names.

At this point, the initial inclusion $C \sqsubseteq_{[n]} D$ can be replaced by $F_C \sqsubseteq_{[n]} D$. To eliminate the nesting in $D$, in the **second step** we replace it with a fresh concept name $F_D$ and add the inclusion $F_D \sqsubseteq D^*$, where $\cdot^*$ is recursively defined as

$$A^* = A \tag{74}$$

$$(C \sqcap D)^* = C^* \sqcap D^* \tag{75}$$

$$(\exists R.H)^* = \exists R.F_H \qquad \text{and add } F_H \sqsubseteq H^*, \text{ where } F_H \text{ is fresh.} \tag{76}$$

Finally, in the **third step**, all inclusions of type $A \sqsubseteq D_1 \sqcap \ldots \sqcap D_h$ are split into $A \sqsubseteq D_i$, $1 \le i \le h$. The resulting knowledge base, that we denote with $\mathcal{KB}^*$, consists of instances of the following axiom schemata:

$$A \sqsubseteq_{[n]} B \qquad A \sqsubseteq \exists P.B \qquad A_1 \sqcap A_2 \sqsubseteq B \qquad \exists P \sqsubseteq B$$

Now we prove the properties *(iii)* and *(iv)* mentioned above.

**Lemma 7.10** *Every model of* $\mathsf{Circ_{var}}(\mathcal{KB})$ *can be extended to a model of* $\mathsf{Circ_{var}}(\mathcal{KB}^*)$ *on the same domain.*

**Proof.** First, note that inclusions (65)–(73) are definitorial, that is every interpretation $\mathcal{I}$ of $\mathcal{KB}$ has exactly one extension that satisfies them. This extension, for simplicity we continue to call it $\mathcal{I}$, is obtained by setting $F_i^{\mathcal{I}} = (A_{i+1} \sqcap \ldots \sqcap A_n \sqcap \exists R_1 \sqcap \ldots \sqcap \exists R_m)^{\mathcal{I}}$, $G_j^{\mathcal{I}} = (\exists R_{j+1} \sqcap \ldots \sqcap \exists R_m)^{\mathcal{I}}$ and $B_j^{\mathcal{I}} = (\exists R_j)^{\mathcal{I}}$.

Then, we can extend $\mathcal{I}$ by recursively setting $F_H^{\mathcal{I}} = H^{\mathcal{I}}$, for each fresh concept $F_H$ introduced in step 2. It is straightforward to see by structural induction on $H$ that $(H^*)^{\mathcal{I}} = H^{\mathcal{I}}$, and hence inclusions in step 2 and 3 are satisfied. Thus, if $\mathcal{I}$ is a classical model of $\mathcal{KB}$, then it is also a classical model of $\mathcal{KB}^*$.

Assume now that $\mathcal{I}$ is a model of $\mathsf{Circ_{var}}(\mathcal{KB})$, we have to show that $\mathcal{I}$ is minimal also with respect to $\mathcal{KB}^*$. Suppose not, and let $\mathcal{J}$ be a classical model of $\mathcal{KB}^*$ such that $\mathcal{J} <_{\mathcal{KB}^*, \mathsf{var}} \mathcal{I}$. By structural induction it is straightforward to see that for all $C \sqsubseteq_{[n]} D$ in $\mathcal{KB}$, $(D^*)^{\mathcal{J}} \subseteq D^{\mathcal{J}}$. Since $F_C^{\mathcal{J}} = C^{\mathcal{J}}$ holds for all fresh concept names $F_C$ occurring in the LHS of a rule, we have that for each inclusion $C \sqsubseteq_{[n]} D$ in $\mathcal{KB}$, $\mathsf{sat}_{\mathcal{J}}(F_C \sqsubseteq_{[n]} F_D) \subseteq \mathsf{sat}_{\mathcal{J}}(C \sqsubseteq_{[n]} D)$ — which implies that $\mathcal{J}$ is a classical model of $\mathcal{KB}$. Concerning $\mathcal{I}$, for every $C \sqsubseteq_{[n]} D$ in $\mathcal{KB}$, we have $(F_C)^{\mathcal{I}} = C^{\mathcal{I}}$ and $(F_D)^{\mathcal{I}} = D^{\mathcal{I}}$ by construction, that is $\mathsf{sat}_{\mathcal{I}}(C \sqsubseteq_{[n]} D) = \mathsf{sat}_{\mathcal{I}}(F_C \sqsubseteq_{[n]} F_D)$.

The previous arguments entail that if $\mathsf{sat}_{\mathcal{I}}(F_C \sqsubseteq_{[n]} F_D) \subseteq \mathsf{sat}_{\mathcal{J}}(F_C \sqsubseteq_{[n]} F_D)$ (respectively $\mathsf{sat}_{\mathcal{I}}(F_C \sqsubseteq_{[n]} F_D) \subset \mathsf{sat}_{\mathcal{J}}(F_C \sqsubseteq_{[n]} F_D)$), then $\mathsf{sat}_{\mathcal{I}}(C \sqsubseteq_{[n]} D) \subseteq \mathsf{sat}_{\mathcal{J}}(C \sqsubseteq_{[n]} D)$ (resp. $\mathsf{sat}_{\mathcal{I}}(C \sqsubseteq_{[n]} D) \subset \mathsf{sat}_{\mathcal{J}}(C \sqsubseteq_{[n]} D)$). Therefore, it would follow that $\mathcal{J} <_{\mathcal{KB},\mathsf{var}} \mathcal{I}$, which contradicts the hypothesis. □

**Lemma 7.11** *All models of* $\mathsf{Circ}_{\mathsf{var}}(\mathcal{KB}^*)$ *are models of* $\mathsf{Circ}_{\mathsf{var}}(\mathcal{KB})$.

**Proof.** The proof is similar to Lemma 7.10, in particular we already know that *(i)* if an individual satisfies an inclusion $F_C \sqsubseteq_{[n]} F_D$ in $\mathcal{KB}^*$, then it satisfies $C \sqsubseteq_{[n]} D$, and *(ii)* a classical model $\mathcal{J}$ of $\mathcal{KB}$ can be extended in such a way that $\mathsf{sat}_{\mathcal{J}}(C \sqsubseteq_{[n]} D) = \mathsf{sat}_{\mathcal{J}}(F_C \sqsubseteq_{[n]} F_D)$, for each (possibly defeasible) inclusion in $\mathcal{KB}$. From this, we have that every classical model of $\mathcal{KB}^*$ is a classical model of $\mathcal{KB}$ and, by assuming that some classical model $\mathcal{J}$ of $\mathcal{KB}$ improves $\mathcal{I}$ (i.e. $\mathcal{J} <_{\mathcal{KB},\mathsf{var}} \mathcal{I}$), it follows that $\mathcal{J}$ can be extended into a classical model of $\mathcal{KB}^*$ such that $\mathcal{J} <_{\mathcal{KB}^*,\mathsf{var}} \mathcal{I}$. □

The following result, whose proof can be found in the Appendix, represents a small model property for $LL_f \mathcal{EL}^{\perp}$, which uses the above transformation of $\mathcal{KB}$ into $\mathcal{KB}^*$.

**Lemma 7.12** *Let* $\mathcal{KB} = \langle \mathcal{K}, \prec_{\mathcal{K}} \rangle$ *be an* $LL_f \mathcal{EL}^{\perp}$ *knowledge base, and* $C, D$ *be* $\mathcal{EL}^{\perp}$ *concepts. For all models* $\mathcal{I} \in \mathsf{Circ}_{\mathsf{var}}(\mathcal{KB})$ *and for all* $x \in C^{\mathcal{I}} \setminus D^{\mathcal{I}}$ *there exists a model* $\mathcal{J} \in \mathsf{Circ}_{\mathsf{var}}(\mathcal{KB})$ *such that* (i) $\Delta^{\mathcal{J}} \subseteq \Delta^{\mathcal{I}}$, (ii) $x \in C^{\mathcal{J}} \setminus D^{\mathcal{J}}$, *and* (iii) $|\Delta^{\mathcal{J}}|$ *is* $O((|\mathcal{KB}|^2 + |C|)^d)$, *where* $d = depth(D) + 1$.

Now we have to extend the above result to *aLL* $\mathcal{EL}^{\perp}$. First, we show under which conditions the concept names defined in $\mathcal{K}_a$ can be removed by unfolding them, using the unf operator defined in Section 2. The proofs of the following two propositions can be found in the Appendix.

**Proposition 7.13** *Let* $\mathcal{KB} = \langle \mathcal{K}_{LL} \cup \mathcal{K}_a, \prec \rangle$ *be an aLL* $\mathcal{EL}^{\perp}$ *knowledge base. Every model of* $\mathsf{Circ}_F(\mathsf{unf}(\mathcal{KB}))$ *can be extended to a model of* $\mathsf{Circ}_F(\mathcal{KB})$.

The converse holds only if defined predicates are variable. The reason is that by adding a definition like $A \equiv \exists P$ where $A$ is fixed, one fixes the expression $\exists P$, too, thereby changing its semantics.

**Proposition 7.14** *Let* $\mathcal{KB} = \langle \mathcal{K}_{LL} \cup \mathcal{K}_a, \prec \rangle$ *be an aLL* $\mathcal{EL}^{\perp}$ *knowledge base and suppose that all the concept names defined in* $\mathcal{K}_a$ *are variable. Then, for all models* $\mathcal{I}$ *of* $\mathsf{Circ}_F(\mathcal{KB})$, *the restriction of* $\mathcal{I}$ *to primitive predicates is a model of* $\mathsf{Circ}_F(\mathsf{unf}(\mathcal{KB}))$.

With these lemmata we can prove:

**Lemma 7.15** *Let* $\mathcal{KB} = \langle \mathcal{K}, \prec_K \rangle$ *be an aLL* $\mathcal{EL}^{\perp}$ *knowledge base (where* $\mathcal{K} = \mathcal{K}_a \cup \mathcal{K}_{LL}$*) and let* $C, D$ *be* $\mathcal{EL}^{\perp}$ *concepts. For all models* $\mathcal{I} \in \mathsf{Circ}_{\mathsf{var}}(\mathcal{KB})$ *and for all* $x \in C^{\mathcal{I}} \setminus D^{\mathcal{I}}$ *there exists a model* $\mathcal{J} \in \mathsf{Circ}_{\mathsf{var}}(\mathcal{KB})$ *such that* (i) $\Delta^{\mathcal{J}} \subseteq \Delta^{\mathcal{I}}$, (ii) $x \in C^{\mathcal{J}} \setminus D^{\mathcal{J}}$, *and* (iii) $|\Delta^{\mathcal{J}}|$ *is* $O((|\mathcal{KB}|^2 + |C|)^d)$, *where* $d = depth(D) + 1 + |\mathcal{K}_a|^2$.

**Proof.** Let $\mathcal{I} \in \mathsf{Circ}_{\mathsf{var}}(\mathcal{KB})$ and $x \in C^{\mathcal{I}} \setminus D^{\mathcal{I}}$. Let $\mathcal{KB}' = \mathsf{unf}(\mathcal{KB})$, $C' = \mathsf{unf}(C, \mathcal{K}_a)$, and $D' = \mathsf{unf}(D, \mathcal{K}_a)$. Notice that $\mathcal{KB}'$ is a $LL_f$ knowledge base. By Proposition 7.14, the restriction $\mathcal{I}'$ of $\mathcal{I}$ to primitive predicates is a model of $\mathsf{Circ}_{\mathsf{var}}(\mathcal{KB}')$. In particular, it holds that $x \in (C')^{\mathcal{I}'} \setminus (D')^{\mathcal{I}'}$. By applying Lemma 7.12 to $\mathcal{KB}'$, $C'$, $D'$, and $\mathcal{I}'$, we obtain that there exists a model $\mathcal{J}$ of $\mathsf{Circ}_{\mathsf{var}}(\mathcal{KB}')$ such that $x \in (C')^{\mathcal{J}} \setminus (D')^{\mathcal{J}}$ and $|\Delta^{\mathcal{J}}|$ is $O((|\mathcal{KB}'|^2 + |C'|)^{d'})$, where $d' = depth(D') + 1$. We have $|\mathcal{KB}'|^2 + |C'| \leq |\mathcal{KB}|^2 + |C|$, since replacing defined terms with their definitions can only decrease the total number of subformulas. Finally, $depth(D') \leq depth(D) + \sum_{(A \equiv E) \in \mathcal{K}_a} depth(E) \leq depth(D) + |\mathcal{K}_a|^2$, hence the thesis. $\qquad\square$

Consequently we have that:

**Theorem 7.16** *In* $\mathsf{Circ}_{\mathsf{var}}(aLL\ \mathcal{EL}^{\perp})$ *concept satisfiability is in* $\Sigma_2^p$*. Moreover, deciding* $\mathcal{EL}^{\perp}$ *subsumptions* $C \sqsubseteq D$ *or instance checking problems* $D(a)$ *with a constant bound on the quantifier depth of* $D$*'s unfolding w.r.t. the given DKB is in* $\Pi_2^p$*.*

**Proof.** Similar to the proof of Theorem 6.2. $\qquad\square$

Currently, we do not know whether the bound on quantifier nesting is necessary to the above upper complexity bounds.

Next we prove that the $\Sigma_2^p$ and $\Pi_2^p$ upper bounds for $\mathsf{Circ}_{\mathsf{var}}$ are tight. Actually, a much simpler fragment suffices to reach that complexity:

**Definition 7.17** An $\mathcal{EL}^{\perp}$ knowledge base is *left local* (LL) if its concept inclusions are instances of the following schemata:

$$A \sqsubseteq_{[n]} \exists P.B \qquad A_1 \sqcap A_2 \sqsubseteq B \qquad \exists P \sqsubseteq B \qquad \exists P_1 \sqsubseteq \exists P_2.B \,,$$

where $A$ and $B$ are either concept names or $\perp$. An *LL* $\mathcal{EL}^{\perp}$ *concept* is any concept that can occur in the above inclusions.

Schema $A \sqsubseteq_{[n]} B$ can be emulated in $LL\ \mathcal{EL}^{\perp}$ by the inclusions $A \sqsubseteq_{[n]} \exists R$, $\exists R \sqsubseteq B$ and $B \sqsubseteq \exists R$, for a fresh role $R$. Note the similarity of $LL$ schemata with the normal form of $\mathcal{EL}$ inclusions (Baader et al., 2005) that, however, would allow the more general inclusions $\exists P.A \sqsubseteq B$ and $\exists P_1.A \sqsubseteq \exists P_2.B$ (that are forbidden by left locality).

Now we prove that reasoning in $\mathsf{Circ}_{\mathsf{var}}(LL\ \mathcal{EL}^{\perp})$ is hard (and hence complete) for $\Sigma_2^p$ and $\Pi_2^p$.

For this purpose, we provide a reduction of minimal entailment over positive, propositional disjunctive logic programs (PDLP), defined in Section 6.1. For each propositional variable $p_i$, $1 \leq i \leq n$, introduce two concept names $P_i$ and $\bar{P}_i$ – where the latter encodes $\neg p_i$. In the following we will denote by $L_j$, $1 \leq j \leq 2n$, a generic $P_i$ or $\bar{P}_i$. For each clause $c_j \in S$ introduce a concept name $C_j$. Then, two other concept names $True$ and $False$ represent the set of true and false literals respectively. Finally, the concept names $\mathsf{Lit}$ and $\mathsf{Min}$ are used to model minimal propositional assignments; we need also an auxiliary role $R$.

First, literals are reified, i.e. modeled as individuals, with the axioms:

$$\top \sqsubseteq \exists R.L_i \qquad\qquad (1 \leq i \leq 2n) \qquad (77)$$

$$L_i \sqcap L_j \sqsubseteq \perp \qquad\qquad (1 \leq i < j \leq 2n) \qquad (78)$$

$$L_i \sqsubseteq_n \perp \qquad\qquad (1 \leq i \leq 2n) \qquad (79)$$

The first axiom makes all $L_i$ nonempty. Axioms (78) make them pairwise disjoint. Finally, axioms (79) minimize the concepts $L_i$ and make them singletons. Then, we represent $S$ by adding for each clause $c_j = l_{j1} \vee \cdots \vee l_{jh}$, $1 \leq j \leq m$, the axioms

$$L_{ji} \sqsubseteq C_j \qquad (1 \leq j \leq m \text{ and } 1 \leq i \leq h) \tag{80}$$

$$C_j \sqsubseteq_n \bot \qquad (1 \leq j \leq m) \tag{81}$$

$$\top \sqsubseteq \exists R.(C_j \sqcap \mathit{True}) \quad (1 \leq j \leq m) \tag{82}$$

By axioms (80) and (81), $C_j$ equals the set of (encodings of) literals in $c_j$. Axioms (82) make sure that each clause holds.

In order to model the concepts $\mathit{True}$ and $\mathit{False}$ and the correct meaning of complementary literals we add the axioms

$$\mathit{True} \sqcap \mathit{False} \sqsubseteq \bot \tag{83}$$

$$P_i \sqcap \mathit{True} \sqsubseteq \exists R.(\bar{P}_i \sqcap \mathit{False}) \quad (1 \leq i \leq n) \tag{84}$$

$$P_i \sqcap \mathit{False} \sqsubseteq \exists R.(\bar{P}_i \sqcap \mathit{True}) \quad (1 \leq i \leq n) \tag{85}$$

$$\bar{P}_i \sqcap \mathit{True} \sqsubseteq \exists R.(P_i \sqcap \mathit{False}) \quad (1 \leq i \leq n) \tag{86}$$

$$\bar{P}_i \sqcap \mathit{False} \sqsubseteq \exists R.(P_i \sqcap \mathit{True}) \quad (1 \leq i \leq n) \tag{87}$$

The axioms defined so far encode the classical semantics of $S$. To minimize models, add the following axioms:

$$\mathrm{Min} \sqcap P_i \sqsubseteq \mathit{False} \quad (1 \leq i \leq n) \tag{88}$$

$$\mathrm{Min} \sqcap \bar{P}_i \sqsubseteq \mathit{True} \quad (1 \leq i \leq n) \tag{89}$$

$$L_i \sqsubseteq \mathrm{Lit} \qquad (1 \leq i \leq 2n) \tag{90}$$

$$C_j \sqsubseteq \mathrm{Lit} \qquad (1 \leq j \leq m) \tag{91}$$

$$\mathrm{Lit} \sqsubseteq_n \mathrm{Min} \tag{92}$$

By (88) and (89), $\mathit{Min}$ collects false positive literals and true negative literals. By (90) and (91), Lit contains all the (representations of) literals and clauses. The purpose of these axioms is giving defeasible inclusions (79) and (81) higher (specificity-based) priority than (92), so that model minimization cannot cause any $L_i$ to be larger than a singleton, nor any $C_j$ to be different from the set of literals of $c_j$. Now (92) *prefers* those models where as many $P_i$ as possible are in $\mathit{False}$.

In the following, given a PDLP $S$, let $\mathcal{KB}_S$ be the Tbox defined above.

**Lemma 7.18** *Given a PDLP $S$, a literal $l$ in $S$'s language, and the encoding $L$ of $l$, the following are equivalent:*

**(minimal entailment)** $S \models_{min} l$;

**(subsumption)** $\mathrm{Circ}_{\mathrm{var}}(\mathcal{KB}_S) \models \top \sqsubseteq \exists R.(\mathit{True} \sqcap L)$;

**(co-sat)** $\mathit{False} \sqcap L$ *is not satisfiable w.r.t* $\mathrm{Circ}_{\mathrm{var}}(\mathcal{KB}_S)$;

**(instance checking)** $\mathrm{Circ}_{\mathrm{var}}(\mathcal{KB}_S) \models (\exists R.(\mathit{True} \sqcap L))(a)$.

This lemma can be proved by analogy with the proof of Lemma 6.5; the details are left to the reader.

The conjunctions ($\sqcap$) nested in $\exists$ can be easily replaced with a new atom $A$ by adding the equivalence $A \equiv \mathit{True} \sqcap L$, that can itself be encoded in $LL\,\mathcal{EL}^{\perp}$, so we have:

**Theorem 7.19** *Subsumption and instance checking over* $\mathsf{Circ}_{\mathsf{var}}(LL\,\mathcal{EL}^\perp)$ *are* $\Pi_2^p$-*hard; concept satisfiability is* $\Sigma_2^p$-*hard. These results hold even if the three reasoning tasks are restricted to LL* $\mathcal{EL}^\perp$ *concepts, and priorities are specificity-based.*

## 7.2 Left Local $\mathcal{EL}^\perp$ and $\mathsf{Circ}_{\mathsf{fix}}$

By a reduction of the validity problem for quantified Boolean formula, we can show that $\mathsf{Circ}_{\mathsf{fix}}(aLL\,\mathcal{EL}^\perp)$ is more complex than $\mathsf{Circ}_{\mathsf{var}}(aLL\,\mathcal{EL}^\perp)$, unless the polynomial hierarchy collapses. Computing the truth of a quantified Boolean formula

$$\psi = Q_1 p_1 \ldots Q_n p_n . \varphi$$

(where the $Q_i$'s are quantifiers) can be reduced in polynomial time to subsumption checking in $\mathsf{Circ}_{\mathsf{fix}}(aLL\,\mathcal{EL}^\perp)$ as follows. Introduce concept names $A_0, \ldots, A_n$, $T_i$ and $F_i$ for $i = 1 \ldots n$, and concept names $E_i^j$ for $1 \leq i < j \leq n$. Introduce role names $R$, $bad$, $good$, and $U_i$ for $i = 1 \ldots n$.

We define a $aLL\mathcal{EL}^\perp$ knowledge base $\langle \mathcal{K}, \prec_\mathcal{K} \rangle$, where $\mathcal{K} = \mathcal{K}_{LL} \cup \mathcal{K}_a$. The left-local part $\mathcal{K}_{LL}$ consists of the following groups of axioms. Notice that, in the following description, $i$ is always an arbitrary index in $\{1, \ldots, n\}$. First, we encode the negation normal form $\bar{\varphi}$ of $\neg\varphi$. Let $B_{p_i} = T_i$ and $B_{\neg p_i} = F_i$. For all subformulas $F \wedge G$ of $\bar{\varphi}$ introduce a new concept name $B_{F \wedge G}$ and add the inclusion $B_F \sqcap B_G \sqsubseteq B_{F \wedge G}$. For all subformulas $F \vee G$ of $\bar{\varphi}$ introduce a new concept name $B_{F \vee G}$ and add the inclusions $B_F \sqsubseteq B_{F \vee G}$ and $B_G \sqsubseteq B_{F \vee G}$.

The second group of axioms of $\mathcal{K}_{LL}$ constrains $T_i$ and $F_i$ to avoid inconsistencies. Intuitively $\exists U_i$ means "$p_i$ is undefined":

$$T_i \sqcap F_i \ \sqsubseteq \ \perp \quad (93) \qquad F_i \sqcap \exists U_i \ \sqsubseteq \ \perp \quad (95)$$

$$T_i \sqcap \exists U_i \ \sqsubseteq \ \perp \quad (94)$$

The third group of axioms of $\mathcal{K}_{LL}$ defines a tree that encodes the truth assignments needed to evaluate the QBF:

$$\text{for all } i \neq j \qquad A_i \sqcap A_j \sqsubseteq \perp \tag{96}$$

$$\text{for all } i \text{ s.t. } Q_i = \forall \qquad A_{i-1} \sqsubseteq \exists R.(T_i \sqcap A_i) \sqcap \exists R.(F_i \sqcap A_i) \tag{97}$$

$$\text{for all } i \text{ s.t. } Q_i = \exists \qquad A_{i-1} \sqsubseteq \exists R.A_i \tag{98}$$

The fourth group of axioms of $\mathcal{K}_{LL}$ detects misrepresentations by forcing role $bad$ to point to the nodes of the evaluation tree where something is going wrong (i.e. the truth assignment is incomplete, or some predicate changes value along a branch, or $\varphi$ is false in a leaf).

$$\top \ \sqsubseteq_n \ \exists bad.\exists U_i \tag{99}$$

$$\top \ \sqsubseteq_n \ \exists bad.E_i^j \tag{100}$$

$$\top \ \sqsubseteq_n \ \exists bad.\hat{E}_i^j \tag{101}$$

$$\top \ \sqsubseteq_n \ \exists bad.(B_{\bar{\varphi}} \sqcap A_n) \tag{102}$$

($E_i^j$ and $\hat{E}_i^j$ are defined in $\mathcal{K}_a$ below). Finally, $good$ captures the absence of $bad$:

$$\exists bad \sqcap \exists good \sqsubseteq \bot. \tag{103}$$

The acyclic terminology $\mathcal{K}_a$ has the only purpose of detecting whether some propositional symbol changes its value along a path:

$$E_i^j \quad \equiv \quad A_{j-1} \sqcap T_i \sqcap \exists R.(A_j \sqcap F_i) \tag{104}$$

$$\hat{E}_i^j \quad \equiv \quad A_{j-1} \sqcap F_i \sqcap \exists R.(A_j \sqcap T_i). \tag{105}$$

**Lemma 7.20** *Let $\mathcal{I}$ be a model of $\mathsf{Circ}_{\mathsf{fix}}(\mathcal{KB})$ that satisfies $A_0 \sqcap \exists good$. For all $i = 1 \ldots n$, all the individuals in $\Delta^{\mathcal{I}}$ are contained in either $T_i^{\mathcal{I}}$ or $F_i^{\mathcal{I}}$.*

**Proof.** First, by contradiction, assume that for some $i = 1 \ldots n$ and $x \in \Delta_i^{\mathcal{I}}$, $x$ is neither in $F_i^{\mathcal{I}}$ nor in $T_i^{\mathcal{I}}$. Since axioms (93)-(95) do not prevent $x$ from satisfying $\exists U_i$ and $\mathcal{I}$ must be minimal w.r.t. axiom (99), then the entire domain $\Delta^{\mathcal{I}}$ satisfies $\exists bad.\exists U_i$. However, by axiom (103), this means that $A_0 \sqcap \exists good$ is unsatisfiable against the hypothesis. $\square$

**Lemma 7.21** *Let $\mathcal{I}$ be a model of $\mathsf{Circ}_{\mathsf{fix}}(\mathcal{KB})$ that satisfies $A_0 \sqcap \exists good$. If for some $i = 1 \ldots n$, $x_i \in (A_i \sqcap T_i)^{\mathcal{I}}$ (respectively, $x \in (A_i \sqcap F_i)^{\mathcal{I}}$), then all paths $\{x_i, x_{i+1}, \ldots, x_n\}$ such that $x_j \in A_j$ and $(x_{j-1}, x_j) \in R^{\mathcal{I}}$, where $i < j \le n$, are contained in $T_i^{\mathcal{I}}$ (resp., $y \in F_i^{\mathcal{I}}$).*

**Proof.** Assume that $x_i \in (A_i \sqcap T_i)^{\mathcal{I}}$ and $\{x_i, x_{i+1}, \ldots, x_n\} \not\subseteq T_i^{\mathcal{I}}$. This means that $x_i \ne x_n$ and for some $i < j \le n$, $x_{j-1} \in (A_{j-1} \sqcap T_i)^{\mathcal{I}}$ and $x_j \in (A_{j-1} \sqcap F_i)^{\mathcal{I}}$. Then, by axiom (104), $x_{j-1} \in E_i^j$ and since $\mathcal{I}$ must be minimal w.r.t. axiom (100) the entire domain $\Delta^{\mathcal{I}}$ satisfies $\exists bad.E_i^j$. However, by axiom (103), this means that $A_0 \sqcap \exists good$ is unsatisfiable against the hypothesis. $\square$

**Theorem 7.22** *Concept satisfiability, subsumption checking, and instance checking are PSPACE-hard in $\mathsf{Circ}_{\mathsf{fix}}(a\mathcal{LL}\,\mathcal{EL}^{\perp})$. The result still holds if the nesting level of existential restrictions is bounded by a constant, and the priority relation is empty.*

**Proof.** In order to prove the theorem it suffices to show that the QBF $\psi$ is true iff $A_0 \sqcap \exists good$ is satisfiable w.r.t. the above $\mathcal{KB}$.

[*if*] Let $\mathcal{I}$ be a model of $\mathsf{Circ}_{\mathsf{fix}}(\mathcal{KB})$ that satisfies $A_0 \sqcap \exists good$. Due to axioms (96)-(98), $\mathcal{I}$ must contain a DAG that starts with $x$ (which is in $(A_0 \sqcap \exists good)^{\mathcal{I}}$) and, following the $R$-edges, proceeds through the concepts $A_i$ of increasing index, up to $A_n$. In this DAG, for all $i = 1 \ldots n$ such that $Q_i = \forall$, individuals belonging to a $A_i^{\mathcal{I}}$ have two successors: one in $A_{i+1}^{\mathcal{I}} \sqcap T_{i+1}^{\mathcal{I}}$ and the other in $A_{i+1}^{\mathcal{I}} \sqcap F_{i+1}^{\mathcal{I}}$. Individuals in $A_i^{\mathcal{I}}$, where $Q_i = \exists$, have only one successor, in $A_{i+1}^{\mathcal{I}}$. Due to Lemma 7.20, such a successor is either in $T_{i+1}^{\mathcal{I}}$ or $F_{i+1}^{\mathcal{I}}$.

Now, consider any truth assignment $v$ to the universally quantified variables of $\psi$. In the DAG, follow the unique path from $x$ to a leaf $z \in A_n^{\mathcal{I}}$, that for each level $i$ corresponding to a $Q_i = \forall$ proceeds with $A_{i+1}^{\mathcal{I}} \sqcap T_{i+1}^{\mathcal{I}}$ or $A_{i+1}^{\mathcal{I}} \sqcap F_{i+1}^{\mathcal{I}}$ in accordance with $v$. By Lemma 7.20, for all $i = 1 \ldots n$ $z$ is in either $T_i$ or $F_i$, moreover, by Lemma 7.21, membership of $z$ in $T_i$ or $F_i$ is consistent with $v$. Therefore, $z$ represents a full truth assignment of the variables in $\psi$ which extends $v$.

Now, since $\mathcal{I}$ minimizes the set of abnormal individuals w.r.t. the defeasible inclusion (99) and in all models $\exists good$ and $\exists bad$ are disjoint, $x \in \exists good^{\mathcal{I}}$ implies that $z \notin B_{\bar{\varphi}}^{\mathcal{I}}$. But then, it is straightforward to conclude that this truth assignment satisfies $\varphi$.

[*only if*]. Assume that $\psi$ is true. Assume w.l.o.g. that odd quantifiers $Q_1, Q_3, \ldots, Q_{n-1}$ are universal and even quantifiers are existential. For each existential quantifier $Q_i$, let $f_i : \{T, F\}^{i/2} \to \{T, F\}$ be the function such that for all values $v_1, v_3, \ldots, v_{n-1}$ of the universally quantified variables, $\varphi(v_1, f_2(v_1), v_3, f_4(v_1, v_3), \ldots, f_n(v_1, v_3, \ldots, v_{n-1}))$ is true.

We define a tree-like model $\mathcal{I}$ of $\mathcal{KB}$ that satisfies $A_0 \sqcap \exists good$. We start with a root individual $x$, such that $x \models^{\mathcal{I}} A_0 \sqcap \exists good$. We proceed inductively as follows. For even $i$ (including 0), each individual $y \in A_i^{\mathcal{I}}$ has two $R$-successors $y', y'' \in A_{i+1}^{\mathcal{I}}$, such that $y' \in T_i^{\mathcal{I}}$ and $y'' \in F_i^{\mathcal{I}}$ (see axioms (97)). For odd $i$, each individual $y \in A_i^{\mathcal{I}}$ has one $R$-successor $y' \in A_{i+1}^{\mathcal{I}}$ (see axioms (98)), such that $y'$ satisfies either $T_i$ or $F_i$, according to the value of $f_i$ when applied to the truth values that can be read along the path from the root to $y$. Along each $R - path$ $x_0 \ldots x_i \ldots x_n$ the same concept $T_i$ or $F_i$ assigned to $x_i$ is assigned to all $x_j$, with $i < j \leq n$, and indifferently either $T_i$ or $F_i$ is assigned to the $x_h$ with $1 \leq h \leq i$. The model is completed by assigning to $x_n$ *(i)* $B_{F \wedge G}$, for all subformulas $F \wedge G$ of $\bar{\varphi}$, such that $F$ and $G$ are assigned to $x_n$, and *(ii)* $B_{F \vee G}$, for all subformulas $F \vee G$ of $\bar{\varphi}$, such that $F$ or $G$ are assigned to $x_n$.

We leave to the reader the proof that the structure just defined satisfies the classical part of $\mathcal{KB}$. Regarding minimality w.r.t. the defeasible inclusions in $\mathcal{KB}$, we remark the following. All the individuals violate inclusions (99). However, due to rules (94) and (95), the situation cannot be improved by simply modifying the roles. Similarly, all the individuals violate inclusions (100)-(101). However, since the $E_i^j$ are all empty these defeasible inclusions cannot be improved.

Finally, since each leaf $z \in A_n$ represents a truth assignment that satisfies $\phi$, then $B_{\bar{\varphi}}$ is empty and hence our model is also minimal w.r.t. the inclusion (102). □

The $LL$ fragment of $\mathcal{EL}^\perp$, unlike $LL_f$, does not fully support unqualified existential. Consequently, Theorem 4.4 cannot by used to transfer the hardness results of Theorem 7.19 from var to fix.[9] The above hardness results hold only for the more general framework $\mathsf{Circ}_{\mathsf{var}}(LL_f\,\mathcal{EL}^\perp)$ and hence, by Theorem 4.4,for $\mathsf{Circ}_{\mathsf{fix}}(LL_f\,\mathcal{EL}^\perp)$:

**Proposition 7.23** *Subsumption and instance checking over* $\mathsf{Circ}_{\mathsf{fix}}(LL_f\,\mathcal{EL}^\perp)$ *are* $\Pi_2^p$*-hard; concept satisfiability is* $\Sigma_2^p$*-hard. These results hold even if queries contain only* $LL\,\mathcal{EL}^\perp$ *concepts, and priorities are specificity-based.*

The following result, whose proof can be found in the Appendix, shows a context in which the above lower bounds are tight: namely, the case in which the priority relation is empty (i.e., DIs are mutually incomparable) and, for subsumption queries $C \sqsubseteq D$ or instance checking queries $D(a)$, the quantifier depth of $D$ is bounded by a constant.

**Lemma 7.24** *Let* $\mathcal{KB} = \langle \mathcal{K}_S \cup \mathcal{K}_D, \emptyset \rangle$ *be an* $LL_f\mathcal{EL}^\perp$ *knowledge base, and* $C, D$ *be* $\mathcal{EL}^\perp$ *concepts. For all models* $\mathcal{I} \in \mathsf{Circ}_{\mathsf{fix}}(\mathcal{KB})$ *and for all* $x \in C^{\mathcal{I}} \setminus D^{\mathcal{I}}$ *there exists a model* $\mathcal{J} \in \mathsf{Circ}_{\mathsf{fix}}(\mathcal{KB})$ *such that (i)* $\Delta^{\mathcal{J}} \subseteq \Delta^{\mathcal{I}}$, *(ii)* $x \in C^{\mathcal{J}} \setminus D^{\mathcal{J}}$ *(iii)* $|\Delta^{\mathcal{J}}|$ *is* $O((|\mathcal{KB}| + |C|)^d)$ *where* $d = depth(D)$.

Going back to the $LL$ fragment, in the following we prove that $\mathsf{Circ}_{\mathsf{fix}}$ is less complex than $\mathsf{Circ}_{\mathsf{var}}$ (unless the polynomial hierarchy collapses). In particular, we show that $\mathsf{Circ}_{\mathsf{fix}}(LL\mathcal{EL}^\perp)$ is tractable. Algorithm 1 takes as input a knowledge base $\mathcal{KB}$ and two concepts $C$ and $D$ (we may assume without loss of generality that $C = A_C \sqcap \prod_{i=1}^n \exists P_i.B_i$) and checks whether $\mathsf{Circ}_{\mathsf{fix}}(\mathcal{KB}) \models C \sqsubseteq D$.

---

9. We will prove below that in $LL\,\mathcal{EL}^\perp$, $\mathsf{Circ}_{\mathsf{fix}}$ is actually less complex than $\mathsf{Circ}_{\mathsf{var}}$.

---

**Algorithm 1:**

---

**Data**: $C = A_C \sqcap \prod_1^n \exists P_i.B_i$, $D$, $\mathcal{KB} = \langle \mathcal{K}, \prec \rangle$.
A := $\{A \sqsubseteq_n \exists P.B \mid C \models A\}$;
X := $C$;
**while** A $\neq \emptyset$ **do**
    remove from A a defeasible inclusion $A \sqsubseteq_n \exists P.B$;
    **if** $\mathsf{SupCls}(\exists P.B) \subseteq \mathsf{SupCls}(C)$ *and* $NonEmpty(\exists P.B, \mathcal{K}_S) \subseteq NonEmpty(C, \mathcal{K}_S)$ **then**
        X := X $\sqcap \exists P.B$;
**return** X $\sqsubseteq_{\mathcal{K}_S} D$;

---

With $\mathsf{SupCls}(H)$ we mean the set of *superclasses* of a concept $H$, i.e., the set of $B \in \mathsf{N_C} \cup \{\bot\}$ such that $H \sqsubseteq_{\mathcal{K}_S} B$.

Given a concept $H$, the operator $NonEmpty(H, \mathcal{K}_S)$ represents the set of concepts that are forced to be *non empty* whenever $H$ is. Note that this set includes some concepts that are forced to be non empty by the ABox in $\mathcal{KB}$, independently from $H$. We write $H \rightsquigarrow A$ iff $H \sqsubseteq_{\mathcal{K}_S} \exists R.A$ for some $R$, and we denote by $\stackrel{+}{\rightsquigarrow}$ the transitive closure of $\rightsquigarrow$. Then, $NonEmpty(H, \mathcal{K}_S)$ is formally defined as follows:

$$NE\_Kernel = \{H\} \cup \bigcup_{a \in \mathsf{N_I}} \{A \mid \mathcal{K}_S \models A(a)\} \cup \bigcup_{a \in \mathsf{N_I}, R \in \mathsf{N_R}} \{A \mid \mathcal{K}_S \models (\exists R.A)(a)\}$$
$$NonEmpty(H, \mathcal{K}_S) = \bigcup_{A \in NE\_Kernel} \{A' \mid A \stackrel{+}{\rightsquigarrow} A'\}.$$

Roughly speaking, the algorithm accumulates the RHS of defeasible inclusions actively satisfied by a witness of $C$. Then, it tries to derive $D$. In particular, a defeasible inclusion $A \sqsubseteq_n \exists R.B$ is actively satisfied just in the case (i) does not entail locally $\bot$ or a concept name not subsumed by $C$, and (ii) does not entail globally the non-emptiness of a concept name that should be empty. The rationale is that concept names are fixed and circumscription cannot change their extension as the application of $A \sqsubseteq_n \exists R.B$ could instead require.

**Lemma 7.25** $\mathsf{Circ_{fix}}(\mathcal{KB}) \models C \sqsubseteq D$ holds iff *Algorithm 1 returns* true.

**Proof.** [*if*] It suffices to show that for all models of $\mathsf{Circ_{fix}}(\mathcal{KB})$, X subsumes $C$, where X is the formula obtained after the while statement. Assume per absurdum that for some model $\mathcal{I}$ and an individual $x \in \Delta^{\mathcal{I}}, x \in C^{\mathcal{I}} \setminus \mathsf{X}^{\mathcal{I}}$. This means that for some defeasible inclusion $A \sqsubseteq_n \exists P.B$ *(i)* $x \notin \mathsf{sat}_{\mathcal{I}}(A \sqsubseteq_n \exists P.B)$ and *(ii)* from line 1, $\mathsf{SupCls}(\exists P.B) \subseteq \mathsf{SupCls}(C)$ and $NonEmpty(\exists P.B, \mathcal{K}_S) \subseteq NonEmpty(C, \mathcal{K}_S)$.

Note that, since $NonEmpty(\exists P.B, \mathcal{K}_S) \subseteq NonEmpty(C, \mathcal{K}_S)$, whenever $\exists R.B \sqsubseteq_{\mathcal{K}_S} \exists S.\bar{B}$ there exists an individual $y_{\bar{B}} \in \bar{B}^{\mathcal{I}}$. Let $\mathcal{I}'$ be the interpretation obtained from $\mathcal{I}$ by adding all such $(x, y_{\bar{B}})$. Clearly, by adding new arcs the set of individuals that satisfied a defeasible inclusion $\delta$ cannot decrease, therefore for all $\delta \in \mathcal{K}_D$, $\mathsf{sat}_{\mathcal{I}}(\delta) \subseteq \mathsf{sat}_{\mathcal{I}'}(\delta)$. Moreover, since $x \in (\exists R.B)^{\mathcal{I}'}$, $\mathsf{sat}_{\mathcal{I}}(A \sqsubseteq_n \exists P.B) \subset \mathsf{sat}_{\mathcal{I}'}(A \sqsubseteq_n \exists P.B)$ and hence $\mathcal{I}' <_{\mathsf{fix}} \mathcal{I}$.

From condition *(ii)* and the fact that defeasible inclusions do not conflict with each other, it is easy to verify that $\mathcal{I}'$ is also a classical model of $\mathcal{KB}$, but this would mean that $\mathcal{I}$ is not a model of $\mathsf{Circ_{fix}}(\mathcal{KB})$ against the hypothesis.

[*only if*] Assume that Algorithm 1 returns false. Let $\mathcal{I}$ be the following interpretation:

- $\Delta^{\mathcal{I}} = \{x_C\} \cup \{x_A \mid A \in NonEmpty(C, \mathcal{K}_S)\} \cup \{x_a \mid a \in \mathsf{N_I}\}$;

- for all $B \in \mathsf{N_C}$, $B^{\mathcal{I}}$ is the union of: *(i)* $\{x_C\}$ if $B \in S_1$, *(ii)* the set of $x_A$, with $A \in S_2$, such that $A \models_{\mathcal{K}_S} B$, and *(iii)* the set of $x_a$, with $a \in \mathsf{N_I}$, such that $K_S \models B(a)$;

- for all $R \in \mathsf{N_R}$, $R^{\mathcal{I}}$ is the union of the pairs *(i)* $(x_A, x_B)$ where $A, B \in S_2$ and $A \sqsubseteq_{\mathcal{K}_S} \exists R.B$, *(ii)* $(x_a, x_b)$ where $a, b \in \mathsf{N_I}$ and $K_S \models R(a, b)$, *(iii)* $(x_a, x_B)$ where $a \in \mathsf{N_I}$, $B \in S_2$ and $\mathcal{K}_S \models \exists R.B(a)$, *(iv)* $(x_C, x_B)$ where $B \in S_2$ and $\mathsf{X} \sqsubseteq_{\mathcal{K}_S} \exists R.B$; other arcs are not relevant.

By construction $\mathcal{I}$ is a (classical) model of $\mathcal{K}_S$ and $x_C \in C^{\mathcal{I}} \setminus D^{\mathcal{I}}$, hence in order to prove $\mathcal{I} \in \mathsf{Circ_{fix}}(\mathcal{KB})$ it remains to show that $\mathcal{I}$ is minimal. Note that, since defeasible inclusions do not contain nested roles on the right side, the set of defeasible inclusions satisfied by an individual does not affect the set of defeasible inclusions satisfied by another individual. Therefore, an interpretation can be *improved* point-wise and we can assume w.l.o.g. that all the individuals in $\mathcal{I}$, except $x_C$, cannot be further improved. Assume now that there exists an interpretation $\mathcal{J}$ that improves $\mathcal{I}$ in $x_C$, this means in particular that for some $\delta = A \sqsubseteq_n \exists P.B$, $x_C \in \mathsf{sat}_{\mathcal{J}}(\delta) \setminus \mathsf{sat}_{\mathcal{I}}(\delta)$.

The assumption $x_C \notin \mathsf{sat}_{\mathcal{I}}(\delta)$ means that $\exists P.B$ does not satisfy the condition in line 1 and, since concept names are fixed, $\delta$ cannot be satisfied in $\mathcal{J}$. □

**Theorem 7.26** *In* $\mathsf{Circ_{fix}}(LL\,\mathcal{EL}^{\perp})$ *DKBs,* $LL\,\mathcal{EL}^{\perp}$ *subsumption, instance checking, and concept consistency are in P.*

**Proof.** Since $\mathsf{SupCls}(H)$ and $NonEmpty(H, \mathcal{K}_S)$ are based on classical reasoning, they can be performed in polynomial time. Moreover, the number of iterations in Algorithm 1 is bounded by the number of defeasible inclusions. Therefore, due to Lemma 7.25, the subsumption problem is tractable. By Theorem 3.9, instance checking and concept inconsistency can be reduced to subsumption. □

Complexity is low under $\mathsf{Circ_{fix}}$ because in this context $LL$ axioms are not general enough to simulate quantifier nesting nor conjunctions of existential restrictions. In $\mathsf{Circ_{var}}$ these features can be simulated by abbreviating compound concepts $C$ with concept names $A$ using equivalences $A \equiv C$ such that $C$ does not depend on qualified existentials (hence the $LL$ restriction is preserved). With $\mathsf{Circ_{fix}}$, such equivalences change the semantics of $C$ whenever $C$ is (or contains) an existential restriction, because $A$ is fixed and prevents $C$ from varying freely. As we reintroduce the missing features, complexity increases again.

Let $LL_2\mathcal{EL}^{\perp}$ support the schemata:

$$A \sqsubseteq_{[n]} \exists P.B \quad \exists P_1 \sqcap \exists P_2 \sqsubseteq \exists P_3.B \quad \exists P \sqsubseteq B$$

One may easily verify that $LL_2\mathcal{EL}^{\perp}$ is equivalent to $LL\,\mathcal{EL}^{\perp}$ plus schema $\exists P_1 \sqcap \exists P_2 \sqsubseteq \exists P_3.B$. The missing axioms can be reformulated using fresh roles $R$ and suitable equivalences $\exists R \equiv C$ (that preserve $C$'s semantics because $R$ is a varying predicate)[10].

With these additional schemata, one can create conflicts between variable concepts, as in $\exists P_1 \sqcap \exists P_2 \sqsubseteq \perp$. Then different defeasible inclusions may block each other, thereby creating a potentially exponential search space.

**Theorem 7.27** *Subsumption and instance checking over* $\mathsf{Circ_{fix}}(LL_2\mathcal{EL}^{\perp})$ *are coNP-hard; concept satisfiability is NP-hard. These results hold even if the three reasoning tasks are restricted to* $LL_2\mathcal{EL}^{\perp}$ *concepts, and priorities are specificity-based.*

---

10. In particular, schema $A_1 \sqcap A_2 \sqsubseteq B$ can be emulated by the inclusions $A_1 \sqsubseteq \exists R_1$, $A_2 \sqsubseteq \exists R_2$, $B \sqsubseteq \exists R_3$, $\exists R_1 \sqsubseteq A_1$, $\exists R_2 \sqsubseteq A_2$, $\exists R_3 \sqsubseteq B$, and $\exists R_1 \sqcap \exists R_2 \sqsubseteq \exists R_3$.

**Proof.** By reduction of SAT. For each propositional variable $p_i$ introduce the concept names $A_i$, $\bar{A}_i$, and role $U_i$, representing $p_i$'s truth value (resp. true, false, and undefined). These alternatives are made mutually inconsistent with:

$$A_i \sqcap \bar{A}_i \sqsubseteq \bot \qquad A_i \sqcap \exists U_i \sqsubseteq \bot \qquad \bar{A}_i \sqcap \exists U_i \sqsubseteq \bot$$

For each given clause $c_j = l_{j,1} \vee \cdots \vee l_{j,n}$, introduce a concept name $\bar{C}_j$ representing $c_j$'s falsity, and for each $\bar{L}_{j,k}$ representing the complement of $l_{j,k}$ add $\bar{L}_{j,1} \sqcap \cdots \sqcap \tilde{L}_{j,n} \sqsubseteq \bar{C}_j$.

Define a concept name $\bar{F}$ representing the falsity of the given set of clauses, and a disjoint concept $F$ with:

$$\bar{C}_j \sqsubseteq \bar{F} \quad \text{(for all input clauses } c_j\text{)} \qquad \bar{F} \sqcap F \sqsubseteq \bot.$$

Now, with a defeasible inclusion, $\exists U_i$ is forced to be true for all individuals that satisfy neither $A_i$ nor $\bar{A}_i$; moreover, a role $U$ detects undefined literals:

$$\top \sqsubseteq_n \exists U_i \qquad \exists U_i \sqsubseteq \exists U.$$

Let $\mathcal{K}$ be the above set of inclusions and $\mathcal{KB} = \langle \mathcal{K}, \prec_{\mathcal{K}} \rangle$. It can be proved that the given set of clauses $S$ is unsatisfiable iff $\mathsf{Circ}_{\mathsf{fix}}(\mathcal{KB}) \models F \sqsubseteq \exists U$, therefore subsumption checking is coNP-hard.

Similarly, it can be proved that $S$ is unsatisfiable iff $\mathsf{Circ}_{\mathsf{fix}}(\mathcal{KB}') \models (\exists U)(a)$, where $\mathcal{KB}' = \langle \mathcal{K}', \prec_{\mathcal{K}'} \rangle$ and $\mathcal{K}' = \mathcal{K} \cup \{F(a)\}$; therefore instance checking is coNP-hard.

Finally, it can be proved that $S$ is satisfiable iff $F \sqcap \exists OK$ is satisfiable w.r.t. $\mathsf{Circ}_{\mathsf{fix}}(\mathcal{KB}'')$, where $\mathcal{KB}'' = \langle \mathcal{K}'', \prec_{\mathcal{K}''} \rangle$ and $\mathcal{K}'' = \mathcal{K} \cup \{\exists U \sqcap \exists OK \sqsubseteq \bot\}$; therefore satisfiability checking is NP-hard.

We are only left to remark that $\mathcal{K}$ can be easily encoded in $LL_2\mathcal{EL}^\perp$. □

We prove that this bound is tight using Algorithm 2. The algorithm non-deterministically looks for an individual $x$ (in some model) that satisfies $C$ and not $D$. $S_1$ guesses any additional fixed concept names satisfied by $x$; $S_2$ guesses the concept names that are satisfied somewhere in the model (not necessarily by $x$) and finally $\prec'$ guesses a total extension of $\prec$ that determines the application order of GDIs.

Similarly to Algorithm 1, Algorithm 2 selects the defeasible inclusions that are *active* in $x$ and accumulates in the formula X the RHS of those that are not blocked, i.e. do not require to change the interpretation of the concept names neither locally nor globally. The major differences are that *(i)* defeasible inclusions are extracted according to $\prec'$ and *(ii)* in line 2 the entire accumulated formula X $\sqcap \exists P.B$ is used to check that a defeasible inclusion is not blocked.

Finally, note that the variable part of $C$ (i.e. $\bigsqcap_{i=1}^n \exists P_i.B_i$) is introduced in X only in line 8, after all defeasible inclusions have been applied, because defeasible inclusions can influence the variable part (e.g. by forcing it to be empty).

**Lemma 7.28** $\mathsf{Circ}_{\mathsf{fix}}(\mathcal{KB}) \models C \sqsubseteq D$ holds *iff all the runs of Algorithm 2 return* true.

**Proof.** [*if*] Assume per absurdum that there exists an interpretation $\mathcal{I} \in \mathsf{Circ}_{\mathsf{fix}}(\mathcal{KB})$ and an individual $x \in \Delta^{\mathcal{I}}$ such that $x \in C^{\mathcal{I}} \setminus D^{\mathcal{I}}$. Let $S_1$ and $S_2$ be the set of concept names in $\mathsf{N_C}$ that $\mathcal{I}$ satisfies respectively locally in $x$ and globally – i.e., for some individual. Let $\prec'$ be a linearization of $\prec$ *compatible* with $\mathcal{I}$, i.e. for all $\delta, \delta' \in \mathcal{K}_{\mathrm{D}}$ *(i)* either $\delta \prec' \delta'$ or $\delta' \prec' \delta$, *(ii)* $\delta \prec \delta'$ implies $\delta \prec' \delta'$, *(iii)* if $\delta$ and $\delta'$ are not comparable according to $\prec$ ($\delta \not\prec \delta'$ and $\delta' \not\prec \delta$) and $x \in \mathsf{sat}_{\mathcal{I}}(\delta) \setminus \mathsf{sat}_{\mathcal{I}}(\delta')$, then $\delta \prec' \delta'$.

---

**Algorithm 2**:

   **Data:** $C = A_C \sqcap \prod_1^n \exists P_i . B_i$, $D$, $\mathcal{KB} = \langle \mathcal{K}, \prec \rangle$.
   Guess $S_1, S_2 \subseteq \mathsf{N_C}$, where $\sqcap S_1 \models A_C$ and $S_1 \subseteq S_2$, and a linearization $\prec'$ of $\prec$;
   $\mathsf{A} := \{A \sqsubseteq_n \exists P.B \mid \sqcap S_1 \models A\}$;
   $\mathsf{X} := \prod S_1$;
   **while** $\mathsf{A} \neq \emptyset$ **do**
       remove from $\mathsf{A}$ the $\prec'$-minimal inclusion $A \sqsubseteq_n \exists P.B$;
       **if** $\mathsf{SupCls}(\mathsf{X} \sqcap \exists P.B) \subseteq S_1$ *and* $NonEmpty(\mathsf{X} \sqcap \exists P.B, \mathcal{K_S}) \subseteq S_2$ **then**
           $\mathsf{X} := \mathsf{X} \sqcap \exists P.B$;
   $\mathsf{X} := \mathsf{X} \sqcap \prod_1^n \exists P_i . B_i$;
   **return** $\mathsf{SupCls}(\mathsf{X}) \nsubseteq S_1$ *or* $NonEmpty(\mathsf{X}, \mathcal{K_S}) \nsubseteq S_2$ or $\mathsf{X} \sqsubseteq_{\mathcal{K_S}} D$;

---

Let $\mathsf{X}$ be the result of running Algorithm 2 on guesses $S_1$, $S_2$, and $\prec'$. It is straightforward to see that for all $\delta = A \sqsubseteq_n \exists P.B \in \mathcal{K_D}$ such that $\exists P.B$ occurs in $\mathsf{X}$, $x \in \mathsf{sat}_{\mathcal{I}}(\delta)$. This, together with the fact that $x \in C^{\mathcal{I}}$, implies that 1) $\mathsf{SupCls}(\mathsf{X}) \subseteq S_1$; 2) $NonEmpty(\mathsf{X}, \mathcal{K_S}) \subseteq S_2$ and, since $x \notin D^{\mathcal{I}}$, 3) $\mathsf{X} \nsqsubseteq_{K_S} D$. But this means that on this run Algorithm 2 should return false.

[*only if*] Assume that for some guess of $S_1$, $S_2$ and $\prec'$ Algorithm 2 returns false. Let $\mathcal{I}$ be defined as in Lemma 7.25. In a similar way it can be proved that $\mathcal{I}$ is a classical model of $\mathcal{K_S}$ and $x_C \in C^{\mathcal{I}} \setminus D^{\mathcal{I}}$. Assume now that $\mathcal{I}$ is improved by an interpretation $\mathcal{J}$, w.l.o.g. we can also assume that *(i)* for some $\delta = A \sqsubseteq_n \exists P.B$, $x_C \in \mathsf{sat}_{\mathcal{J}}(\delta) \setminus \mathsf{sat}_{\mathcal{I}}(\delta)$ and *(ii)* for all $\delta'$ with higher priority than $\delta$ or not comparable with it, we have $x_C \in \mathsf{sat}_{\mathcal{J}}(\delta')$ iff $x_C \in \mathsf{sat}_{\mathcal{I}}(\delta')$.

If $\mathsf{X}'$ is the value of $\mathsf{X}$ when $\delta$ is extracted on line 2 of Algorithm 2, since $\delta' \prec \delta$ implies $\delta' \prec' \delta$, all the $\delta'$ already extracted have higher priority or are not comparable with $\delta$. Since *(ii)* holds and by construction $x_C \in \mathsf{X}'^{\mathcal{I}}$, $x_C \in \mathsf{X}'^{\mathcal{J}}$. However, the assumption $x_C \notin \mathsf{sat}_{\mathcal{I}}(\delta)$ means that $\mathsf{X}' \sqcap \exists P.B$ does not satisfy the condition in line 2 and since concept names are fixed $\delta$ cannot be satisfied in $\mathcal{J}$. $\qquad\square$

**Theorem 7.29** $LL_2\mathcal{EL}^{\perp}$ *subsumption and instance checking over* $\mathsf{Circ_{fix}}(LL_2\mathcal{EL}^{\perp})$ *are in coNP;* $LL_2\mathcal{EL}^{\perp}$ *concept satisfiability is in NP.*

**Proof.** It is analogous to Theorem 7.26 and left to the reader. $\qquad\square$

It can be verified that the $LL_2$ fragment does not support quantifier nesting. With quantifier nesting, one would obtain $LL_f\mathcal{EL}^{\perp}$ (i.e. full $LL$).

# 8. Related Work

DLs have been extended with nonmonotonic constructs such as default rules (Straccia, 1993; Baader & Hollunder, 1995a, 1995b), autoepistemic operators (Donini et al., 1997, 2002), and circumscription (Cadoli, Donini, & Schaerf, 1990; Bonatti et al., 2009, 2009; Bonatti, Faella, & Sauro, 2010). An articulated comparison of these approaches can be found in the work of Bonatti, Lutz, and Wolter (2009).

Most of these approaches concern logics whose reasoning tasks' complexity lies beyond PH (unless the hierarchy collapses). For example, the logics considered by Donini et al. (1997, 2002) range from PSPACE to 3-ExpTime. The circumscribed DLs studied by Cadoli et al. (1990) as well

Table 1: Main complexity results. The corresponding decision problems for the classical versions of the considered logics are solvable in polynomial time.

| | DL-lite$_R$ | | $\mathcal{EL}$ | | $\mathcal{EL}^{\perp}$ | | | | |
| | var | fix$^{(\natural)}$ | var | fix | var | fix | | | |
| | | | | | $LL - aLL^{(\dagger)}$ | $LL$ | $LL_2$ | $LL_f$ | $aLL$ |
| concept satisfiability | $\Sigma_2^p$ | $\leq \Sigma_2^p$ | trivial | | $\Sigma_2^p$ | $\leq$P | NP | $\geq \Sigma_2^p$ $^{(\#)}$ | $\geq$PSPACE |
| subsumption & instance checking | $\Pi_2^p$ | $\leq \Pi_2^p$ | $\leq$P | $\geq$ExpTime | $\Pi_2^p$ $^{(\ddagger)}$ | $\leq$P | co-NP | $\geq \Pi_2^p$ $^{(\#)}$ | $\geq$PSPACE |

(†) with specificity-based priorities; general priorities make var at least as complex as fix

(‡) if quantifier nesting is bounded in the r.h.s. of subsumptions and in instance checking problems

(♮) if DIs are left-fixed or the priority relation is empty

(#) membership holds if the priority relation is empty and condition (‡) holds as Bonatti et al. (2009) range from NP$^{\text{NExp}}$ to NExp$^{\text{NP}}$. Some logics are undecidable (Baader & Hollunder, 1995a; Bonatti et al., 2009).

A pioneering approach to low-complexity, circumscribed description logics was presented by Cadoli et al. (1990). That approach applies non-prioritized circumscription to a fragment of the description logic $\mathcal{ALE}$. Decidability of reasoning is shown by a reduction to propositional reasoning under the Extended Closed World Assumption (ECWA), which is in $\Pi_2^p$. To the best of our knowledge, that was the first effective reasoning method for a nonmonotonic description logic.

A hybrid of Circ$_{\text{fix}}(\mathcal{EL}^{\perp})$ and closed world assumption has been proved to be in PTIME (Bonatti et al., 2010). On the one hand, that approach imposes less restrictions on the structure of inclusions; on the other hand, it cannot be fully extended to variable predicates without affecting tractability.

A recent approach that is similar in spirit to circumscription has been taken by Giordano et al. (2008). They extend $\mathcal{ALC}$ with a modal operator $T$ representing typicality, and maximize $T$'s extension to achieve nonmonotonic inferences. Decidability is proved via a tableau algorithm that also establishes a co-NExpTime$^{\text{NP}}$ upper bound for subsumption. No matching lower bounds are given; it is proved that reasoning in the underlying monotonic logic is NP-hard.

Finally, an approach based on *rational closures* and $\mathcal{ALC}$ can be found in the work of Casini and Straccia (2010). An appealing feature of this approach is that reasoning can be reduced to classical inference. Complexity is not increased by nonmonotonic reasoning: it ranges from PSPACE to ExpTime.

## 9. Discussion and Future Work

The main complexity results of this paper are summarized in Table 1. By restricting circumscribed KBs to Circ$_{\text{var}}$(DL-lite$_R$), complexity decreases significantly (from (co)-NExpTime$^{\text{NP}}$ to the second level of PH). The same complexity upper bounds hold in Circ$_{\text{fix}}$(DL-lite$_R$) whenever the priority relation is empty or the defeasible inclusions admit only concept names on the LHS. However, the general case is still an open question.

On the contrary, restricting the language to $\mathcal{EL}$ or $\mathcal{EL}^\perp$ does not in general suffice to bring complexity within PH. In particular, it can be proved that reasoning tasks are undecidable in $\mathrm{Circ}_F^*(\mathcal{EL})$ (i.e., when roles can be fixed) and that reasoning in $\mathrm{Circ}_{\mathrm{fix}}(\mathcal{EL})$ and $\mathrm{Circ}_{\mathrm{var}}(\mathcal{EL}^\perp)$ is in general ExpTime-hard.

The main source of the higher complexity of the $\mathcal{EL}$ family (w.r.t. DL-lite$_R$) has been identified by introducing a further restriction called *full left locality* ($LL_f$) that suffices to confine complexity within the second level of PH under $\mathrm{Circ}_{\mathrm{var}}$ with specificity-based priorities, provided that the quantifier nesting level in subsumption queries and instance checking queries is suitably bounded (no restrictions are needed on concept satisfiability).

Since the left locality restriction rules out acyclic terminologies (which are commonly used in ontologies), it has been relaxed to *almost left local* ($aLL$) knowledge bases, that support acyclic terminologies with the restriction that unfolding (i.e., the process of replacing the atoms defined in the acyclic terminology with their definition) should yield a $LL_f$ knowledge base. Reasoning becomes PSPACE-hard, in general; however in the $aLL$ fragment of $\mathrm{Circ}_{\mathrm{var}}$ (and under the same assumptions needed for $LL_f$), reasoning remains complete for the second level of PH. In particular, the assumption that the priorities are determined by specificity is essential: By Theorem 4.5, general priorities make $\mathrm{Circ}_{\mathrm{var}}$ at least as complex as $\mathrm{Circ}_{\mathrm{fix}}$, that is, PSPACE-hard.

We have also analyzed the complexity of several fragments lying between $LL$ and $aLL$ under $\mathrm{Circ}_{\mathrm{fix}}$. These results provide some further information about the complexity sources in circumscribed DLs. For example, quantifier nesting in the KB is partially responsible for complexity (presumably because it enables conflicts between the default properties of different individuals): in particular, by removing quantifier nesting (i.e., by restricting KBs to the $LL_2$ fragment) complexity drops to the first level of PH. The other source of complexity, of course, is due to the conflicts between defeasible inclusions concerning each individual in isolation; in $\mathrm{Circ}_{\mathrm{fix}}(LL\mathcal{EL}^\perp)$ a defeasible inclusion can never block another inclusion (because fixed predicates prevent this) and—as a consequence—complexity drops within PTIME.

We have also proved that in all fragments that fully support unqualified existential restrictions, variable concept names can be eliminated. Moreover, in $\mathcal{EL}^\perp$ and its various left local fragments, compound concepts can be replaced with concept names in the left-hand side of defeasible inclusions, without affecting expressiveness. In the same fragments, general priorities can be simulated using only specificity-based priorities.

We have to leave several interesting questions open: First, it is not clear whether general priorities are necessary to the hardness results for DL-lite$_R$; in particular, it would be interesting to establish the exact complexity of DL-lite$_R$ with specificity-based priorities. Other gaps in the complexity of circumscribed DL-lite$_R$ concern the complexity of $\mathrm{Circ}_{\mathrm{fix}}$ with unrestricted GDIs or nonempty priority relations, and the complexity of reasoning with fixed roles. The next interesting question is whether the bound on quantifier nesting in the queries is actually needed to confine complexity of circumscribed $\mathcal{EL}^\perp$ within the second level of PH. Finally, there is no exact charcterization of the complexity of $\mathrm{Circ}_{\mathrm{fix}}(LL_f\mathcal{EL}^\perp)$ and of the fragments whose complexity lies beyond PH.

For the fragments that do belong to the second level of PH, we see an interesting opportunity of encoding reasoning in ASP and use some well-engineered engine such as DLV (Eiter, Leone, Mateis, Pfeifer, & Scarcello, 1997) to test scalability. In order to evaluate implementations experimentally, it is necessary to set up suitable benchmarks that, in a first stage, must necessarily be synthetic problems, since nonmonotonic KBs have not been supported so far. Of course, identifying meaningful criteria for problem generation is a nontrivial issue. Therefore, systematic experimental evaluations still require a significant body of work.

## Acknowledgments

The authors wish to thank Frank Wolter for granting the permission to publish his undecidability proof. Moreover, they are grateful to Frank Wolter and Carsten Lutz for many stimulating discussions and feedback. This work has been carried out in the framework of project LoDeN, that has been (very) partially supported by the Italian Ministry for Research.

## Appendix A. Additional Lemmas and Proofs

### A.1 Proofs for Section 6

**Lemma 6.3.** *Given a PDLP $S$ over $PV = \{p_1, \ldots, p_n\}$ and a truth assignment $I \subseteq PV$, $I$ is a minimal model of $S$ iff the interpretation $model(S, I, \Delta)$ is a model of $\mathsf{Circ}_{\mathsf{var}}(\mathcal{KB}_S)$, for all domains $\Delta$ with $|\Delta| = 2n + 1$.*

**Proof.** [*only if*] Let $\mathcal{I} = model(S, I, \Delta)$, we first show that $\mathcal{I}$ is a model of the classical part of $\mathcal{KB}_S$. Since I-V, the Abox (18) and all axioms (20-24) are satisfied. Whereas, axioms (33-36) follow directly from VII.

Since $I$ is an interpretation, VI assures that if $TrueP^{\mathcal{I}} \neq \emptyset$, then $False\bar{P}^{\mathcal{I}} \neq \emptyset$. Together with VIII, axioms (31-32) are satisfied, whereas together with VII, $True$ and $False$ reflect in $\mathcal{I}$ the truth values of $I$; therefore $True^{\mathcal{I}} \cap False^{\mathcal{I}} = \emptyset$ and hence axiom (30) is satisfied. Moreover, as $I$ is a model of $S$, for each $c \in S$ there exists at least one literal $l_i$ occurring in $c$ such that $I \models l_i$. Due to V-VII, $C_i^{\mathcal{I}} \cap True^{\mathcal{I}} \neq \emptyset$ and, due to VII and VIII (where $X = True$ and $Y = C_j$), $\mathcal{I}$ satisfies axioms (26-29).

It remains to prove that there exists no interpretation $\mathcal{J}$ such that $\mathcal{J} <_{\mathsf{var}} \mathcal{I}$. As $\Delta^{\mathcal{I}}$ is finite, we can assume w.l.o.g. that $\mathcal{J}$ is a model of $\mathcal{KB}_S$. Assume that for some $1 \leq i \leq 2n$, $\mathsf{sat}_{\mathcal{I}}(L_i \sqsubseteq_n \neg L_i) \subset \mathsf{sat}_{\mathcal{J}}(L_i \sqsubseteq_n \neg L_i)$, this is equivalent to saying that $L_i^{\mathcal{J}} \subset L_i^{\mathcal{I}}$, but since $L_i^{\mathcal{I}}$ is a singleton, $L_i^{\mathcal{J}}$ would be empty contradicting axioms (18-23). Similarly, $\mathsf{sat}_{\mathcal{I}}(C_j \sqsubseteq_n \neg C_j) \subset \mathsf{sat}_{\mathcal{J}}(C_j \sqsubseteq_n \neg C_j)$ iff $C_j^{\mathcal{J}} \subset C_j^{\mathcal{I}}$. Thus, due to V some axiom $L_{ji} \sqsubseteq C_j$ would not be satisfied in $\mathcal{J}$. Therefore the defeasible inclusions with highest priority cannot be improved.

Now assume that for each literal and clause concept it holds $L_i^{\mathcal{J}} = L_i^{\mathcal{I}} = \{d_i\}$ and $C_j^{\mathcal{J}} = C_j^{\mathcal{I}}$. Since $\mathcal{I}$ reflects the truth values of $I$, all the $d_i$'s that are not included in $FalseL_i^{\mathcal{I}}$ are included in $TrueL_i^{\mathcal{I}}$. Thus, if for all $1 \leq i \leq 2n$, $\mathsf{sat}_{\mathcal{I}}(P_i \sqsubseteq_n FalseP_i)$ were equal to $\mathsf{sat}_{\mathcal{J}}(P_i \sqsubseteq_n FalseP_i)$ then there would be no way for $\mathcal{J}$ to improve a defeasible inclusion $P_i \sqsubseteq_n TrueP_i$. Therefore, the only possibility so far is that $\mathcal{J}$ improves some instance of (40).

Note that $True^{\mathcal{J}}$ and $False^{\mathcal{J}}$ are a partition of $\bigcup_i P_i^{\mathcal{J}}$. Otherwise, we could set a $P_i^{\mathcal{J}}$ without truth value (i.e., $d_i \notin TrueP_i^{\mathcal{J}} \cup FalseP_i^{\mathcal{J}}$) to $FalseP_i$ — since no classical inclusion is jeopardized we would obtain an improvement of $\mathcal{J}$ according to (40), against the hypothesis that $\mathcal{J}$ is a model. Due to (31-36), $True^{\mathcal{J}}$ and $False^{\mathcal{J}}$ are a partition of all $\{d_1, \ldots, d_{2n}\}$.

Thus, we can consider the propositional assignment $J$ such that $p_i \in J$ iff $P_i^{\mathcal{J}} \subseteq True^{\mathcal{J}}$.

First, for all clauses $c_j$, since $\mathcal{J}$ satisfies axioms (26-29), for some $1 \leq i \leq 2n$ we have $L_i^{\mathcal{J}} \subseteq C_j^{\mathcal{J}} \cap True^{\mathcal{J}}$. As $L_i^{\mathcal{J}} = L_i^{\mathcal{I}}$ and $C_j^{\mathcal{J}} = C_j^{\mathcal{I}}$, $l_{jl}$ occurs in $c_j$. But this means that $J \models l_i$, and hence $J \models c_j$. Thus, $J$ is a model of $S$. Finally, as said before, if $\mathcal{J} <_{\mathsf{var}} \mathcal{I}$, then the intersection of $\bigcup_i P_i^{\mathcal{I}}$ and $\mathit{False}^{\mathcal{I}}$ is strictly contained in the intersection of $\bigcup_i P_i^{\mathcal{J}}$ and $\mathit{False}^{\mathcal{J}}$. This implies that $J \subset I$, against the hypothesis that $I$ is a minimal model.

[*if*] Assume that $\mathcal{I} = model(S, I, \Delta)$ is a model of $\mathsf{Circ}_{\mathsf{var}}(\mathcal{KB}_S)$. First we show that $I$ is a model of $S$, i.e., for all $c_j \in S$, $I \models c_j$. As $\mathcal{I}$ satisfies axioms (26-29), for some $1 \leq i \leq 2n$ it holds $d_i \in C_j^{\mathcal{I}} \cap \mathit{True}^{\mathcal{I}}$. Due to IV and V, $d_i$ belongs to some $L_i^{\mathcal{I}}$ such that the corresponding literal $l_i$ occurs in $c_j$. According to VI, this implies that $I \models l_i$, and hence $I \models c_j$.

It remains to show that $I$ is a minimal model. Assume that there exists a model $J$ of $S$ such that $J \subset I$, without loss of generality we can assume that $J$ is minimal model. Let $\mathcal{J} = model(S, J, \Delta)$, from the above arguments $\mathcal{J}$ is a model of $\mathcal{KB}_S$. Furthermore, $\mathsf{sat}_{\mathcal{J}}(\delta) = \mathsf{sat}_{\mathcal{I}}(\delta)$ for each $\delta$ of type $L_i \sqsubseteq_n \neg L_i$ or $C_j \sqsubseteq_n \neg C_j$. Finally, as $\mathit{True}^{\mathcal{J}}$ and $\mathit{False}^{\mathcal{J}}$ reflect the truth values of $J$, $\bigcup_i P_i^{\mathcal{I}} \cap \mathit{False}^{\mathcal{I}} \subset \bigcup_i P_i^{\mathcal{J}} \cap \mathit{False}^{\mathcal{J}}$ and hence $\mathcal{J} <_{\mathsf{var}} \mathcal{I}$ due to the improvement of DIs (40). □

**Lemma 6.4.** *If $\mathcal{I}$ is a model of $\mathsf{Circ}_{\mathsf{var}}(\mathcal{KB}_S)$, then there exist a minimal model $I$ of $S$ such that $p_i \in I$ iff $P_i^{\mathcal{I}} \subseteq \mathit{True}^{\mathcal{I}}$ iff $\bar{P}_i^{\mathcal{I}} \subseteq \mathit{False}^{\mathcal{I}}$, for all $i = 1, \ldots, n$.*

**Proof.** Let $\mathcal{I}$ be a model of $\mathsf{Circ}_{\mathsf{var}}(\mathcal{KB}_S)$. First, we show that $L_i^{\mathcal{I}}$ is a singleton, for all $1 \leq i \leq 2n$. Assume the contrary. Clearly, to satisfy (18–21), each $L_i^{\mathcal{I}}$ has to be nonempty. Therefore, for some $1 \leq k \leq n$, $L_k^{\mathcal{I}}$ contains at least two individuals. We will show that there exists an interpretation $\mathcal{I}'$ that improves $\mathcal{I}$.

For all $1 \leq i \leq 2n$, let $d_i$ be an arbitrary element of $L_i^{\mathcal{I}}$, and let $\Delta = \{d_1, \ldots, d_{2n}\} \cup \{a^{\mathcal{I}}\}$. As the $L_i^{\mathcal{I}}$ are disjoint with each other (see axioms (22)) and $\mathit{NonEmpty}^{\mathcal{I}}$ is disjoint with any $L_i^{\mathcal{I}}$, we have $|\Delta| = 2n + 1$.

All PDLP are satisfiable, thus there exists a model $\hat{I}$ of $S$. Let $\widehat{\mathcal{I}} = model(S, \hat{I}, \Delta)$. Let $\mathcal{I}'$ be an interpretation such that: *(i)* $\Delta^{\mathcal{I}'} = \Delta^{\mathcal{I}}$, *(ii)* for all roles $R$, $R^{\mathcal{I}'} = \Delta^{\mathcal{I}} \times \Delta^{\mathcal{I}}$, *(iii)* $\mathcal{I}'$ coincides with $\widehat{\mathcal{I}}$ on $\Delta$ with respect to all concept names, and *(iv)* all the other individuals $d \in \Delta^{\mathcal{I}} \setminus \Delta$ do not belong to any concept name. It is straightforward to see that $\mathcal{I}'$ satisfies the classical part of $\mathcal{KB}_S$. Furthermore, by construction, *(a)* for all $1 \leq i \leq 2n$, $L_i^{\mathcal{I}'} \subseteq L_i^{\mathcal{I}}$; *(b)* for all $1 \leq j \leq m$, $C_j^{\mathcal{I}'} \subseteq C_j^{\mathcal{I}}$; *(c)* for some $1 \leq l \leq 2n$, $L_l^{\mathcal{I}'} \subset L_l^{\mathcal{I}}$. Thus, $\mathcal{I}' <_{\mathsf{var}} \mathcal{I}$, due to the improvement of DI (23).

By the above argument, we have $L_i^{\mathcal{I}} = \{d_i\}$. Define the truth valuation $I = \{p_i \mid d_i \in \mathit{True}^{\mathcal{I}}\}$. It remains to prove that $I$ is a minimal model of $S$. The fact that $I$ is a model of $S$ is ensured by axioms (24–29). Then, assume by contradiction that there exists a model $J$ of $S$ that is smaller than $I$ (i.e., $J \subset I$), and let $\mathcal{J} = model(S, J, \Delta)$. From $\mathcal{J}$ we can build an interpretation $\mathcal{J}'$ with $\Delta^{\mathcal{J}'} = \Delta^{\mathcal{I}}$ and such that $\mathcal{J}'$ is a classical model of $\mathcal{KB}_S$ and $\mathcal{J}' <_{\mathsf{var}} \mathcal{I}$, thus contradicting the hypothesis that $\mathcal{I}$ is a model of $\mathsf{Circ}_{\mathsf{var}}(\mathcal{KB}_S)$. We define $\mathcal{J}'$ by copying from $\mathcal{J}$ all the properties (concepts and roles) of the individuals in $\Delta^{\mathcal{J}} = \Delta$, and by leaving all the individuals in $\Delta^{\mathcal{I}} \setminus \Delta$ out of concept and role extensions. □

## A.2 Proofs for Section 7

Given a KB $\mathcal{K}$, an interpretation $\mathcal{I}$, and an individual $z$, recall the definition of $\mathcal{KB}^{[z]}$ from Section 6.2. Redefine the notion of "support" as follows: $\mathsf{supp}_{\mathcal{I}}(C)$ is the set of individuals $z \in \Delta^{\mathcal{I}}$ such that $\top \sqsubseteq_{\mathcal{KB}^{[z]}} C$ holds.

**Lemma 7.12.** *Let $\mathcal{KB} = \langle \mathcal{K}, \prec_{\mathcal{K}} \rangle$ be an $LL_f \mathcal{EL}^{\perp}$ knowledge base, and $C, D$ be $\mathcal{EL}^{\perp}$ concepts. For all models $\mathcal{I} \in \mathsf{Circ}_{\mathsf{var}}(\mathcal{KB})$ and for all $x \in C^{\mathcal{I}} \setminus D^{\mathcal{I}}$ there exists a model $\mathcal{J} \in \mathsf{Circ}_{\mathsf{var}}(\mathcal{KB})$ such that* (i) $\Delta^{\mathcal{J}} \subseteq \Delta^{\mathcal{I}}$, (ii) $x \in C^{\mathcal{J}} \setminus D^{\mathcal{J}}$, *and* (iii) $|\Delta^{\mathcal{J}}|$ *is* $O((|\mathcal{KB}|^2 + |C|)^d)$, *where* $d = depth(D) + 1$.

**Proof.** Given two individuals $x$ and $y$ in $\Delta^{\mathcal{I}}$, the *distance* $d(x, y)$ is the minimal length of role-paths from $x$ to $y$. Let $\mathcal{KB}^*$ be the knowledge base obtained from $\mathcal{KB}$ by applying the transformation presented in Section 7. Notice that $|\mathcal{KB}^*| \leq |\mathcal{KB}|^2$.

By Lemma 7.10, $\mathcal{I}$ can be extended into a model of $\mathsf{Circ}_{\mathsf{var}}(\mathcal{KB}^*)$, which we continue to call $\mathcal{I}$ for convenience. We define a *small* model $\mathcal{J}$ of $\mathsf{Circ}_{\mathsf{var}}(\mathcal{KB}^*)$ such that $x \in C^{\mathcal{J}} \setminus D^{\mathcal{J}}$. Then, we obtain the thesis by Lemma 7.11.

We start from an initial domain $\Delta^{\mathcal{J}}$ that contains *(i)* $x$; *(ii)* all $a^{\mathcal{I}}$, where $a \in \mathsf{N}_\mathsf{I}$ occurs in $\mathcal{KB}^*$; *(iii)* for all concepts $H \in \mathsf{cl}(\mathcal{KB}^*) \cup \mathsf{cl}(C)$ such that $H^{\mathcal{I}} \neq \emptyset$, a witness $y_H \in H^{\mathcal{I}}$; *(iv)* for all concepts $H \in \mathsf{cl}(\mathcal{KB}^*) \cup \mathsf{cl}(C)$ such that $\mathsf{supp}_{\mathcal{I}}(H) \neq \emptyset$, a witness $w_H \in \mathsf{supp}_{\mathcal{I}}(H)$.

We expand $\mathcal{J}$ by exhaustively applying the following rule (where $\exists P$ is a special case of $\exists P.H$ with $H = \top$):

*Let $y \in \Delta^{\mathcal{J}}$ and $\exists P.H \in \mathsf{cl}(\mathcal{KB}^*) \cup \mathsf{cl}(C)$ be such that $y \in (\exists P.H)^{\mathcal{I}}$ and $y \notin (\exists P.H)^{\mathcal{J}}$. If $d(x, y) < d$, then add $z$ to $\Delta^{\mathcal{J}}$ and $(y, z)$ to $P^{\mathcal{J}}$, where $z$ is such that $(y, z) \in P^{\mathcal{I}}$ and $z \in H^{\mathcal{I}}$. Otherwise, add $(y, y_H)$ to $P^{\mathcal{J}}$.*

Finally, for each concept name $A$, set $A^{\mathcal{J}} = \Delta^{\mathcal{J}} \cap A^{\mathcal{I}}$.

With respect to the cardinality of $\Delta^{\mathcal{J}}$, note that initially the number of individuals in $\Delta^{\mathcal{J}}$ is $O(|\mathcal{KB}^*| + |C|)$. During the expansion, for each individual whose distance from $x$ is less than $d$, at most $O(|\mathcal{KB}^*| + |C|)$ new individuals are added. This means that $|\Delta^{\mathcal{J}}| = O((|\mathcal{KB}^*| + |C|)^d) = O((|\mathcal{KB}|^2 + |C|)^d)$.

By construction for each individual $y \in \Delta^{\mathcal{J}}$ and $H \in \mathsf{cl}(\mathcal{KB}^*) \cup \mathsf{cl}(C)$ if $y \in H^{\mathcal{I}}$, then $y \in H^{\mathcal{J}}$. In particular, in case $H = \exists P$, also the inverse holds, if $y \in \exists P^{\mathcal{J}}$, then $y \in \exists P^{\mathcal{I}}$. From the previous two facts it immediately follows that $\mathcal{J}$ is a classical model of $\mathcal{KB}^*$ and $x \in C^{\mathcal{J}}$. Moreover, since up to a distance $d$ from $x$, $P^{\mathcal{J}}$ is contained in $P^{\mathcal{I}}$, for all $P \in \mathsf{N}_\mathsf{R}$, it is easy to see that $x \notin D^{\mathcal{J}}$.

It remains to show that $\mathcal{J}$ is minimal. Assume by contradiction that for some classical model $\mathcal{J}'$ of $\mathcal{KB}^*$, it holds $\mathcal{J}' <_{\mathsf{var}} \mathcal{J}$, we show that there exists a classical model $\mathcal{I}'$ of $\mathcal{KB}^*$ such that $\mathcal{I}' <_{\mathsf{var}} \mathcal{I}$ — against the hypothesis that $\mathcal{I} \in \mathsf{Circ}_{\mathsf{var}}(\mathcal{KB}^*)$.

We distinguish two cases: in the first cas, all individuals $w_H$ introduced in clause (iv) still satisfy the corresponding concept $H$ in $\mathcal{J}'$; in the second case, at least one $w_H$ does not satisfy its concept $H$. We define $\mathcal{I}'$ as follows. In both cases, individual names are interpreted as in $\mathcal{I}$ and concept names for individuals in $\Delta^{\mathcal{J}}$ are interpreted as in $\mathcal{J}'$.

In the first case, an individual $z \in \Delta^{\mathcal{I}} \setminus \Delta^{\mathcal{J}}$ satisfies a concept name $A$, that is $z \in A^{\mathcal{I}'}$, if and only if $z \in \mathsf{supp}_{\mathcal{I}}(A)$. Moreover, for each $P \in \mathsf{N}_\mathsf{R}$, $P^{\mathcal{I}'}$ is the minimal set such that:

1. $P^{\mathcal{J}'} \subseteq P^{\mathcal{I}'}$;

2. if $z \in \Delta^{\mathcal{I}} \setminus \Delta^{\mathcal{J}}$ and $z \in \mathsf{supp}_{\mathcal{I}}(\exists P.H)$, and $y \in H^{\mathcal{J}'}$ then $(z, y) \in P^{\mathcal{I}'}$.

We prove that $\mathcal{I}'$ is a classical model of $\mathcal{KB}^*$. Since $\mathcal{I}'$ is a *copy* of $\mathcal{J}'$ over $\Delta^{\mathcal{J}}$ and $\mathcal{J}'$ is a classical model, we only need to show that the individuals $z \in \Delta^{\mathcal{I}} \setminus \Delta^{\mathcal{J}}$ satisfy all the strong inclusions in $\mathcal{K}_S$. Note that if $z$ satisfies in $\mathcal{I}'$ the LHS $H$ of a strong inclusion, then $z$ supports $H$ in $\mathcal{I}$. By definition, $z$ supports also the RHS in $\mathcal{I}$. If the RHS is a concept name $B$, then $z \in B^{\mathcal{I}'}$ by construction. Otherwise, i.e., if the RHS is $\exists P.H$, by step 2 above, it suffices to show that $H^{\mathcal{J}'}$ is not empty. However, by assumption, the witness of $\exists P.H$ introduced in clause (iv) still satisfies $\exists P.H$ in $\mathcal{J}'$. Therefore, there exists an individual $y$ satisfying $H$ in $\mathcal{J}'$.

Next, we prove that $\mathcal{I}' <_{\mathsf{var}} \mathcal{I}$. Since $\mathcal{J}' <_{\mathsf{var}} \mathcal{J}$, it suffices to show that an individual $z \in \Delta^{\mathcal{I}} \setminus \Delta^{\mathcal{J}}$ satisfies in $\mathcal{I}'$ all the defeasible inclusions it satisfies in $\mathcal{I}$. Assume that a DI $\delta = (A \sqsubseteq_n B)$ is satisfied by $z$ in $\mathcal{I}$. If $z \in A^{\mathcal{I}'}$, then by construction $z \in \mathsf{supp}_{\mathcal{I}}(A)$. Clearly, if $z \in \mathsf{supp}_{\mathcal{I}}(A)$ and $z \in \mathsf{sat}_{\mathcal{I}}(A \sqsubseteq_n B)$, then $z \in \mathsf{supp}_{\mathcal{I}}(B)$. Therefore, $z \in B^{\mathcal{I}'}$.

We are left to prove the theorem for the second case, i.e.: at least one $w_H$ does not satisfy its concept $H$. Clearly, $w_H$ does not support $H$ in $\mathcal{J}'$ anymore. In particular, there must be a DI $\delta$ such that $w_H \in \mathsf{sat}_{\mathcal{J}}(\delta) \setminus \mathsf{sat}_{\mathcal{J}'}(\delta)$. From $\mathcal{J}' <_{\mathsf{var}} \mathcal{J}$, it follows that there must be a DI $\delta'$ such that $\delta' \prec_{\mathcal{K}} \delta$ and $\mathsf{sat}_{\mathcal{J}}(\delta') \subset \mathsf{sat}_{\mathcal{J}'}(\delta')$. Now, in $\mathcal{I}'$ we can safely violate all DIs whose priority is lower than $\delta'$, and in particular all DIs whose LHS classically subsumes $\top$. Then, complete the definition of $\mathcal{I}'$ as follows. Each basic concept $A$ holds in an individual $z \in \Delta^{\mathcal{I}} \setminus \Delta^{\mathcal{J}}$ if and only if $\top \sqsubseteq_{\mathcal{K}_{\mathsf{S}}} A$. For each $P \in \mathsf{N}_{\mathsf{R}}$, $P^{\mathcal{I}'}$ is the minimal set such that

1. $P^{\mathcal{J}'} \subseteq P^{\mathcal{I}'}$;

2. if $z \in \Delta^{\mathcal{I}} \setminus \Delta^{\mathcal{J}}$, $\top \sqsubseteq_{\mathcal{K}_{\mathsf{S}}} \exists P.H$, and $y \in H^{\mathcal{J}'}$ then $(z, y) \in P^{\mathcal{I}'}$.

It is easy to verify that $\mathcal{I}'$ is a classical model of $\mathcal{KB}^*$. In order to prove that $\mathcal{I}' <_{\mathsf{var}} \mathcal{I}$, note that the following two facts hold.

First, an individual $z \in \Delta^{\mathcal{I}} \setminus \Delta^{\mathcal{J}}$ satisfies all the DIs whose priority is not minimal. Assume that $\delta_1 \prec_{\mathcal{K}} \delta_2$, for some DIs $\delta_1$ and $\delta_2$, this means that the LHS of $\delta_2$ subsumes the LHS of $\delta_1$ but not the vice versa. Then, the LHS of $\delta_1$ does not subsume $\top$ and hence, by construction, $z$ vacuously satisfies $\delta_1$.

Second, if $z$ violates a DI $\delta'' = (A \sqsubseteq_n B)$, then $\delta' \prec_{\mathcal{K}} \delta''$. As before, since $\delta' \prec_{\mathcal{K}} \delta$, its LHS does not subsume $\top$. However, since $z$ violates $\delta''$, $z \in A^{\mathcal{I}'}$ and hence $A$ subsumes $\top$. Therefore, $\delta' \prec_{\mathcal{K}} \delta''$.

From the first fact it immediately follows that $\mathsf{sat}_{\mathcal{I}}(\delta') \subset \mathsf{sat}_{\mathcal{I}'}(\delta')$. Assume now, that for some $\delta''$, $\mathsf{sat}_{\mathcal{I}}(\delta'') \not\subseteq \mathsf{sat}_{\mathcal{I}'}(\delta'')$. If there exists an individual $z \in \Delta^{\mathcal{I}} \setminus \Delta^{\mathcal{J}}$ such that $z \in \mathsf{sat}_{\mathcal{I}}(\delta'') \setminus \mathsf{sat}_{\mathcal{I}'}(\delta'')$, then by the second fact $\delta' \prec_{\mathcal{K}} \delta''$. Otherwise, there must exist an individual $w \in \Delta^{\mathcal{J}}$ such that $w \in \mathsf{sat}_{\mathcal{I}}(\delta'') \setminus \mathsf{sat}_{\mathcal{I}'}(\delta'')$. However, in $\Delta^{\mathcal{J}}$ the set of DIs that an individual satisfies in $\mathcal{I}$ (resp. $\mathcal{I}'$) is the same set of DIs it satisfies in $\mathcal{J}$ (resp. $\mathcal{J}'$). This means that there exists a defeasible inclusion $\delta'''$ such that $\delta''' \prec_{\mathcal{K}} \delta''$ and $\mathsf{sat}_{\mathcal{J}}(\delta''') \subset \mathsf{sat}_{\mathcal{J}'}(\delta''')$. Due to the first fact, $\delta'''$ is satisfied in all $\Delta^{\mathcal{I}} \setminus \Delta^{\mathcal{J}}$, and hence $\mathsf{sat}_{\mathcal{I}}(\delta''') \subset \mathsf{sat}_{\mathcal{I}'}(\delta''')$. □

**Proposition 7.13.** *Let* $\mathcal{KB} = \langle \mathcal{K}_{LL} \cup \mathcal{K}_a, \prec \rangle$ *be an aLL* $\mathcal{EL}^\perp$ *knowledge base. Every model of* $\mathsf{Circ}_F(\mathsf{unf}(\mathcal{KB}))$ *can be extended to a model of* $\mathsf{Circ}_F(\mathcal{KB})$.

**Proof.** Let $\mathcal{K}_{\mathsf{unf}} = \mathsf{unf}(\mathcal{KB}) = \langle \mathcal{K}', \prec' \rangle$. Let $\mathcal{I}$ be any model of $\mathsf{Circ}_F(\mathcal{K}_{\mathsf{unf}})$. Extend it to a classical model $\mathcal{J}$ of $\mathcal{K}' \cup \mathcal{K}_a$ by setting $A^{\mathcal{J}} = D^{\mathcal{J}}$ for all definitions $A \equiv D$ in $\mathcal{K}_a$. Note that $\mathcal{K}' \cup \mathcal{K}_a$ is classically equivalent to $\mathcal{K}_{LL} \cup \mathcal{K}_a$. Now suppose that $\mathcal{J}$ is not a model of $\mathsf{Circ}_F(\mathcal{KB})$. Since by construction $\mathcal{J}$ is a classical model of $\mathcal{K}_a$ and the strong axioms of $\mathcal{K}_{LL}$, there must be a classical model $\mathcal{J}'$ of the same axioms such that $\mathcal{J}' <_F \mathcal{J}$. By restricting $\mathcal{J}'$ to primitive predicates (i.e., predicates that are not defined in $\mathcal{K}_a$), we obtain a classical model $\mathcal{I}'$ of $\mathcal{K}'$. Note that for all defeasible inclusions $\delta \in \mathcal{K}_{LL}$, it holds $\mathsf{sat}_{\mathcal{J}'}(\delta) = \mathsf{sat}_{\mathcal{J}'}(\mathsf{unf}(\delta, \mathcal{K}_a)) = \mathsf{sat}_{\mathcal{I}'}(\mathsf{unf}(\delta, \mathcal{K}_a))$ and $\mathsf{sat}_{\mathcal{J}}(\delta) = \mathsf{sat}_{\mathcal{J}}(\mathsf{unf}(\delta, \mathcal{K}_a)) = \mathsf{sat}_{\mathcal{I}}(\mathsf{unf}(\delta, \mathcal{K}_a))$. It follows that $\mathcal{I}' <_F \mathcal{I}$, a contradiction. □

**Proposition 7.14.** *Let* $\mathcal{KB} = \langle \mathcal{K}_{LL} \cup \mathcal{K}_a, \prec \rangle$ *be an aLL* $\mathcal{EL}^\perp$ *knowledge base and suppose that all the concept names defined in* $\mathcal{K}_a$ *are variable. Then, for all models* $\mathcal{I}$ *of* $\mathsf{Circ}_F(\mathcal{KB})$*, the restriction of* $\mathcal{I}$ *to primitive predicates is a model of* $\mathsf{Circ}_F(\mathsf{unf}(\mathcal{KB}))$.

**Proof.** Let $\mathcal{K}_{\mathsf{unf}} = \mathsf{unf}(\mathcal{KB}) = \langle \mathcal{K}', \prec' \rangle$. Let $\mathcal{J}$ be the restriction of $\mathcal{I}$ to primitive predicates. In other words, $\mathcal{J}$ is obtained from $\mathcal{I}$ by dropping the interpretation of all concept names defined in $\mathcal{K}_a$, which are all variable. It can be easily verified that $\mathcal{J}$ is a classical model of $\mathcal{K}'$. Suppose by contradiction that $\mathcal{J}$ is not a model of $\mathsf{Circ}_F(\mathcal{K}_{\mathsf{unf}})$; then there exists a classical model $\mathcal{J}'$ of $\mathcal{K}'$ such that $\mathcal{J}' <_F \mathcal{J}$. Now extend $\mathcal{J}'$ to a model $\mathcal{I}'$ of $\mathcal{K}_{LL} \cup \mathcal{K}_a$ by setting $A^{\mathcal{I}'} = D^{\mathcal{I}'}$ for all definitions $A \equiv D$ in $\mathcal{K}_a$. Since the predicates defined in $\mathcal{K}_a$ are variable, all fixed predicates preserve their extensions across $\mathcal{I}$, $\mathcal{J}$, $\mathcal{J}'$, and $\mathcal{I}'$. Moreover, for all defeasible inclusions $\delta \in \mathcal{K}_{LL}$, we have $\mathsf{sat}_{\mathcal{I}'}(\delta) = \mathsf{sat}_{\mathcal{I}'}(\mathsf{unf}(\delta, \mathcal{K}_a)) = \mathsf{sat}_{\mathcal{J}'}(\mathsf{unf}(\delta, \mathcal{K}_a))$ and $\mathsf{sat}_{\mathcal{I}}(\delta) = \mathsf{sat}_{\mathcal{I}}(\mathsf{unf}(\delta, \mathcal{K}_a)) = \mathsf{sat}_{\mathcal{J}}(\mathsf{unf}(\delta, \mathcal{K}_a))$. It follows that $\mathcal{I}' <_F \mathcal{I}$, a contradiction. $\qquad\square$

Given a knowledge base $\mathcal{KB}$, an interpretation $\mathcal{I}$ and a concept $D$, again we have to override the notion of support; $\mathsf{supp}_{\mathcal{I}}(D)$ is the set of $z \in \Delta^{\mathcal{I}}$ such that

$$\Big( \bigcap_{z \in A^{\mathcal{I}}} A \Big) \sqsubseteq_{\mathcal{KB}^{[z]}} D \,.$$

Clearly, if $\mathcal{I}$ is a classical model of $\mathcal{KB}$ and $z \in \mathsf{supp}_{\mathcal{I}}(D)$, then $z \in D^{\mathcal{I}}$.

**Lemma 7.24.** *Let* $\mathcal{KB} = \langle \mathcal{K}_S \cup \mathcal{K}_D, \emptyset \rangle$ *be an* $LL_f \mathcal{EL}^{\perp}$ *knowledge base, and* $C, D$ *be* $\mathcal{EL}^{\perp}$ *concepts. For all models* $\mathcal{I} \in \mathsf{Circ}_{\mathsf{fix}}(\mathcal{KB})$ *and for all* $x \in C^{\mathcal{I}} \setminus D^{\mathcal{I}}$ *there exists a model* $\mathcal{J} \in \mathsf{Circ}_{\mathsf{fix}}(\mathcal{KB})$ *such that (i)* $\Delta^{\mathcal{J}} \subseteq \Delta^{\mathcal{I}}$, *(ii)* $x \in C^{\mathcal{J}} \setminus D^{\mathcal{J}}$ *(iii)* $|\Delta^{\mathcal{J}}|$ *is* $O((|\mathcal{KB}| + |C|)^d)$ *where* $d = depth(D)$.

**Proof.** Define the "small" model $\mathcal{J}$ as in the proof of Lemma 7.12, using the new definition of support. Regarding the size of $\mathcal{J}$ and the fact that it is a classical model of $\mathcal{KB}$, the same arguments as in the proof of Lemma 7.12 apply. In particular, it holds $|\Delta^{\mathcal{J}}| = O((|\mathcal{KB}| + |C|)^d)$.

It remains to show that $\mathcal{J}$ is $<_{\mathsf{fix}}$-minimal. Assume by contradiction that for some interpretation $\mathcal{J}'$, it holds $\mathcal{J}' <_{\mathsf{fix}} \mathcal{J}$; as usual, we show that there exists $\mathcal{I}'$, such that $\mathcal{I}' <_{\mathsf{fix}} \mathcal{I}$. Let $\mathcal{I}'$ be defined as follows: $\Delta^{\mathcal{I}'} = \Delta^{\mathcal{I}}$; $a^{\mathcal{I}'} = a^{\mathcal{I}}$, for all $a \in \mathsf{N}_\mathsf{I}$; $A^{\mathcal{I}'} = A^{\mathcal{I}}$, for all $A \in \mathsf{N}_\mathsf{C}$; $P^{\mathcal{I}'}$ is the minimal set such that:

- $P^{\mathcal{J}'} \subseteq P^{\mathcal{I}'}$, and

- for all $z \in \Delta^{\mathcal{I}'} \setminus \Delta^{\mathcal{J}'}$, for all $y \in \Delta^{\mathcal{J}'}$ and for all $\exists P.H \in \mathsf{cl}(\mathcal{KB})$ such that $z \in \mathsf{supp}_{\mathcal{I}}(\exists P.H)$, if $y \in H^{\mathcal{I}'}$, then $(z, y) \in P^{\mathcal{I}'}$ ($\exists P$ can be seen as a special case where $H = \top$).

First, we prove that $\mathcal{I}'$ is a classical model of $\mathcal{KB}$. In particular, it suffices to show that classical inclusions are satisfied in $\Delta^{\mathcal{I}'} \setminus \Delta^{\mathcal{J}'}$. Given a classical inclusion $C_1 \sqsubseteq D_1$ of $\mathcal{KB}$, assume that $z \in C_1^{\mathcal{I}'}$, and recall that $C_1$ is of the type

$$A_1 \sqcap \ldots \sqcap A_n \sqcap \exists R_1 \sqcap \ldots \sqcap \exists R_m.$$

It suffices to show that there exists an individual $w \in \Delta^{\mathcal{J}'}$ that satisfies $C_1$ as well. By construction, for all $\exists R$ occurring in $C_1$, $z \in \mathsf{supp}_{\mathcal{I}}(\exists R)$, therefore $z \in \mathsf{supp}_{\mathcal{I}}(C_1)$. This means that there exists a witness $w \in \mathsf{supp}_{\mathcal{I}}(C_1)$ in $\Delta^{\mathcal{J}}$. Since for each concept $E$, $w \in E^{\mathcal{I}}$ implies $w \in E^{\mathcal{J}}$, it follows that $w \in \mathsf{supp}_{\mathcal{J}}(C_1)$. Since the priority relation is empty, for each DI $\delta$ in $\mathcal{K}_D$ it holds $\mathsf{sat}_{\mathcal{J}}(\delta) \subseteq \mathsf{sat}_{\mathcal{J}'}(\delta)$. As a consequence, for each concept $E$ it holds $\mathsf{supp}_{\mathcal{J}}(E) \subseteq \mathsf{supp}_{\mathcal{J}'}(E)$. In particular, $w \in \operatorname{supp}_{\mathcal{J}'}(C_1)$ and hence $w \in C_1^{\mathcal{J}'}$ and $w \in D_1^{\mathcal{J}'}$. By construction, we obtain that $z \in D_1^{\mathcal{I}'}$.

It remains to prove that $\mathcal{I}'$ improves $\mathcal{I}$ according to $<_{\mathsf{fix}}$. Let $\delta = (C_1 \sqsubseteq_n D_1)$ be a defeasible inclusion in $\mathcal{KB}$. Since $\mathcal{J}' <_{\mathsf{fix}} \mathcal{J}$, it suffices to prove that for all $z \in \Delta^{\mathcal{I}'} \setminus \Delta^{\mathcal{J}'}$, if $z \in \mathsf{sat}_{\mathcal{I}}(\delta)$ then $z \in \mathsf{sat}_{\mathcal{I}'}(\delta)$.

Suppose first that $z$ vacuously satisfies $\delta$ in $\mathcal{I}$ (i.e., it violates $C_1$). Atomic concepts have the same extension in $\mathcal{I}$ and $\mathcal{I}'$, and $z$ satisfies an unqualified existential in $\mathcal{I}'$ only if it satisfies the same existential in $\mathcal{I}$. Hence, $z$ vacuously satisfies $\delta$ in $\mathcal{I}'$ as well.

Suppose instead that $z$ actively satisfies $\delta$ in $\mathcal{I}$. If $z \notin \operatorname{supp}_{\mathcal{I}}(C_1)$, then $z \notin C_1^{\mathcal{I}'}$, and so $z$ vacuously satisfies $\delta$ in $\mathcal{I}'$. Otherwise, $z \in \operatorname{supp}_{\mathcal{I}}(C_1)$ and $z \in \operatorname{supp}_{\mathcal{I}}(D_1)$. By construction, there is a witness $w \in \Delta^{\mathcal{J}'}$ such that $w \in \operatorname{supp}_{\mathcal{J}}(D_1) \subseteq \operatorname{supp}_{\mathcal{J}'}(D_1)$. This implies that $w \in D_1^{\mathcal{J}'}$ and, considering the construction of $P^{\mathcal{I}'}$, $z \in D_1^{\mathcal{J}'}$. Therefore, $z \in \mathsf{sat}_{\mathcal{J}'}(\delta)$ and we obtain the thesis. □